\documentclass[review,sort]{elsarticle}
\usepackage{lineno}
\usepackage{subcaption}
\usepackage{amsmath, amssymb, latexsym}
\usepackage{mathrsfs}
\usepackage[scr=boondox]{mathalfa}
\usepackage{scalerel}
\usepackage{babel}
\usepackage[hidelinks]{hyperref}
\usepackage{xurl}
\usepackage{paralist}
\usepackage{mathtools}
\usepackage{siunitx}
\usepackage{threeparttable}
\usepackage{multirow}

\usepackage{booktabs}
\usepackage{makecell}
\usepackage{siunitx}
\usepackage{tabu}
\usepackage{float}
\usepackage{enumitem}
\usepackage{placeins}
\usepackage{algorithm}
\usepackage{algpseudocode}

\usepackage[nolist]{acronym}
\begin{acronym}
	\acro{PDE}[PDE]{partial differential equation}
	\acro{FE}[FE]{finite element}
	\acro{FEM}[FEM]{\ac{FE} method}
	\acro{RB}[RB]{reduced basis}
	\acro{CRB}[CRB]{certified \ac{RB}}
	\acro{EIM}[EIM]{empirical interpolation method}
	\acro{SCM}[SCM]{successive constraint method}
	\acro{SCRBE}[SCRBE]{static condensation reduced basis element}
	\acro{SC}[SC]{static condensation}
	\acro{PR}[PR]{port reduction}
	\acro{PR-SCRBE}[PR-SCRBE]{\ac{PR}-\ac{SCRBE}}
	\acro{PR-RB}[PR-RB]{port-reduced \ac{RB}}
	\acro{DOF}{degrees of freedom}
	\acro{ML}{machine learning}
	\acro{POD}{proper orthogonal decomposition}
	\acro{ROM}{reduced-order modelling}
	\acro{RSM}{response surface methodology}
	\acro{CAD}{computer-aided design}
	\acro{MA}{manifold-alignment}
	\acro{CRM}{common research model}
	\acro{RSM}{response surface methodology}
	\acro{NN}{neural network}
	\acro{RBF}{radial basis function}
	\acro{MA}{manifold-alignment}
	\acro{AC}{archetype component}
	\acro{IC}{instantiated component}
	\acro{MDO}{multidisciplinary design optimization}
	\acro{MUMPS}{MUltifrontal Massively Parallel sparse direct Solver}
    \acro{GPR}{Gaussian process regression}
    \acro{GP}{Gaussian process}
    \acro{QoIs}{quantities of interest}
\end{acronym}
\journal{arXiv}

\begin{document}
\begin{frontmatter}

\title{Molecular Dynamics Investigation of Mass Transport During Evaporation for the Binary System of n-Dodecane and Nitrogen}
\author[1]{Suman Chakraborty \fnref{fn1}}
\ead{chakra30@purdue.edu}

\author[2]{Bongseok Kim \fnref{fn1}}
\ead{kim4853@purdue.edu}

\fntext[fn1]{These two authors contributed equally to this work.}

\author[1,2]{Li Qiao\corref{cor}}
\ead{lqiao@purdue.edu}

\address[1]{School of Aeronautics and Astronautics, Purdue University, West Lafayette}
\address[2]{School of Mechanical Engineering, Purdue University, West Lafayette}

\cortext[cor]{Corresponding author}
	
\begin{abstract}
The study of interfacial fluxes under evaporative or condensation processes are ubiquitous in thermal systems, propulsion devices, and many other engineering applications. Most continuum scale models fail to capture the true nature of thermodynamic property variation across the interface, particularly under high-temperature and high-pressure conditions. An improvement over the sharp interface assumption of such continuum scale models is the consideration of a diffused interface and using Kinetic Boundary Conditions (KBCs) to model the mass-transport across the liquid vapor interface. Prior studies on KBCs mainly address monoatomic fluids. Two of the main ingredients required to form KBCs are: density and mass flux. Here, we study a Type-III binary mixture of n-dodecane and nitrogen using non-equilibrium molecular dynamics at near-critical temperatures. Interfacial properties such as thickness, density gradient, and surface tension were analyzed. 
A key result is the temporal evolution of the evaporation and reflected mass fluxes across the vapor--liquid interface. 
We observe that both the evaporation and reflection fluxes increase with increasing temperature, 
indicating enhanced molecular activity and mass transport across the interface at higher $T_r$.
\pagebreak
In contrast, the evaporation coefficient $\alpha_{\mathrm{evap}}$ decreases from about $\alpha \approx 0.978$ at $T_r = 0.70$ to $\alpha \approx 0.905$ at $T_r = 0.95$ 
because the reflected-out flux increases along with the evaporation flux, which reduces the net efficiency of molecular evaporation across the interface.
To the authors' knowledge, this is one of the very few studies estimating mass transport coefficients for Type~III binary systems, laying the foundation for KBCs in hydrocarbon/nitrogen mixtures.
\end{abstract}
\end{frontmatter}

\section{Introduction}\label{sec:Introduction}
Accurate prediction of evaporation and condensation phenomena necessitates the coupled treatment of heat and mass transport across the vapor–liquid interface. 
The coupled simulations involving interfaces play a key role in numerous engineering applications, including diesel and liquid rocket engine combustion, drying~\cite{Ambrosio2011,Fang2019}, distillation, and extraction~\cite{Wang2001,Tadano2015}.
In the context of methodologies handling the interface systems, standard macroscopic models typically treat the interface as a sharp, two-dimensional boundary with a discontinuous density profile, employing a coupled mass and heat transport framework~\cite{Wang2001}. These models further assume local thermodynamic equilibrium, implying the absence of temperature and chemical potential jumps across the interface. However, these assumptions begin to fail under high pressure and temperature (P–T) conditions, particularly near the mixture critical point.

\begin{figure}[hbt!]
	\centering 
	\begin{subfigure}[H]{0.8\linewidth}
		\includegraphics[width=\linewidth]{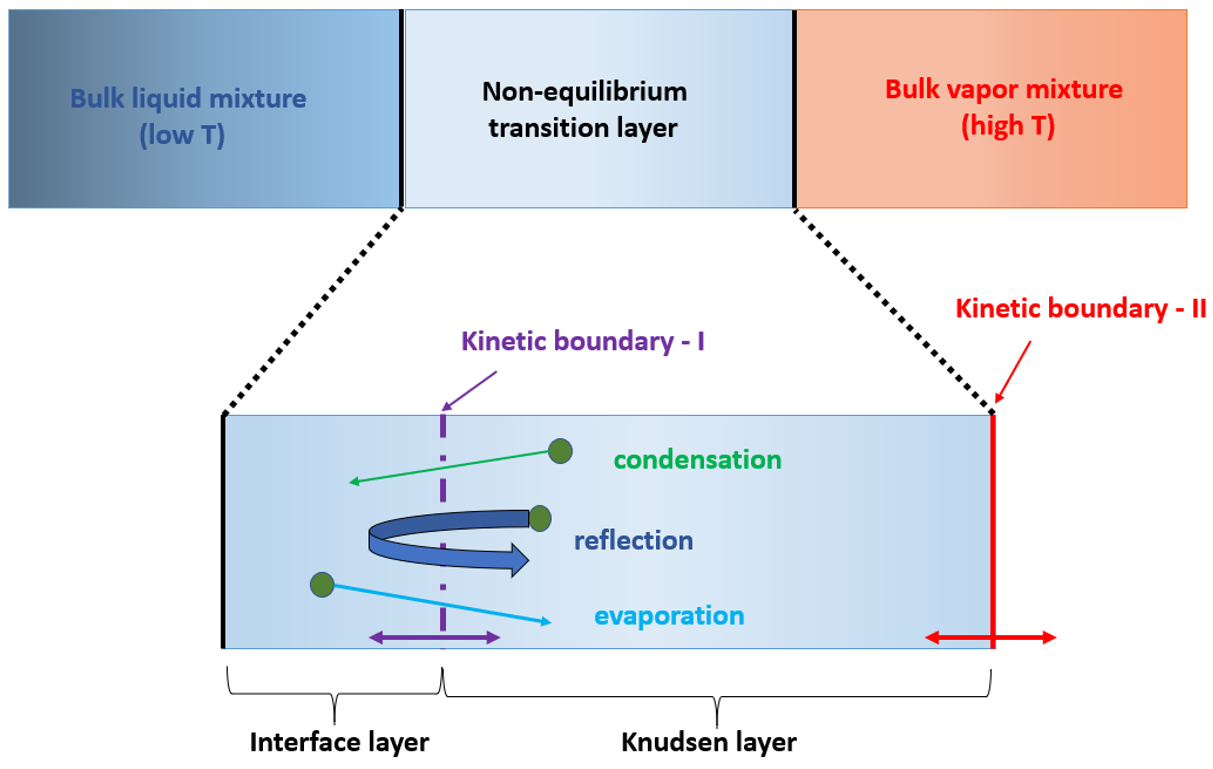}
	\end{subfigure}
	\caption{
Schematic representation of the vapor--liquid interfacial structure, showing the interfacial layer, the Knudsen layer, and the corresponding kinetic boundaries. Kinetic boundary~I marks the transition from the interfacial region to the Knudsen layer, where molecular-level mass transport is dominant, while kinetic boundary~II denotes the interface with the vapor bulk region.
}
	\label{fig:schematic_knudsen}
\end{figure}

To address this limitation, Langmuir~\cite{Langmuir1918} introduced the concept of the Knudsen layer as a corrective mechanism for evaporation and condensation models. The Knudsen layer is a thin region on the order of a few molecular mean free paths adjacent to the bulk liquid, where molecular collisions dominate transport and classical diffusion-based descriptions no longer apply.
The complex transport mechanisms within this nonequilibrium transition layer between the bulk liquid and vapor phases can be effectively handled by introducing kinetic boundary conditions (KBCs), which are commonly classified as KBC-I and KBC-II. As illustrated in Fig.~\ref{fig:schematic_knudsen}, KBC-I denotes the boundary separating the interfacial region from the Knudsen layer.

As interfacial transport is inherently a nanoscale phenomenon, a molecular-level perspective may offer accurate physical quantities of the non-equilibrium transition layer. Several studies~\cite{Ishiyama2004,Yasuoka1994,Kon2006,Watanabe2004,Ishiyama2005} have employed molecular dynamics (MD) simulations to investigate vapor–liquid two-phase systems and have proposed boundary conditions at the KBC-I interface. The Boltzmann equation, which governs molecular kinetics within the Knudsen layer, can be coupled with KBC-I to derive KBC-II. 
A representative formulation of the kinetic boundary condition is expressed as

\begin{linenomath*}
\begin{equation}
    f_{\text{out}} = \left[ \alpha_{\text{evap}} \rho(T_L) + (1 - \alpha_{\text{cond}}) \sigma \right] \hat{f}, \quad \xi_x > 0.
\end{equation}
\end{linenomath*}

Here, $f_{\text{out}}$ is the velocity distribution function of molecules traveling from the interface layer to the Knudsen layer. The term $\rho(T_L)$ denotes the saturated vapor density as a function of the bulk liquid temperature $T_L$. The coefficients $\alpha_{\text{evap}}$ and $\alpha_{\text{cond}}$ represent the evaporation and condensation coefficients, respectively. The quantity $\xi_x$ is the velocity component normal to the interface, and $\xi_x > 0$ indicates molecular motion directed toward the Knudsen layer. The parameter $\sigma$ is associated with the velocity distribution of molecules moving from the Knudsen layer to the interface. The normalized Maxwellian distribution $\hat{f}$ at temperature $T_L$ is defined as

\begin{linenomath*}
\begin{equation}
    \hat{f}(T_L) = \frac{1}{(2\pi R T_L)^{3/2}} \exp\left(-\frac{\xi_i^2}{2 R T_L}\right),
\end{equation}
\end{linenomath*}

where $R$ is the specific gas constant, and $\xi_i$ is the molecular velocity vector in three dimensions.

While most existing studies have focused on vapor--liquid interface systems under equilibrium conditions or on simple monoatomic fluids such as argon and helium, long chain hydrocarbons such as n-dodecane are widely used in propellant applications due to their high energy density, low vapor pressure, and stable liquid phase behavior under high pressure conditions~\cite{boust2020}. 
Recent equilibrium MD studies of alkane and N$_2$ mixtures have highlighted the role of cross-interactions in reproducing interfacial properties such as surface tension~\cite{Morrow2022}, but their scope remains limited to Therefore, despite their practical relevance, KBCs for these complex fluids have not yet been investigated beyond equilibrium conditions.


To address this limitation, we investigate the non equilibrium interfacial behavior of \textit{n}-dodecane in a binary mixture with nitrogen using molecular dynamics simulations. 
Previous studies by the authors ~\cite{Mo_Qiao_2017_sub_super_MD, CHAKRABORTY2019118629, CHAKRABORTY2021}  have explored the equilibrium and non-equilibrium dynamics of different \textit{n}-alkanes/nitrogen systems, developing a better understanding of interfacial properties under near-critical conditions at molecular scales.
Wei et al.~\cite{Wei_et_al_2022_MD_sub_super_vaporization} further examined sub and supercritical fuel vaporization and mixing characteristics, highlighting temperature dependent structural changes in the interfacial region. 
More recently, Zhang et al.~\cite{Zhang_et_al_2024_MD_sub_super_phase_change_alcohol_nheptane} investigated sub and supercritical phase transitions in alcohol and \textit{n}-heptane mixtures, emphasizing the strong coupling between density gradient, interfacial thickness, and molecular diffusion. 
In this work, we focus on the non equilibrium regime of an \textit{n}-dodecane/nitrogen system undergoing net evaporation, where interfacial properties including interface thickness, density gradient, surface tension, and liquid core length are analyzed over time as functions of ambient temperature. 
The primary objective is to quantify the interfacial molecular mass flux, for which two distinct approaches are employed to estimate the evaporation flux.
The corresponding evaporation coefficients are presented as temperature dependent quantities.

In contrast to previous approaches, this study offers several key contributions, as summarized below:
\begin{enumerate}
\item This research investigates the interfacial behavior of an n-dodecane and nitrogen binary system under high-temperature and high-pressure conditions.
\item This research proposes a post-processing algorithm for evaluating evaporation, condensation, and reflection fluxes with high robustness against noise in nonequilibrium MD data.
\item This research employs Gaussian Process Regression (GPR) to effectively handle noisy NEMD data and quantify uncertainty through 95\% confidence intervals, providing a statistical measure of the QoIs variability.
\item This research proposes a data-driven evaporation coefficient model that can facilitate computational fluid dynamics (CFD) simulations by providing high-fidelity, physically consistent boundary conditions for interfacial problems.
\end{enumerate}

Following the introduction, the remainder of this paper is organized as follows. Section~2 outlines the molecular dynamics formulation and simulation setup, including the domain configuration and conditions for the NEMD simulations. Section~3 details the methodology for computing molecular mass flux, introducing two approaches: the fixed boundary method and the two-boundary method. Section~4 presents the results and discussion, where Sec.~4.1 examines microscopic interfacial properties and Sec.~4.2 analyzes interfacial molecular mass transport using both flux-evaluation approaches. Finally, Sec.~5 summarizes the main findings and concludes the study.

\section{Molecular dynamics simulation}
\subsection{Molecular dynamics}
Molecular dynamics (MD) simulations are widely utilized to investigate the time evolution of particle systems across various scientific disciplines, including fluid mechanics~\cite{chio2025fluid}, materials science~\cite{steinhauser2009review}, chemistry~\cite{islam2025nanomaterials}, and biophysics~\cite{hollingsworth2018md}.
In MD, the motion of each particle is governed by Newton’s second law:
\begin{equation}
    m_i \frac{d^2 \mathbf{x}_i}{dt^2} = \mathbf{F}_i,
\end{equation}
where \(m_i\) and \(\mathbf{x}_i\) denote the mass and position of the \(i\)-th particle, and \(\mathbf{F}_i\) represents the net force acting on it. 
Given a potential energy function \(U(\mathbf{x}_i, \mathbf{x}_j)\), the total potential experienced by particle \(i\) and its corresponding force are expressed as
\begin{align}
    U_i &= \sum_{\substack{j=1 \\ j \neq i}}^{N} U(\mathbf{x}_i, \mathbf{x}_j), \\
    \mathbf{F}_i &= -\nabla_{\mathbf{x}_i} U_i.
\end{align}

The temporal evolution of the system is obtained through numerical integration of the equations of motion.
Among the available integration schemes, the velocity Verlet algorithm is one of the most robust and computationally efficient methods for MD simulations. 
This scheme updates particle positions and velocities at each time step \(\Delta t\) according to:
\begin{align}
    \mathbf{x}(t+\Delta t) &= \mathbf{x}(t) + \mathbf{v}(t)\Delta t + \frac{\mathbf{F}(t)}{2m}\Delta t^2, \\
    \mathbf{v}(t+\Delta t) &= \mathbf{v}(t) + \frac{\mathbf{F}(t+\Delta t) + \mathbf{F}(t)}{2m}\Delta t.
\end{align}
Here, \(\mathbf{x}\) and \(\mathbf{v}\) denote the particle position and velocity, respectively. 
This symplectic integrator conserves energy over long simulations and requires only the current positions and forces, making it both memory-efficient and easy to implement.
Algorithm~\ref{alg:velocity_verlet} summarizes the procedure of the velocity Verlet method, 
which is widely implemented in major MD frameworks such as Desmond~\cite{bowers2006desmond}, 
GROMACS~\cite{abraham2015gromacs}, LAMMPS~\cite{plimpton1995lammps,thompson2022lammps}, 
NAMD~\cite{phillips2005namd}, and OpenMM~\cite{eastman2017openmm}. 


\begin{algorithm}[h!]
\caption{Velocity Verlet Algorithm}
\label{alg:velocity_verlet}
\begin{algorithmic}[1]
\small
\State \textbf{Input:} Initial positions $\{\mathbf{x}_i\}_{i=1}^{N}$, velocities $\{\mathbf{v}_i\}_{i=1}^{N}$
\State \textbf{Parameters:} Potential $U$, mass $m$, time step $\Delta t$, total steps $M$
\State \textbf{Output:} Updated positions and velocities after $M$ iterations
\State (1) Compute initial forces $\mathbf{F}_i = -\nabla_{\mathbf{x}_i} U_i$ for all $i$
\State (2) Repeat the following steps for each time step $t = 1,2,\ldots,M$:
\begin{enumerate}[label*=\arabic*. , leftmargin=3em, itemsep=2pt]
    \item \textit{Position update:} 
    $\displaystyle \mathbf{x}_i \gets \mathbf{x}_i + \mathbf{v}_i \Delta t + \frac{\mathbf{F}_i}{2m}\Delta t^2$
    \item \textit{Half-step velocity:} 
    $\displaystyle \mathbf{v}_i \gets \mathbf{v}_i + \frac{\mathbf{F}_i}{2m}\Delta t$
    \item \textit{Force update:} 
    Recompute $\displaystyle \mathbf{F}_i = -\nabla_{\mathbf{x}_i} U_i$ using the updated $\mathbf{x}_i$
    \item \textit{Velocity completion:} 
    $\displaystyle \mathbf{v}_i \gets \mathbf{v}_i + \frac{\mathbf{F}_i}{2m}\Delta t$
\end{enumerate}
\State (3) \textbf{Return:} $\{\mathbf{x}_i\}_{i=1}^{N},\ \{\mathbf{v}_i\}_{i=1}^{N}$
\end{algorithmic}
\end{algorithm}

In this study, the equations of motion were integrated using the velocity Verlet scheme 
implemented in the \textsc{LAMMPS} package~\cite{plimpton1995lammps,thompson2022lammps}. 
The Verlet-type schemes preserve the time-reversibility and symplectic properties of Hamiltonian dynamics, which makes them suitable for long-term molecular simulations. 
These integrators are thus standard in modern MD codes, ensuring numerical stability while accurately capturing interatomic motion and potential energy evolution~\cite{cai2022jcp}.

Intermolecular interactions were modeled using two empirical potential models suited for the molecular characteristics of the working fluids. 
n-Dodecane was represented by the united-atom SKS potential of Smit \textit{et al.}~\cite{Smit1995SKS}, 
in which each CH$_3$ and CH$_2$ group is treated as a single interaction site. 
Bonded interactions are described using harmonic bond, angle, and torsional potentials, 
while nonbonded interactions follow the Lennard--Jones form. 
This coarse-grained representation reproduces the conformational flexibility and thermophysical behavior of long-chain alkanes with significantly reduced computational cost. 
Nitrogen was modeled using the two-center Lennard--Jones (2CLJ) potential developed by Rivera~\cite{Rivera1999N2}, 
where two LJ sites separated by the experimental bond length capture the anisotropy and rotational dynamics of the diatomic molecule. 
The combination of the SKS model for alkanes and the 2CLJ model for N$_2$ has been widely validated for simulations of vapor--liquid hydrocarbon--nitrogen systems.

\subsection{Gaussian process regression}
In this section, we introduce \ac{GPR} to quantify the statistical variability of \ac{QoIs} obtained from non-equilibrium MD simulations.
Because the non-equilibrium states exhibit temporal oscillations, we use a fitting technique to quantitatively describe their statistical evolution over time.
Specifically, we first calculate the \ac{QoIs}, such as interfacial mass fluxes, interface thickness, and liquid core length, and then apply \ac{GPR} to quantify their temporal statistical variation.

The \ac{GP} is defined as a collection of random variables whose any finite subset follows a joint Gaussian distribution.
The joint Gaussian distribution is characterized by its mean function \(m(x)\) and covariance function \(\kappa(x,x')\),
and we write \(f \sim \mathcal{GP}(m,\kappa)\) to denote that the function \(f(x)\) is distributed as a \ac{GP}.
In regression problems, the objective is to approximate an unknown function from a set of input–output data.
\ac{GPR} achieves this by assuming that the data are realizations of a \ac{GP}
and then conditioning on the observations to obtain a posterior \ac{GP}, which serves as the predictive model.

Let the training dataset be
\(\mathcal{D}=\{(x_i, y_i)\,|\, i=1,2,\dots,n_s\}\),
where each input \(x_i \in \mathcal{P}\subset\mathbb{R}\)
and each observation \(y_i\in\mathbb{R}\).
The underlying latent function \(f:\mathcal{P}\rightarrow\mathbb{R}\)
is modeled as a Gaussian process with additive Gaussian noise:
\begin{equation}
    y = f(x) + \epsilon,
    \qquad
    \epsilon \sim \mathcal{N}(0,\sigma^2),
    \qquad
    f(x) \sim \mathcal{GP}\big(m(x),\,\kappa(x,x')\big).
\end{equation}
Here, \(m(x)\) is the mean function, often chosen as a linear combination of basis functions
\(\{H_k(x)\}_{k=1}^K\),
for example \(m(x)=\theta_1+\theta_2x\).
A widely used covariance kernel is the squared exponential kernel:
\begin{equation}
    \kappa(x,x')
    = \sigma_f^2
      \exp\!\left(
        -\frac{(x - x')^2}{2\ell^2}
      \right),
\end{equation}
where \(\ell\) represents the correlation length and \(\sigma_f^2\) is the signal variance.

Given the training inputs
\(\mathbf{X}=[x_1,\dots,x_{n_s}]^\top\)
and outputs
\(\mathbf{y}=[y_1,\dots,y_{n_s}]^\top\),
the joint distribution of the observed outputs \(\mathbf{y}\)
and the test outputs \(\mathbf{y}_*\) at new locations \(\mathbf{X}_*\)
is multivariate Gaussian:
\begin{equation}
\begin{bmatrix}
\mathbf{y}\\[2pt]
\mathbf{y}_*
\end{bmatrix}
\sim
\mathcal{N}\!\left(
\begin{bmatrix}
m(\mathbf{X})\\[2pt]
m(\mathbf{X}_*)
\end{bmatrix},
\begin{bmatrix}
K_y & \kappa(\mathbf{X},\mathbf{X}_*)\\[2pt]
\kappa(\mathbf{X}_*,\mathbf{X}) & \kappa(\mathbf{X}_*,\mathbf{X}_*)
\end{bmatrix}
\right),
\end{equation}
where \(K_y = \kappa(\mathbf{X},\mathbf{X}) + \sigma^2 I_{n_s}\)
accounts for measurement noise (in the noise-free case, \(\sigma^2=0\)).

Conditioning on the training data yields the posterior predictive distribution:
\begin{equation}
\mathbf{y}_*\,|\,\mathbf{y}
\sim
\mathcal{N}\!\big(
m_*(\mathbf{X}_*),\,\kappa_*(\mathbf{X}_*,\mathbf{X}_*)\big),
\end{equation}
with
\begin{align}
m_*(\mathbf{X}_*) &=
m(\mathbf{X}_*)
+ \kappa(\mathbf{X}_*,\mathbf{X})
\,K_y^{-1}\!\left(\mathbf{y}-m(\mathbf{X})\right),
\\
\kappa_*(\mathbf{X}_*,\mathbf{X}_*)
&=
\kappa(\mathbf{X}_*,\mathbf{X}_*)
- \kappa(\mathbf{X}_*,\mathbf{X})
\,K_y^{-1}\!
\kappa(\mathbf{X},\mathbf{X}_*).
\end{align}
The posterior variance \(\kappa_*\) quantifies the uncertainty in the prediction
and is always smaller than the prior variance due to the conditioning on data.

The model hyperparameters
\(\boldsymbol{\theta}=(\theta_1,\theta_2,\ell,\sigma,\sigma_f)\)
are obtained by maximizing the log marginal likelihood:
\begin{equation}
\log p(\mathbf{y}\,|\,\mathbf{X},\boldsymbol{\theta})
=
-\frac{1}{2}
(\mathbf{y}-m(\mathbf{X}))^\top
K_y^{-1}(\mathbf{y}-m(\mathbf{X}))
-\frac{1}{2}\log|K_y|
-\frac{n_s}{2}\log(2\pi).
\end{equation}
This Bayesian regression framework not only provides the mean prediction as a smooth interpolant
but also yields principled uncertainty estimates through the posterior covariance.
For more details on \ac{GPR}, readers are referred to~\cite{rasmussen2006gaussian, rasmussen2003lectures, sullivan2015uq}.

\subsection{MD Simulation configuration}
\label{sec2:simul_config}
\begin{figure}[hbt!]
	\centering 
    	\begin{subfigure}[H]{0.95\linewidth}
		\includegraphics[width=\linewidth]{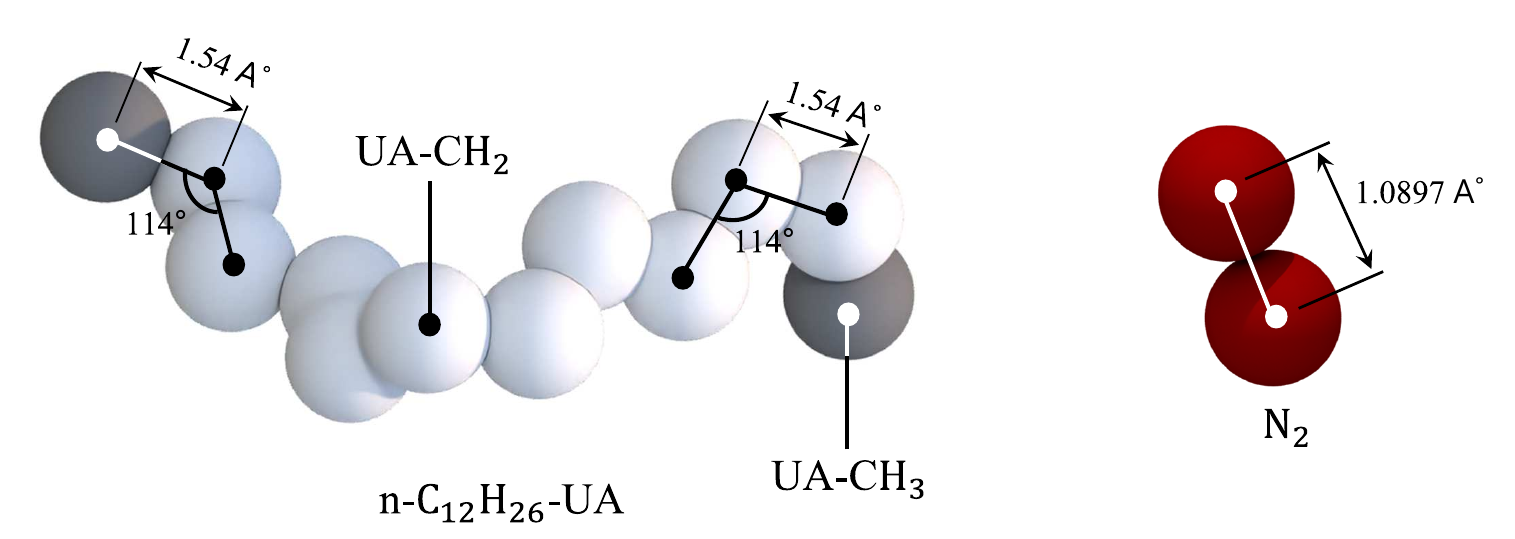}
        \subcaption{Atomistic configuration of liquid \textit{n}-dodecane and nitrogen gas.}
	\end{subfigure} \\[2mm]
	\begin{subfigure}[H]{0.95\linewidth}
		\includegraphics[width=\linewidth]{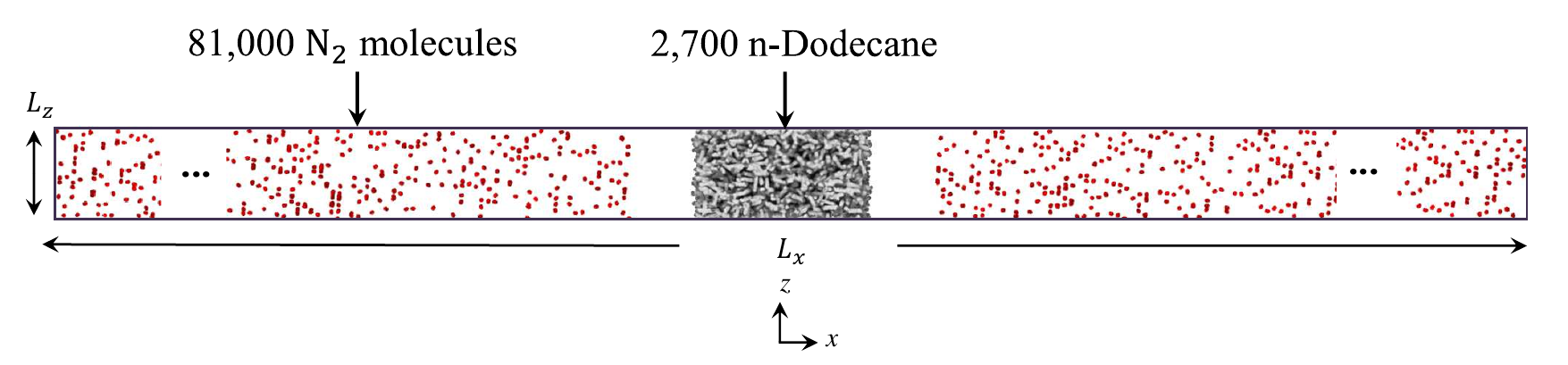}
        \subcaption{Front view of the simulation setup.}
	\end{subfigure}
	\caption{
Atomistic configuration and schematic of the simulation domain used for the NEMD calculations.
The liquid phase consists of \textit{n}-dodecane molecules modeled with the SKS united-atom potential, 
while nitrogen is represented using the 2CLJ potential.
A liquid slab is placed at the center of the domain with nitrogen gas on both sides.
The simulation box has lateral dimensions of $L_x = L_y = 8~\mathrm{nm}$ and periodic boundary conditions are applied in all directions.
The gas regions are thermally regulated to drive evaporation from the liquid surface under controlled non-equilibrium conditions.
}
	\label{fig:simulation_config}
\end{figure}

The computational domain is rectangular and employs periodic boundary conditions in all directions. A liquid slab of n-dodecane is placed at the center of the domain, flanked by nitrogen gas on both sides. The cross-sectional area of the simulation box is fixed at $L_y \times L_z = 8\ \text{nm} \times 8\ \text{nm}$ across all cases. The liquid slab comprises 2700 n-dodecane molecules, initialized at a temperature of 363~K.

To avoid a significant pressure rise caused by evaporated n-dodecane molecules, a sufficiently large number of nitrogen molecules is used in the ambient region. An atom ratio of 5:1 between the ambient and liquid core atoms is maintained throughout all simulations, resulting in 81,000 nitrogen molecules in the surrounding vapor phase. The ambient pressure and temperature are held constant during each simulation.

For minimization of direct thermal effects on liquid–gas interfacial dynamics, the ambient region is heated at a distance of $50\sigma$ from the liquid interface, where $\sigma$ is the molecular diameter of nitrogen~\cite{sengers2000equations}. The ambient temperature is varied to examine its influence on evaporation and interfacial transport fluxes. As a result, the total domain length ranges from 8,158~nm to 10,458~nm depending on the thermal conditions considered.

\begin{table}[hbtp!]
\renewcommand{\arraystretch}{1.0}
\centering
\begin{threeparttable}
\caption{Summary of molecular dynamics system setup and simulation parameters.}
\begin{tabular}{l c}
\hline
Item & Value \\
\hline
Number of n-dodecane molecules   & 2,700 \\
Number of nitrogen molecules     & 81,000 \\
Total number of molecules        & 83,700 \\
Cross-sectional area             & $L_y \times L_z = 8~\text{nm} \times 8~\text{nm}$ \\
Domain length in $x$-direction   & 8,158--10,458 nm \\
MD timestep                      & 2 fs \\
Total simulation length          & 50 ns (25,000,000 steps) \\
\hline
\end{tabular}

\end{threeparttable}
\label{table:md_setup}
\end{table}

\section{Flux evaluation methodology}
\label{sec3:flux_computation}
In this section, we introduce two methods for computing molecular mass flux: the fixed boundary method and the two-boundary method. The former estimates net flux from changes in molecular count within a fixed region, while the latter identifies microscopic events such as evaporation and reflection near the interface. By employing both, we enable cross-validation and capture complementary aspects of interfacial transport, balancing robustness and detailed resolution in non-equilibrium molecular dynamics simulations.

\subsection{Fixed boundary method}
\label{sec: fixed boundary method}

\begin{figure}[hbt!]
	\centering 
	\begin{subfigure}[H]{0.8\linewidth}
		\includegraphics[width=\linewidth]{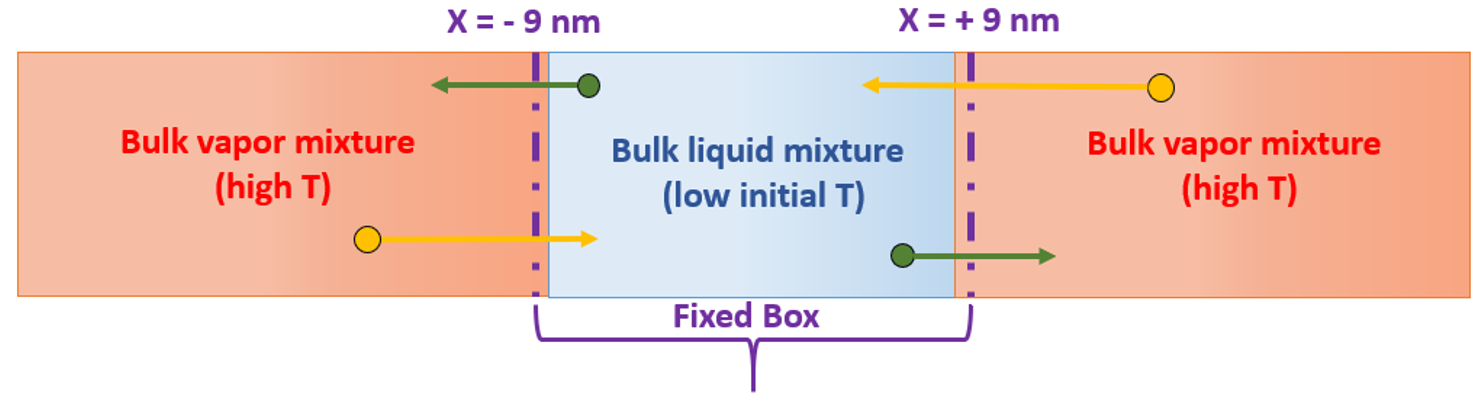}
	\end{subfigure}
	\caption{
Schematic of the computational domain for the fixed boundary method. 
The fixed box enclosing the initial liquid slab is used to track the number of molecules over time, 
allowing evaluation of the net evaporation flux.
}
	\label{fig:fixed_box}
\end{figure}

To delineate the fixed boundary method, we present a schematic of the full computational domain in Fig.~\ref{fig:fixed_box}.
The central region is occupied by liquid n-alkane, specifically n-dodecane, and hot ambient gas composed of nitrogen is placed on both sides. Two fixed cross-sectional planes define the boundaries of a control region referred to as the “fixed box,” which fully encloses the initial length of the liquid film.
In this study, the liquid film of n-dodecane has an initial length of approximately 17.2~nm in all test cases. Accordingly, the fixed boundaries are located at $x = -9$~nm and $x = +9$~nm. 

The evaporation flux is estimated by analyzing the change in the number of molecules within a fixed control volume at two distinct time instances, using trajectory data obtained from molecular dynamics simulations. First, the molecular species of interest, such as n-dodecane or nitrogen, is selected. The number of molecules of the selected species contained within the fixed control region is then counted at two different time points, denoted by \( t_1 \) and \( t_2 \), yielding \( N_1 \) and \( N_2 \), respectively. The difference between these counts, defined as \( \Delta N_{\text{flux}} = N_1 - N_2 \), represents the net number of molecules that have crossed the control boundaries during the time interval \( \Delta t = t_2 - t_1 \).
The corresponding mass flux is computed as
\begin{equation} \label{eq:j}
    \langle J \rangle = \frac{1}{n_s} \sum_{n_s} \frac{m \, \Delta N_{\text{flux}}}{S \, \Delta t},
\end{equation}
where \( m \) is the molecular mass, \( n_s \) is the number of independent samples, and \( S\) is the cross-sectional area of the control region. This method estimates the net molecular mass flux based on molecular counts alone, without requiring the tracking of individual molecular trajectories. It effectively captures the net transfer of mass across the designated control surface, whether molecules are entering or leaving the volume.

\subsection{Two boundary method}
\label{sec:two boundary method}

\begin{figure}[hbt!]
	\centering 
	\begin{subfigure}[H]{0.8\linewidth}
		\includegraphics[width=\linewidth]{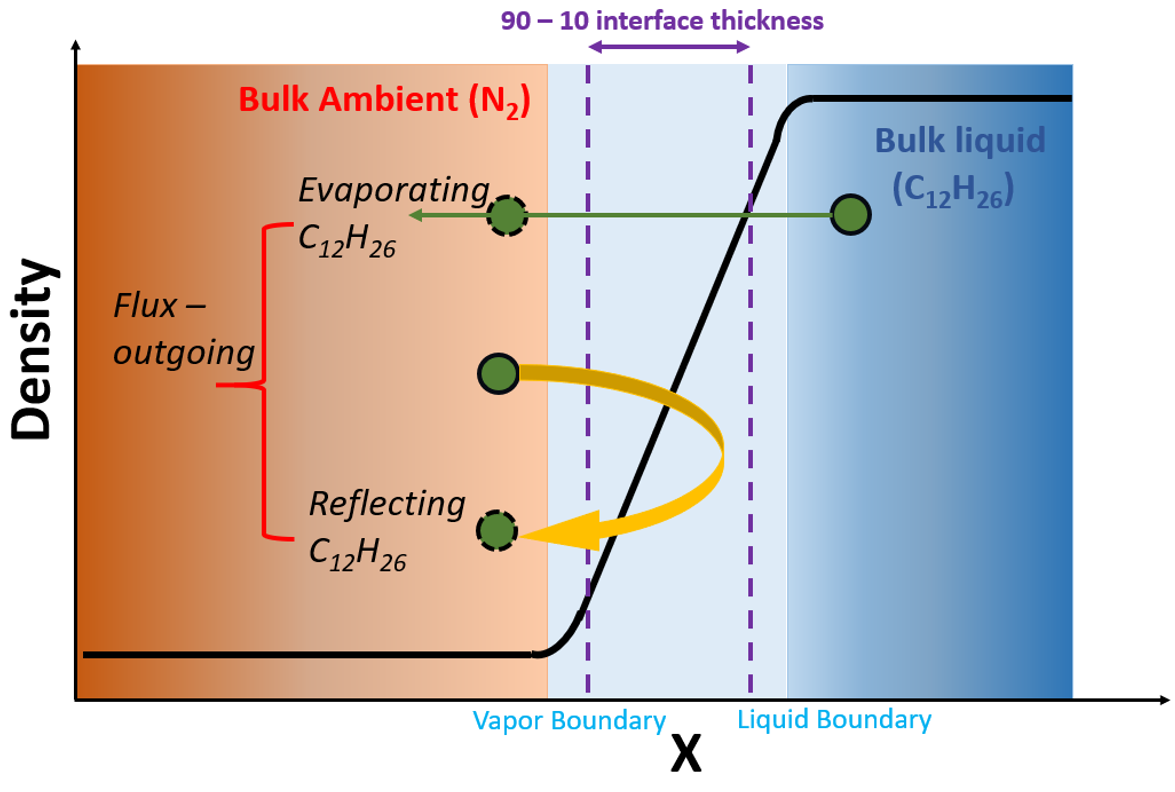}
	\end{subfigure}
	\caption{Schematic of the left vapor--liquid interface region used in the two-boundary method. Two reference planes are defined: one near the vapor side and the other near the liquid core, with the separation corresponding to the interface thickness determined by the 90--10 rule.}
	\label{fig:two_boundaries}
\end{figure}

We present a schematic of the vapor--liquid interface on the left side of the binary system, together with the corresponding density profile in the $x$-direction, in Fig.~\ref{fig:two_boundaries}.
In previous works, the two-boundary method has been previously employed in interfacial transport studies~\cite{kobayashi2016molecular,meland2004nonequilibrium,gu2010molecular}. In this approach, two reference planes are defined: the vapor boundary adjacent to the ambient region and the liquid boundary adjacent to the liquid core.
In this study, the distance between these boundaries is set equal to the interface thickness, calculated using the 90--10 rule~\cite{lekner1977surface}. Applying the two-boundary method to MD trajectory data enables the classification of molecular fluxes across the interface into three categories: evaporation, condensation, and reflection.

This chapter focuses on two primary flux components of n-alkane molecules: those that evaporate into the ambient region and those that reflect off the vapor boundary and return to the ambient. Both processes are illustrated in Fig.~\ref{fig:two_boundaries}. The evaporating flux $J_{\text{evap}}$ and reflecting flux $J_{\text{ref}}$ are evaluated using Eq.~\eqref{eq:j}. Here, $\Delta N_{\text{evap}}$ denotes the number of n-alkane molecules originating from the bulk liquid that cross both boundaries, while $\Delta N_{\text{ref}}$ accounts for ambient-originating molecules that traverse the vapor boundary twice.
The total outgoing molecular mass flux is defined as
\begin{equation} \label{eq:evaporation_coeff}
J_{\text{out}} = J_{\text{evap}} + J_{\text{ref}}.
\end{equation}
The corresponding evaporation coefficient is then calculated as
\begin{equation}
\alpha_{\text{evap}} = \frac{J_{\text{evap}}}{J_{\text{out}}}.
\end{equation}


The evaporation mass flux is evaluated using the procedure in Algorithm~\ref{algorithm_box_evaporation}. 
The molecular dynamics trajectory is post-processed to obtain the density profile along the interface-normal direction. 
From this profile, the bulk liquid and vapor densities are identified, and the 90--10 reference levels are used to locate the left and right vapor--liquid interfaces. 
The region between these interfaces defines the instantaneous liquid region, and the total mass within this region is computed for each snapshot. 
The temporal change of this mass, together with the cross-sectional area $A=L_yL_z$, yields the evaporation mass flux.

\begin{algorithm}[H]
\small
\caption{Evaporation Mass Flux Evaluation using the Two Boundary Method}
\label{algorithm_box_evaporation}
\begin{algorithmic}[1]

\State Set cross-sectional area $A=L_yL_z$.

\For{each snapshot $i$}
    \State Identify vapor--liquid interfaces and define $S(t_i)$.
    \State Compute mass in $S(t_i)$:
    \[
    M_i=\sum_{j:\,x_j\in S(t_i)} m_j.
    \]
\EndFor

\For{$i>0$}
    \State Flux:
    \[
    J_M(t_i)=\frac{M_{i-1}-M_i}{\Delta t\,A}.
    \]
\EndFor

\end{algorithmic}
\end{algorithm}






Condensation of N$_2$ is evaluated by tracking the liquid-region control volume defined at each snapshot. 
Following the interface identification procedure in Algorithm~\ref{algorithm_box_evaporation}, 
the liquid-region interval $S(t)=(x_L(t),x_R(t))$ is determined from the instantaneous density profile. 
A molecule is counted as condensed at time $t$ if it is located outside $S(t-\Delta t)$ and inside $S(t)$. 
The number and total mass of these molecules yield the condensation number flux $J_N$ and mass flux $J_M$ over the interval $\Delta t$.

\begin{algorithm}[H]
\small
\caption{Condensation Mass Flux Evaluation using the Two Boundary Method}
\label{algorithm_box_condensation_n2}
\begin{algorithmic}[1]

\State Set cross-sectional area $A=L_yL_z$.

\For{each snapshot $i$}
    \State Identify vapor--liquid interfaces and define $S(t_i)$.
    \State Compute N$_2$ mass in $S(t_i)$:
    \[
    M_i=\sum_{j:\,x_j\in S(t_i)} m_j.
    \]
\EndFor

\For{$i>0$}
    \State Flux:
    \[
    J_M^{\mathrm{cond}}(t_i)=\frac{M_i-M_{i-1}}{\Delta t\,A}.
    \]
\EndFor

\end{algorithmic}
\end{algorithm}






Now we track the reflection of $n$-dodecane molecules by monitoring domain transitions across five regions, gas(left), interface(left), liquid, interface(right), and gas(right). Following the interface detection procedure consistent with Algorithm~\ref{algorithm_box_evaporation}, a reflection event is recorded when a molecule leaves the left interface (domains 1–2) and subsequently re-enters through the opposite interface (domains 4–5), or vice versa. The total reflected mass is accumulated to evaluate the reflection flux $J_M$ over the given interval.

\begin{algorithm}[htp!]
\small
\caption{Reflection Mass Flux Evaluation using the Two Boundary Method}
\label{algorithm_box_reflection}
\begin{algorithmic}[1]

\State Set cross-sectional area $A=L_yL_z$.
\State Identify vapor--liquid interfaces from the density profile and classify each molecular position
       into vapor-side boundary, liquid boundary, or liquid region.

\State For each C$_{12}$ molecule, construct a domain sequence $d_j(t_n)$ over all snapshots.

\For{each molecule $j$}
    \State Count reflection events by detecting sequences in $d_j(t_n)$
           where a molecule approaches an interface boundary and subsequently returns 
           to the original region.
    \State Record the number of reflection events in time group $k$ as $N_{j,k}$.
\EndFor

\State Group snapshots into blocks of ten consecutive timesteps 
       $\{t_{k,0},\dots,t_{k,9}\}$ and set $\Delta t_k = t_{k,9}-t_{k,0}$.

\For{each time group $k$}
    \State Compute reflection mass flux
    \[
        J_M(t_k)=\frac{\sum_j N_{j,k}\,m_j}{A\,\Delta t_k}.
    \]
\EndFor

\end{algorithmic}
\end{algorithm}

\section{MD simulation and post-processing}
We conducted molecular dynamics simulations for the binary system consisting of n-dodecane and nitrogen under five different ambient temperature conditions. The critical temperature and pressure of n-dodecane are 658.2 K and 18 bar, respectively~\cite{heidemann1980calculation}. In all cases, the initial temperature of the liquid core was maintained at 363 K, while the ambient gas temperature was varied to study the effect of thermal gradients on interfacial transport.

These temperature variations span from subcritical to near-critical regimes, where notable physical transitions occur. As the ambient temperature approaches the critical point of n-dodecane, the density difference between the liquid and vapor phases diminishes, the vapor–liquid interface becomes increasingly diffuse, and the surface tension progressively decreases, eventually approaching zero. This reflects the gradual disappearance of a well-defined interface, leading to a supercritical-like behavior where phase distinction is no longer meaningful. Capturing such phenomena is crucial for understanding evaporation and condensation processes in propulsion applications operating under extreme thermodynamic conditions.

The simulation conditions and the corresponding reduced temperatures are listed in
Table~\ref{table:simulation_cases}. The reduced temperature is defined as
\begin{equation}
    T_r = \frac{T_{\mathrm{ab}}}{T_{\mathrm{c}}}.
\end{equation}

Case~1 \((T_r = 0.70)\) lies in a more subcritical regime and is therefore expected to exhibit 
interfacial behavior distinct from the higher-temperature cases. 
Throughout the remainder of this paper, n-dodecane and nitrogen are referred to as C$_{12}$ 
and N$_2$, respectively.

\begin{table}[hbt!]
    \centering
    \caption{Simulation conditions for different reduced temperatures (\(T_r\)). 
    The critical temperature of n-dodecane is 658.2~K~\cite{NISTWebBookDodecane}.}
    \label{table:simulation_cases}
    \begin{tabular}{c c c}
        \toprule
        Case & Ambient Temperature [K] & Reduced Temperature $T_r$ \\
        \midrule
        Case 1 & 460.7 & 0.70 \\
        Case 2 & 526.6 & 0.80 \\
        Case 3 & 559.5 & 0.85 \\
        Case 4 & 592.4 & 0.90 \\
        Case 5 & 625.3 & 0.95 \\
        \bottomrule
    \end{tabular}
\end{table}

\begin{table}[hbtp!]
\renewcommand{\arraystretch}{1.05}
\centering
\caption{Computation time for each simulation case.
Simulations were performed using two NVIDIA RTX 4090 GPUs in parallel, coupled with 16 CPU cores.}
\begin{tabular}{l c}
\hline
Case & Computation time \\
\hline
Case 1 & 75 hr 12 min 34 s \\
Case 2 & 74 hr 45 min 09 s \\
Case 3 & 75 hr 59 min 50 s \\
Case 4 & 74 hr 03 min 27 s \\
Case 5 & 75 hr 31 min 42 s \\
\hline
\end{tabular}
\label{table:computation_times}
\end{table}

\begin{figure}[hbt!]
	\centering
	\begin{subfigure}[H]{0.48\linewidth}
		\includegraphics[width=\linewidth]{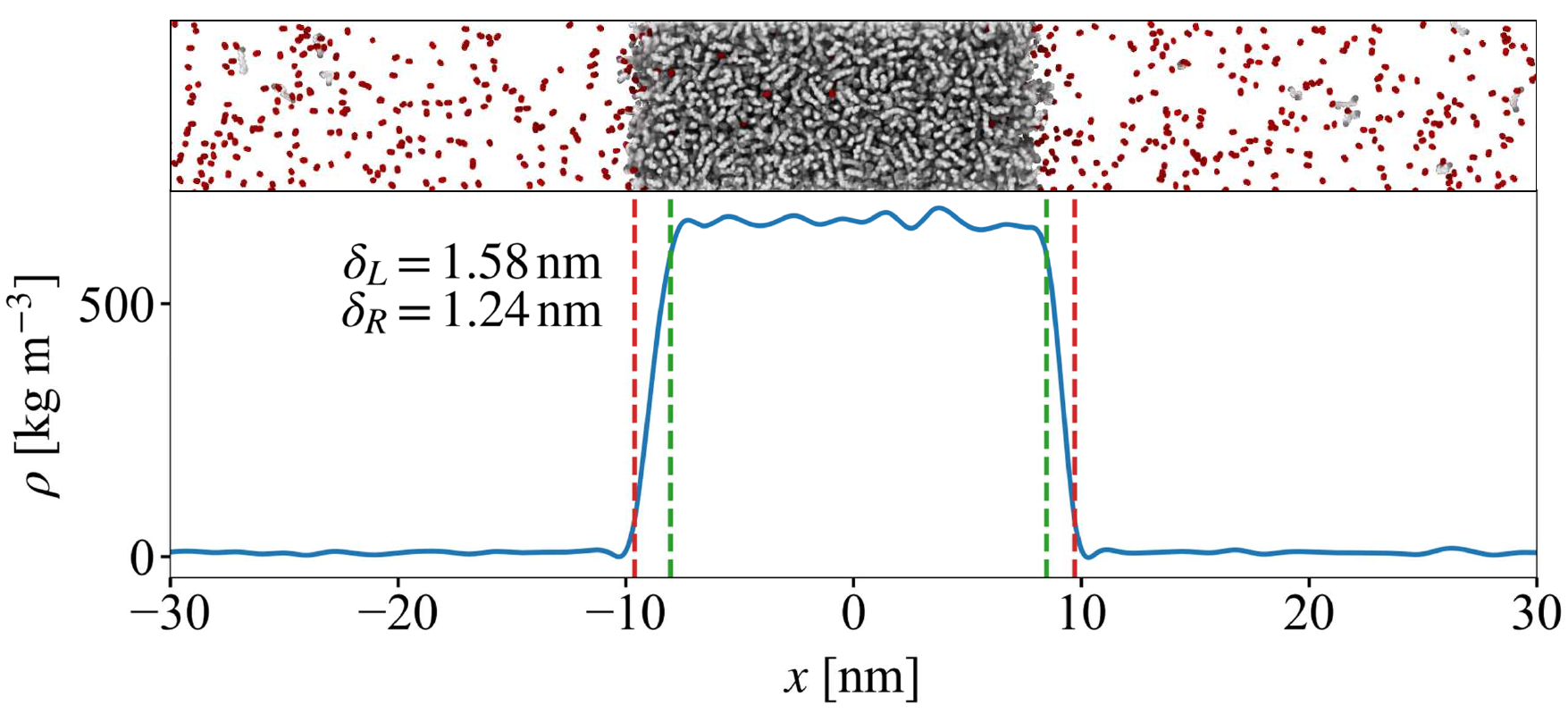}
		\subcaption{Case 1 ($T_r=0.70$, 10 ns)}
	\end{subfigure}
	\begin{subfigure}[H]{0.48\linewidth}
		\includegraphics[width=\linewidth]{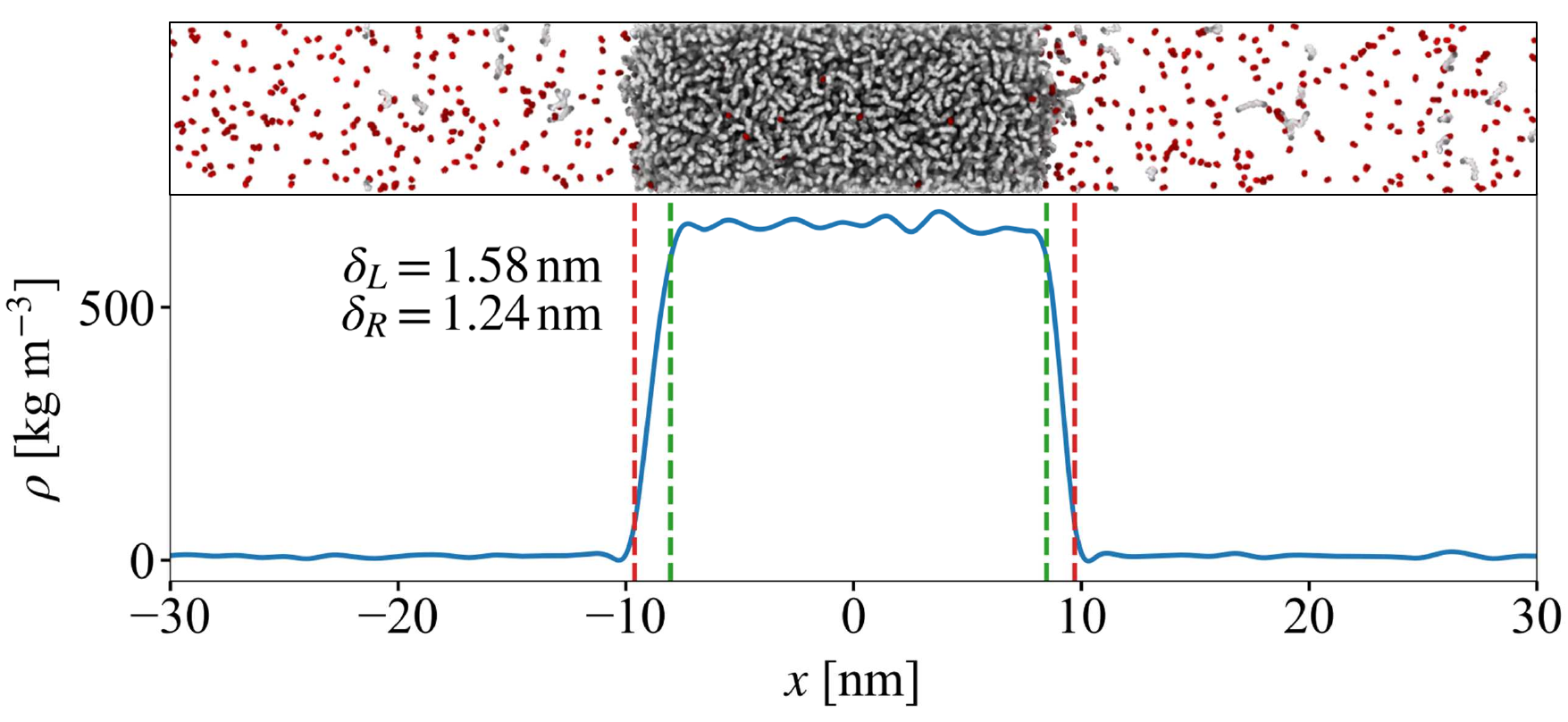}
		\subcaption{Case 5 ($T_r=0.95$, 10 ns)}
	\end{subfigure}\\[2mm]
	\begin{subfigure}[H]{0.48\linewidth}
		\includegraphics[width=\linewidth]{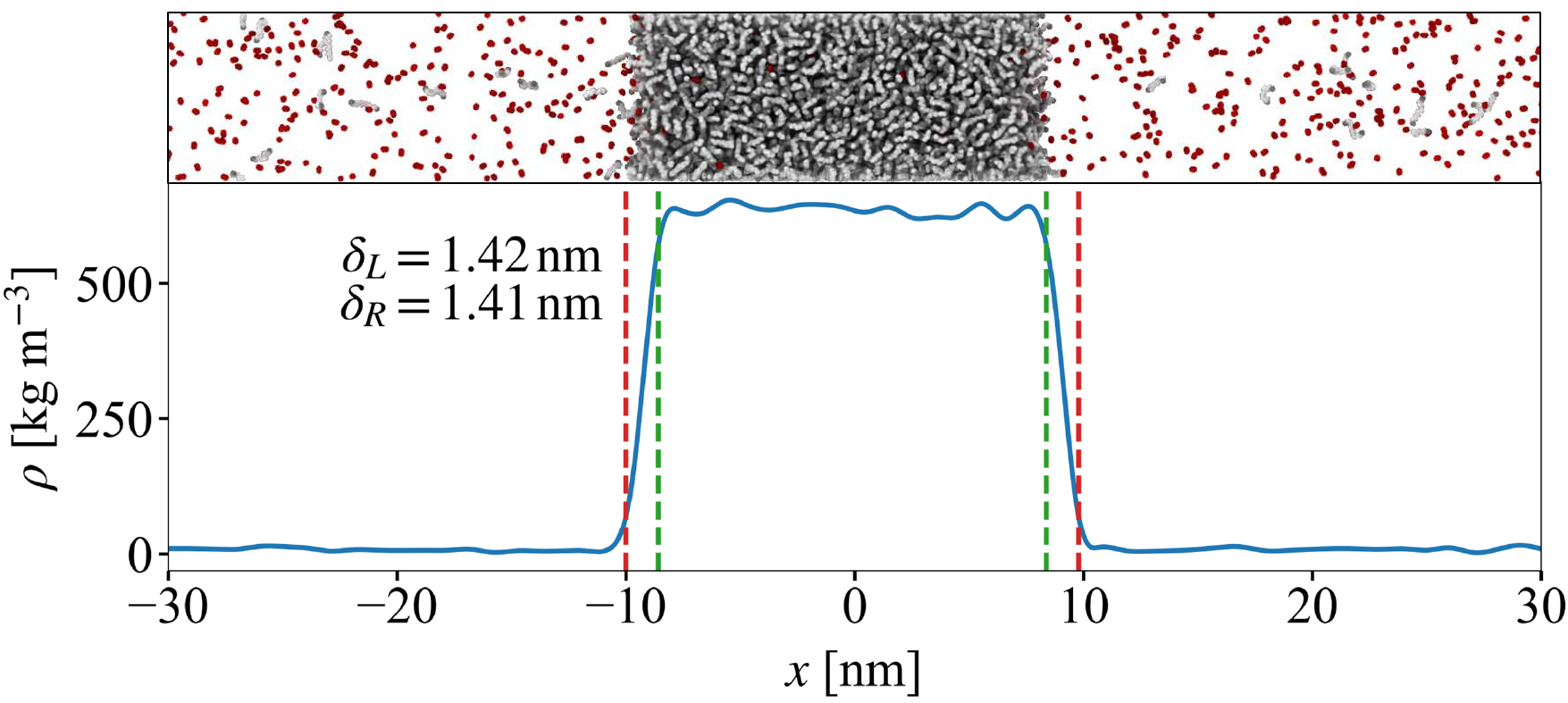}
		\subcaption{Case 1 ($T_r=0.70$, 20 ns)}
	\end{subfigure}
	\begin{subfigure}[H]{0.48\linewidth}
		\includegraphics[width=\linewidth]{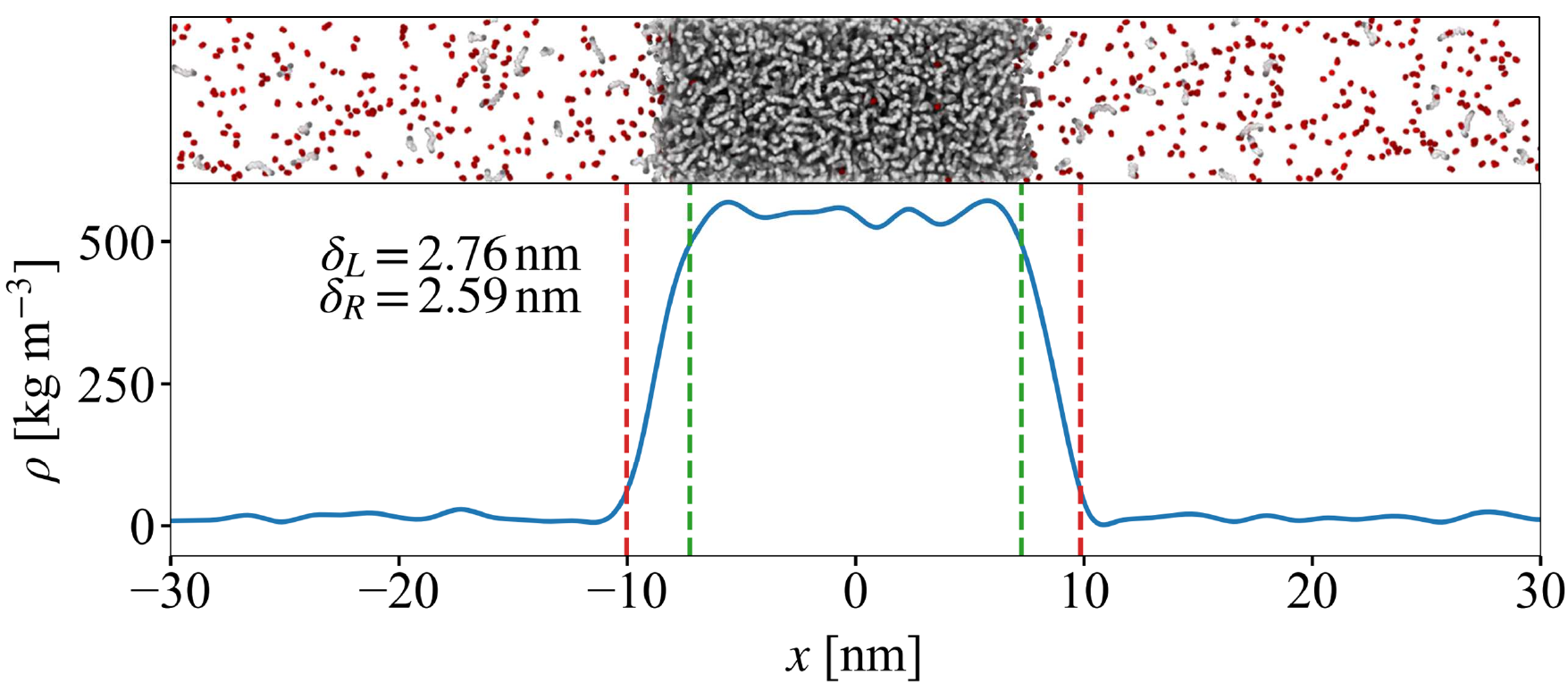}
		\subcaption{Case 5 ($T_r=0.95$, 20 ns)}
	\end{subfigure}\\[2mm]
	\begin{subfigure}[H]{0.48\linewidth}
		\includegraphics[width=\linewidth]{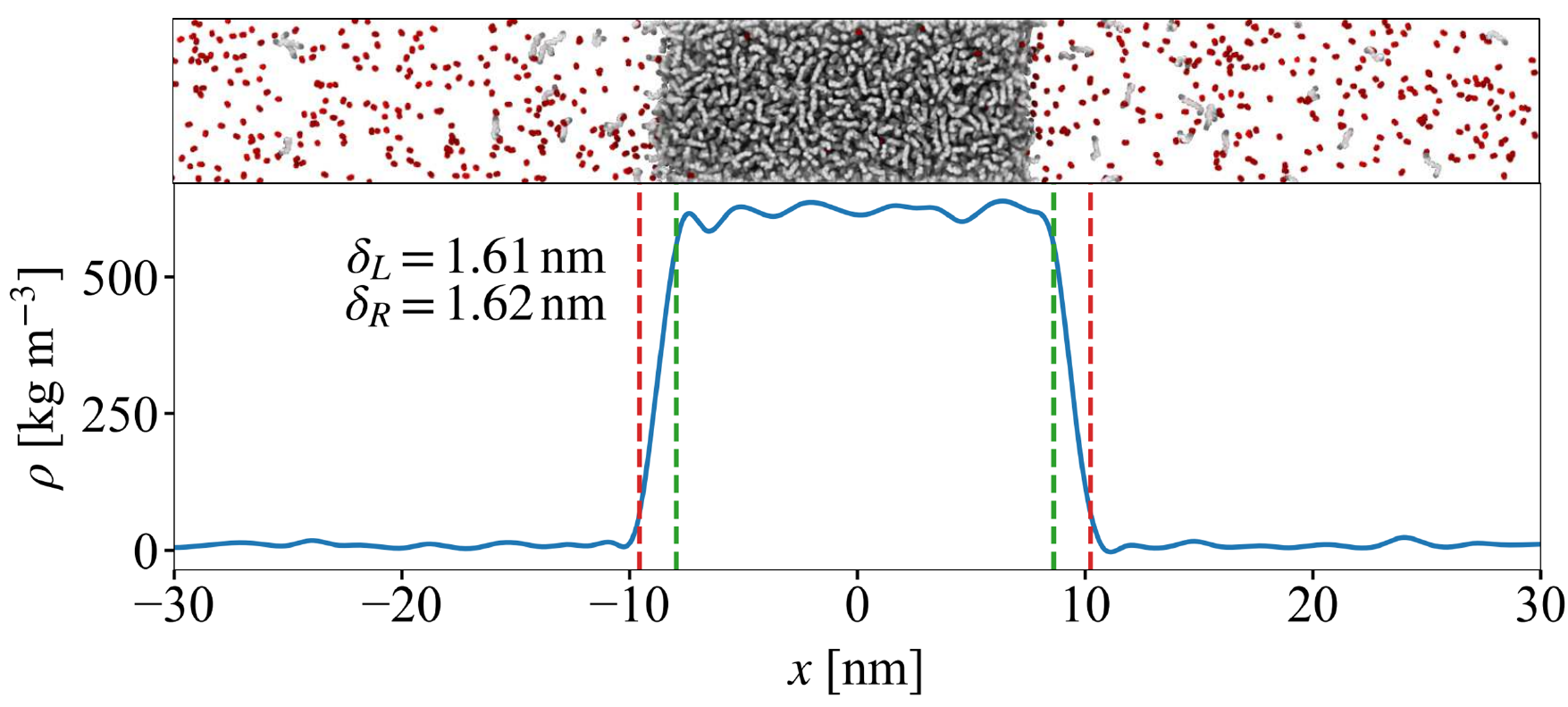}
		\subcaption{Case 1 ($T_r=0.70$, 40 ns)}
	\end{subfigure}
	\begin{subfigure}[H]{0.48\linewidth}
		\includegraphics[width=\linewidth]{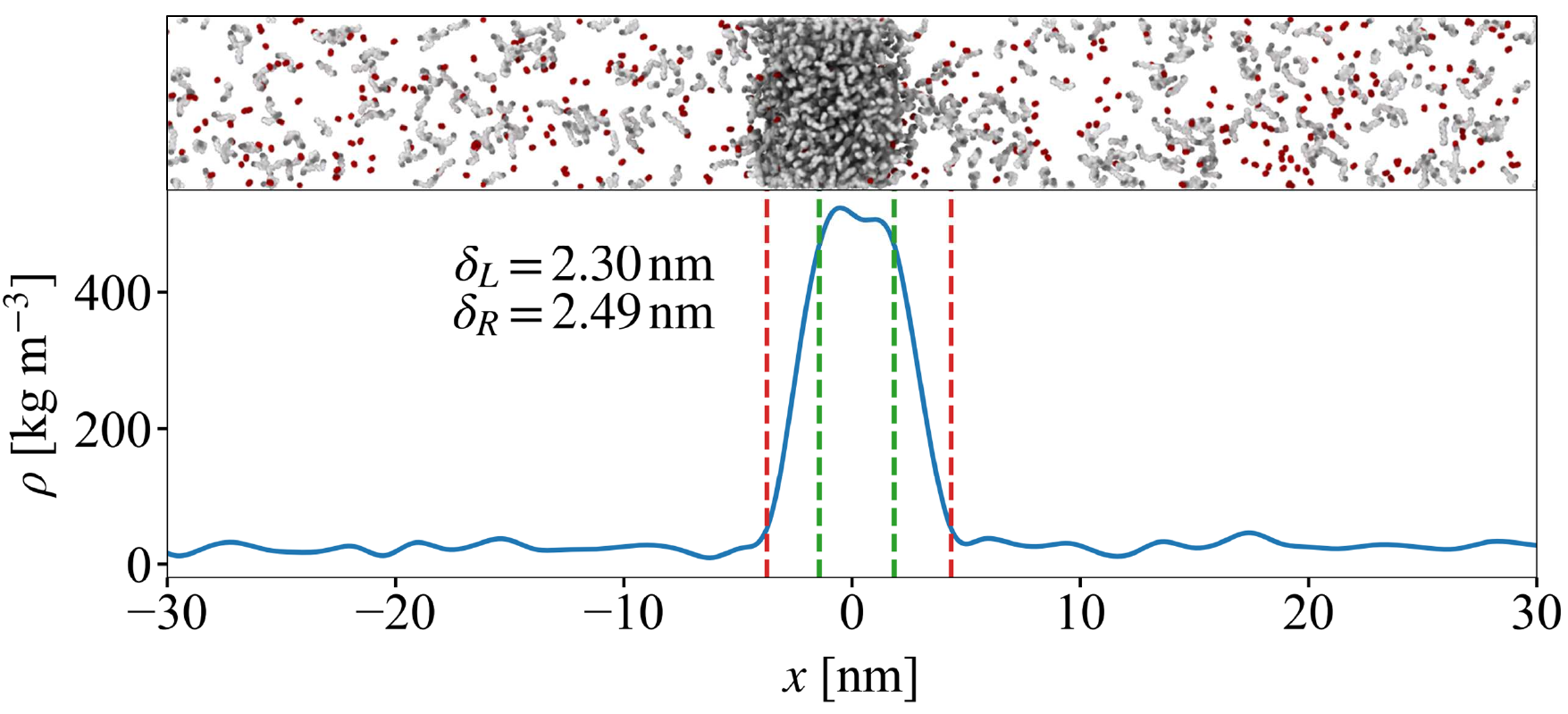}
		\subcaption{Case 5 ($T_r=0.95$, 40 ns)}
	\end{subfigure}\\[2mm]
	\begin{subfigure}[H]{0.48\linewidth}
		\includegraphics[width=\linewidth]{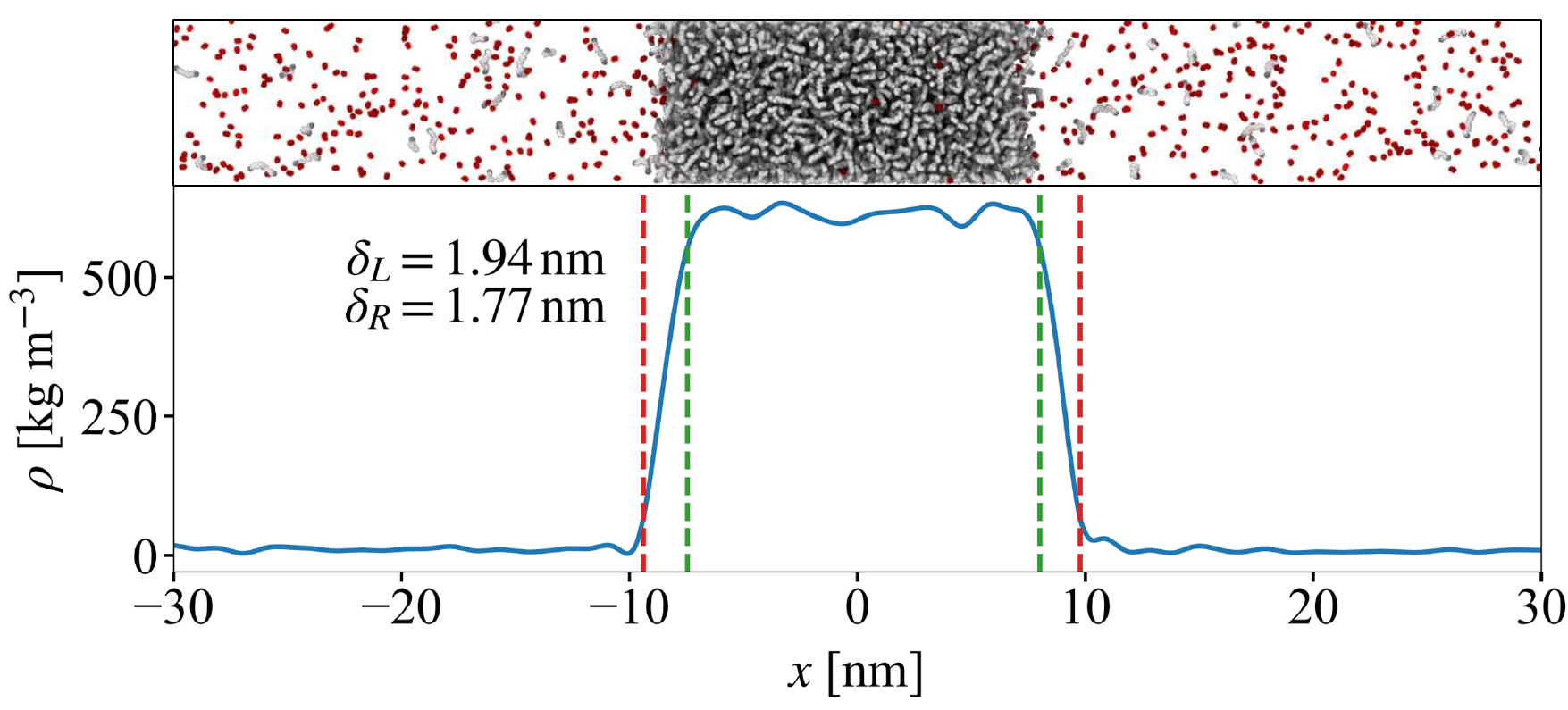}
		\subcaption{Case 1 ($T_r=0.70$, 46.70 ns)}
	\end{subfigure}
	\begin{subfigure}[H]{0.48\linewidth}
		\includegraphics[width=\linewidth]{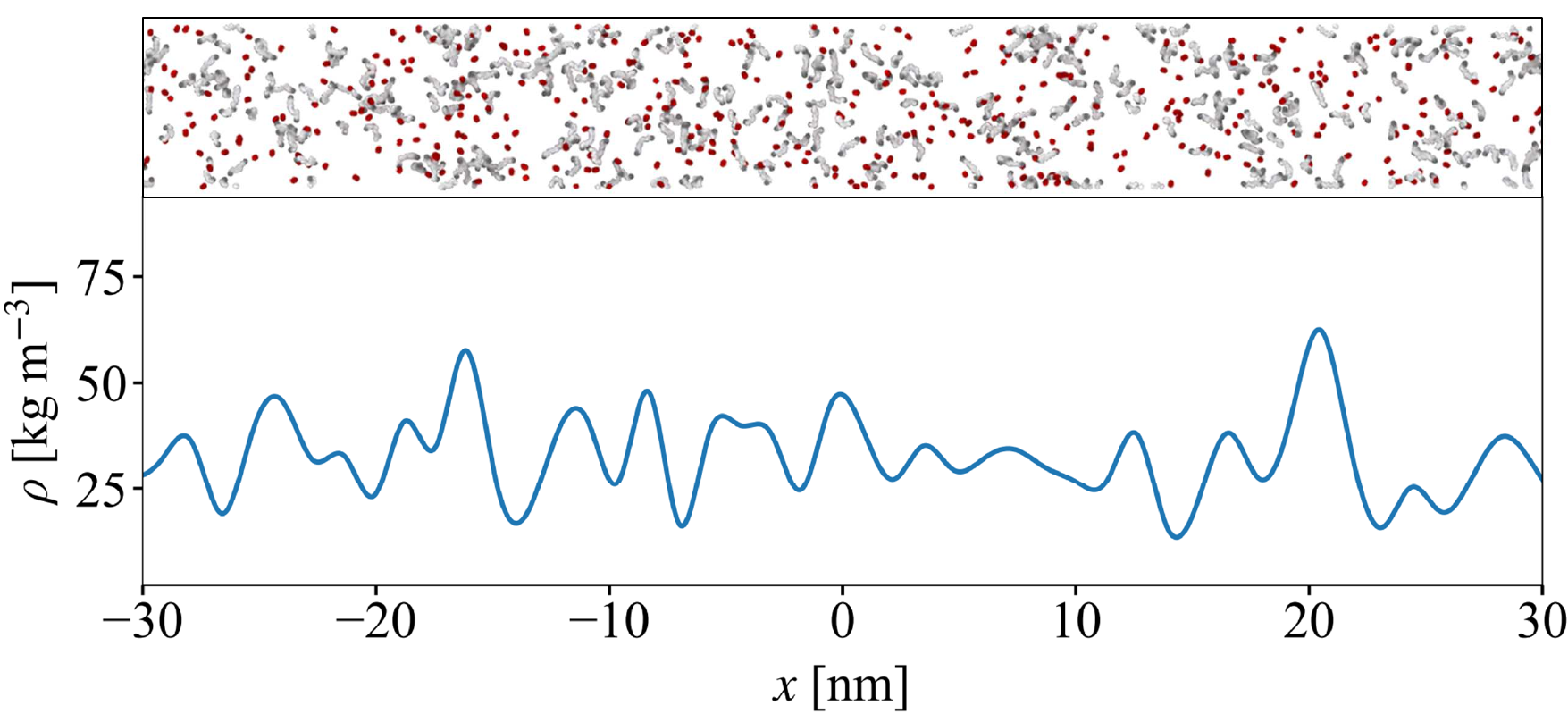}
		\subcaption{Case 5 ($T_r=0.95$, 46.70 ns)}
	\end{subfigure}

\caption{
Temporal evolution of the vapor–liquid interfaces for \textit{n}-dodecane/N$_2$ at 
$T_r=0.70$ (Case~1) and $T_r=0.95$ (Case~5). 
Interfaces are identified using the 90–10 rule. 
The interfacial region broadens over time, and in Case~5 the interface disappears once 
the liquid \textit{n}-dodecane becomes depleted.
}
	\label{fig:simulation_config}
\end{figure}

\subsection{Thermophysical properties}

\begin{figure}[hbt!]
	\centering 
	\begin{subfigure}[H]{0.5\linewidth}
		\includegraphics[width=\linewidth]{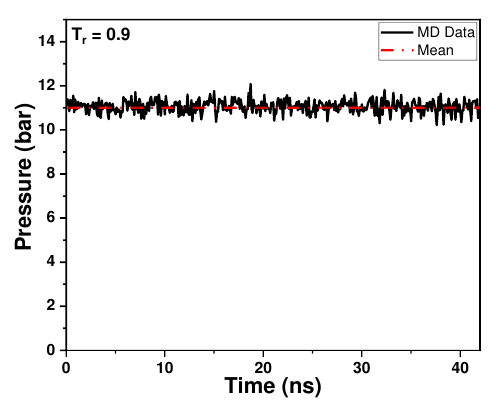}
	\end{subfigure}
	\caption{
Time history of pressure for Case~4 ($T_r=0.9$). 
Instantaneous MD pressure (black) fluctuates around a stable mean of approximately 11.6~bar (red).
}
	\label{fig:pressure}
\end{figure}

We present the temporal evolution of pressure in Fig.~\ref{fig:pressure}. 
The black curve denotes the instantaneous pressure from the MD trajectory, and the red curve 
shows the corresponding time-averaged value. Across all four cases, the pressure remains 
approximately constant at 11.59~bar, with fluctuations of about $\pm 0.5$~bar.

\begin{figure}[hbt!]
	\centering 
	\begin{subfigure}[H]{0.5\linewidth}
		\includegraphics[width=\linewidth]{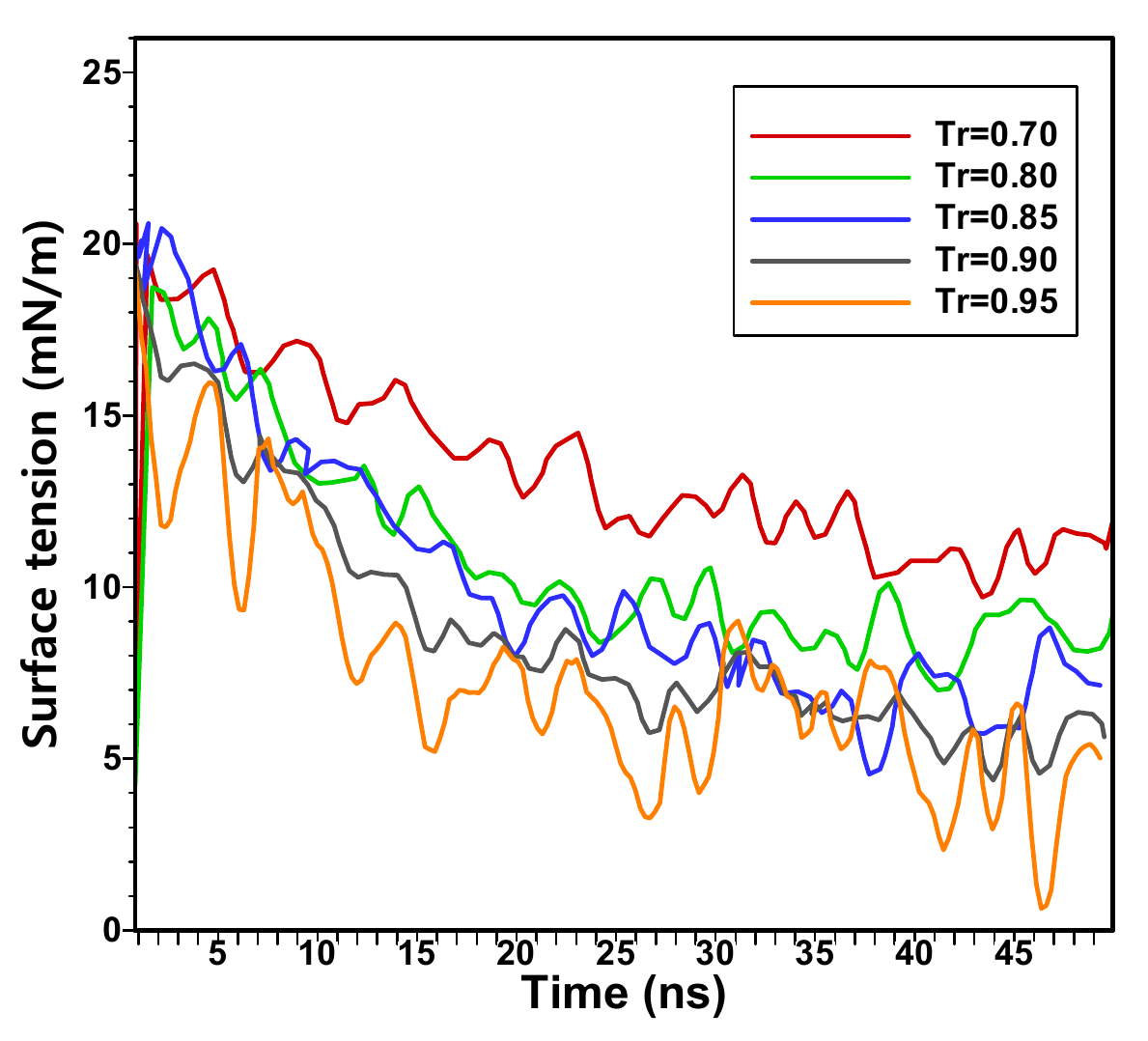}
	\end{subfigure}
	\caption{Temporal evolution of surface tension for the C$_{12}$/N$_2$ system.
Surface tension decreases during the initial heat-up and then stabilizes.
Higher ambient temperatures yield lower surface tension throughout the process.}
	\label{fig:surface_tension}
\end{figure}

\begin{figure}[hbt!]
    \centering

    \begin{subfigure}[H]{0.48\linewidth}
        \centering
        \includegraphics[width=\linewidth]{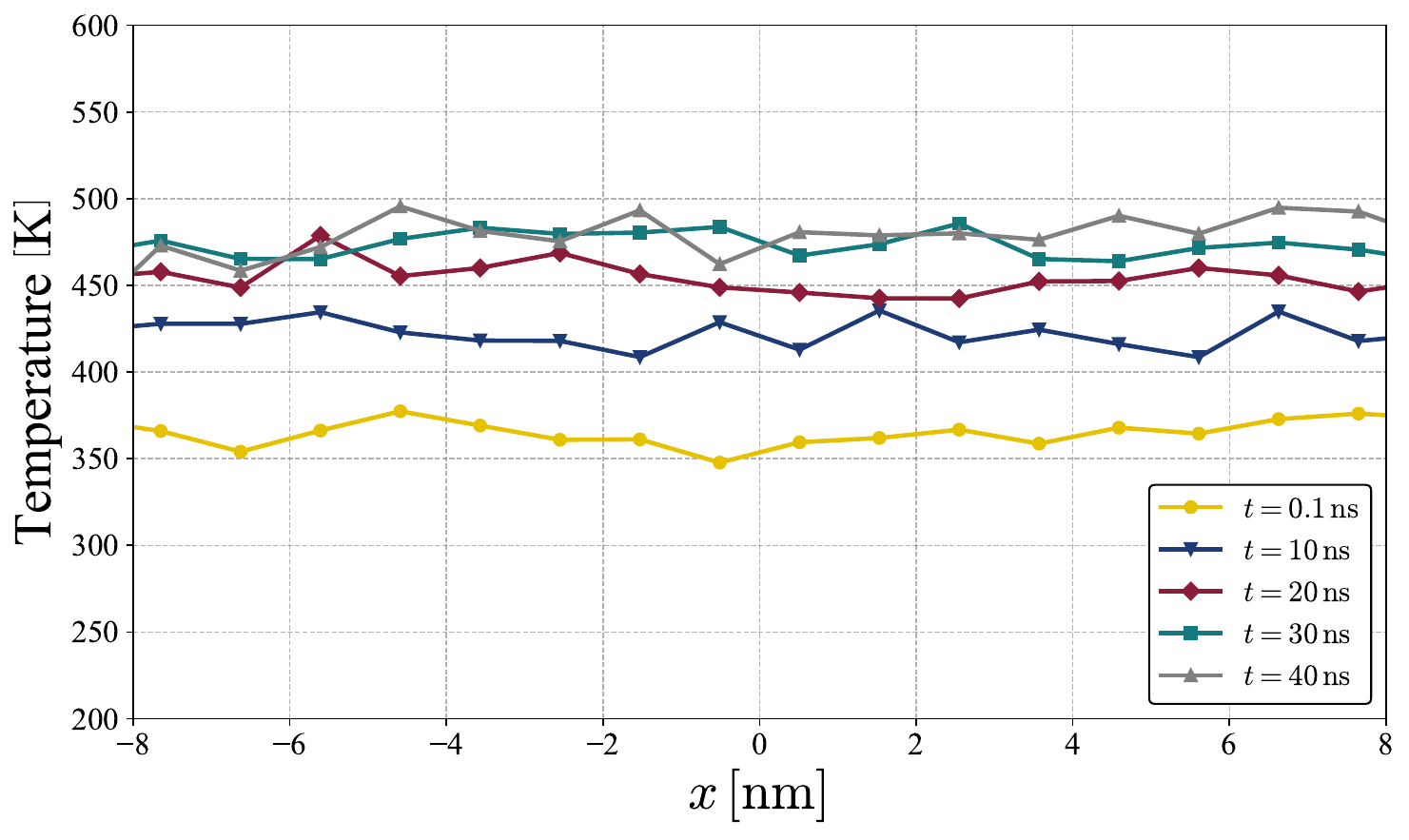}
        \subcaption{Case 1}
    \end{subfigure}
    \hfill
    \begin{subfigure}[H]{0.48\linewidth}
        \centering
        \includegraphics[width=\linewidth]{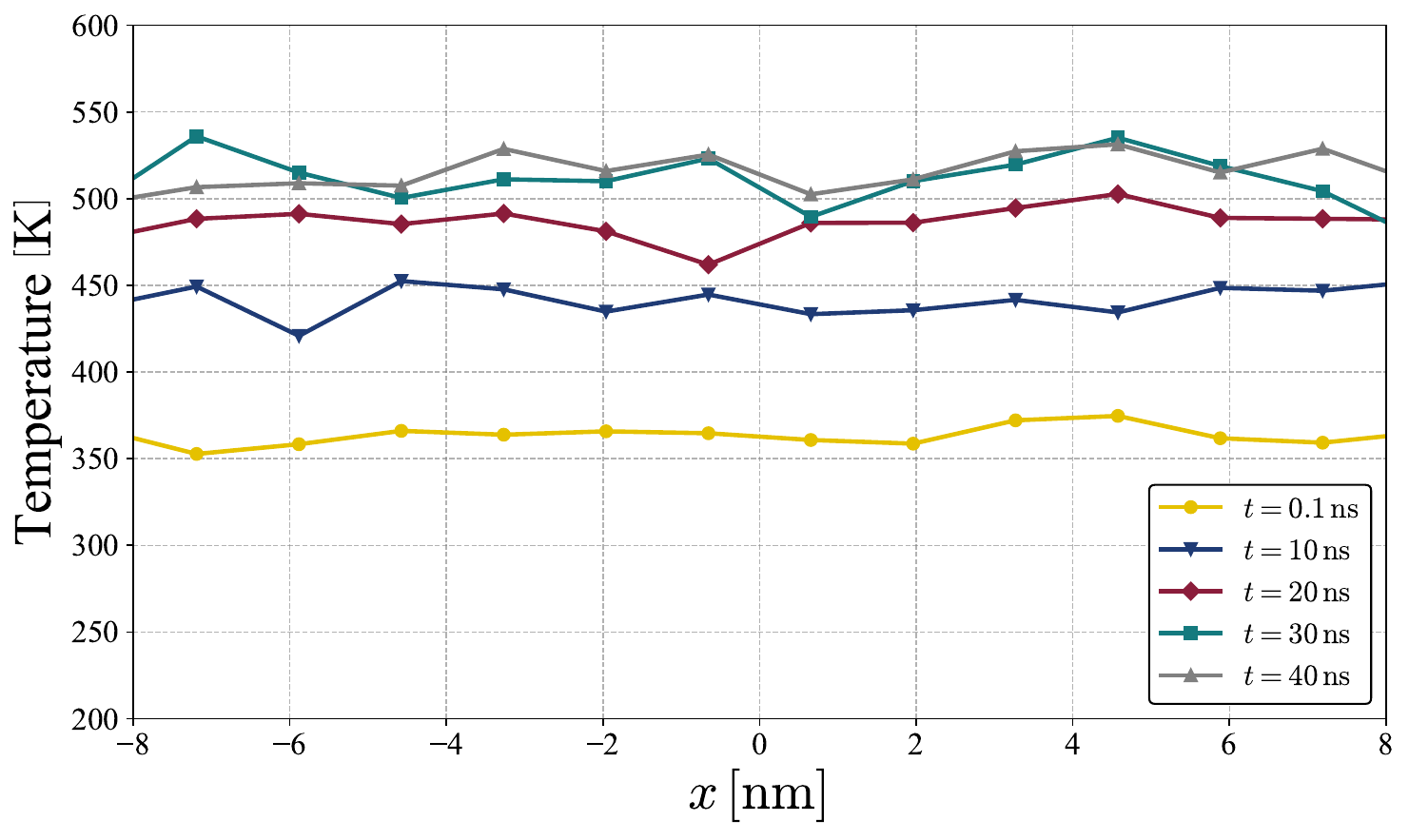}
        \subcaption{Case 2}
    \end{subfigure}

    \begin{subfigure}[H]{0.48\linewidth}
        \centering
        \includegraphics[width=\linewidth]{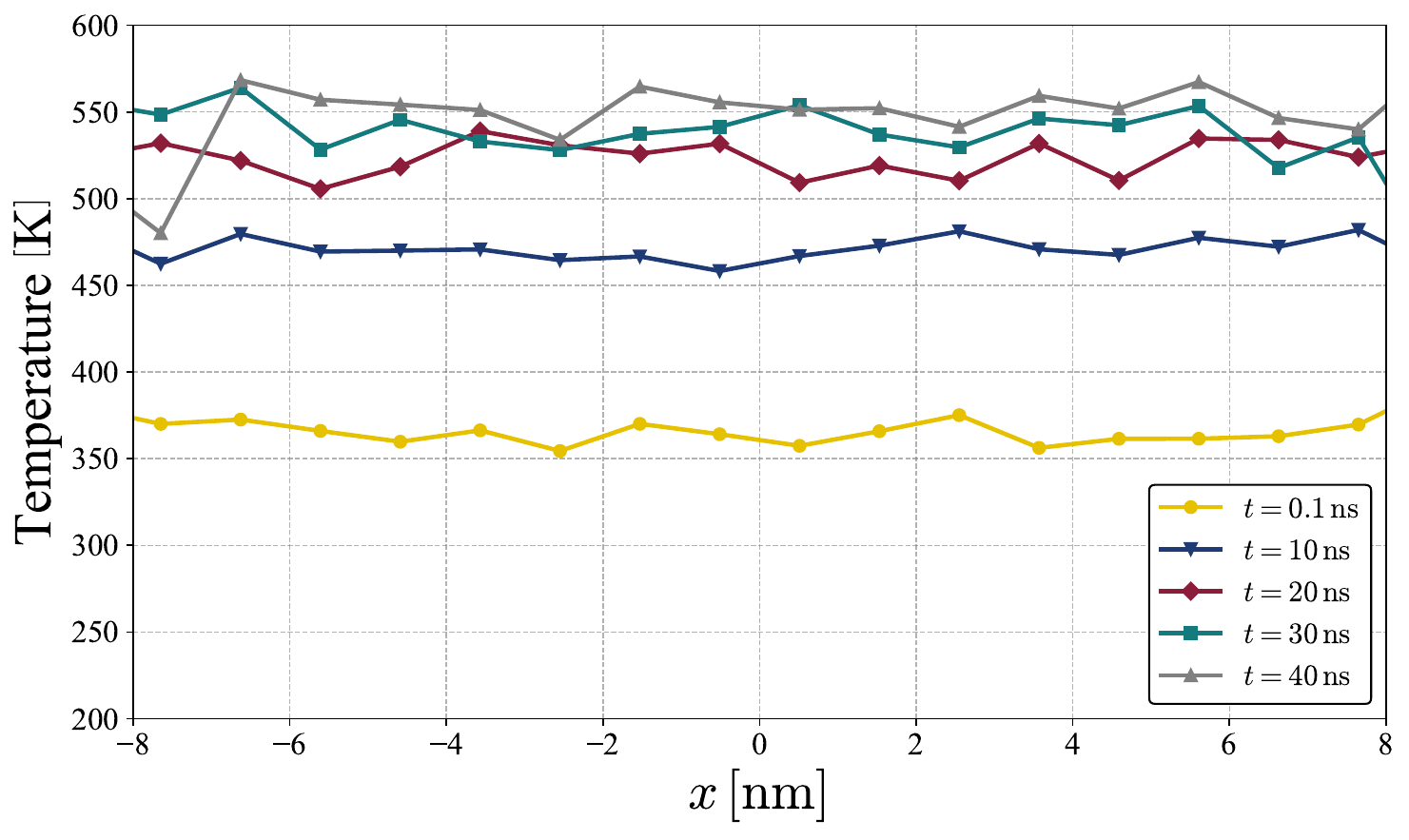}
        \subcaption{Case 3}
    \end{subfigure}
    \hfill
    \begin{subfigure}[H]{0.48\linewidth}
        \centering
        \includegraphics[width=\linewidth]{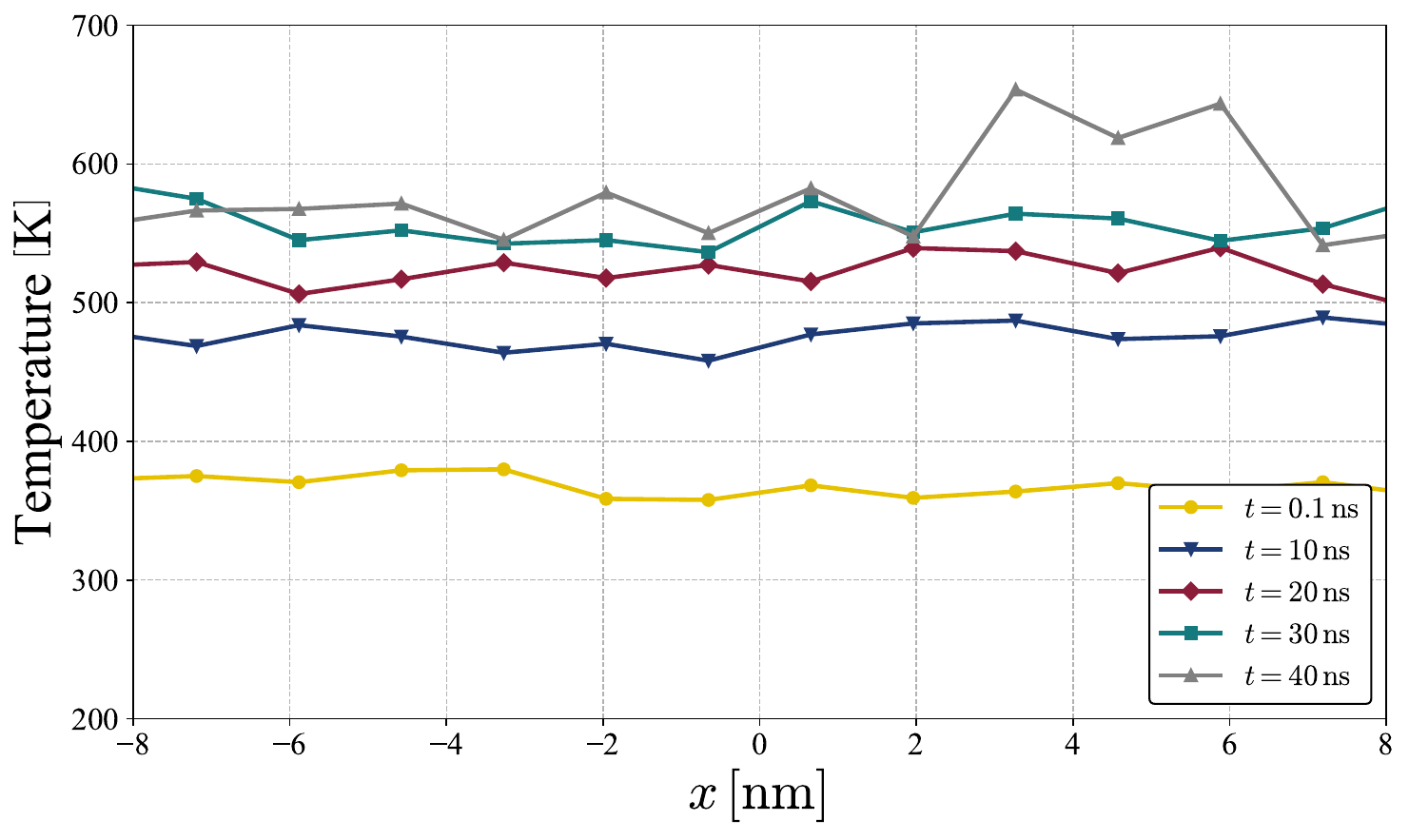}
        \subcaption{Case 4}
    \end{subfigure}

    \begin{subfigure}[H]{0.48\linewidth}
        \centering
        \includegraphics[width=\linewidth]{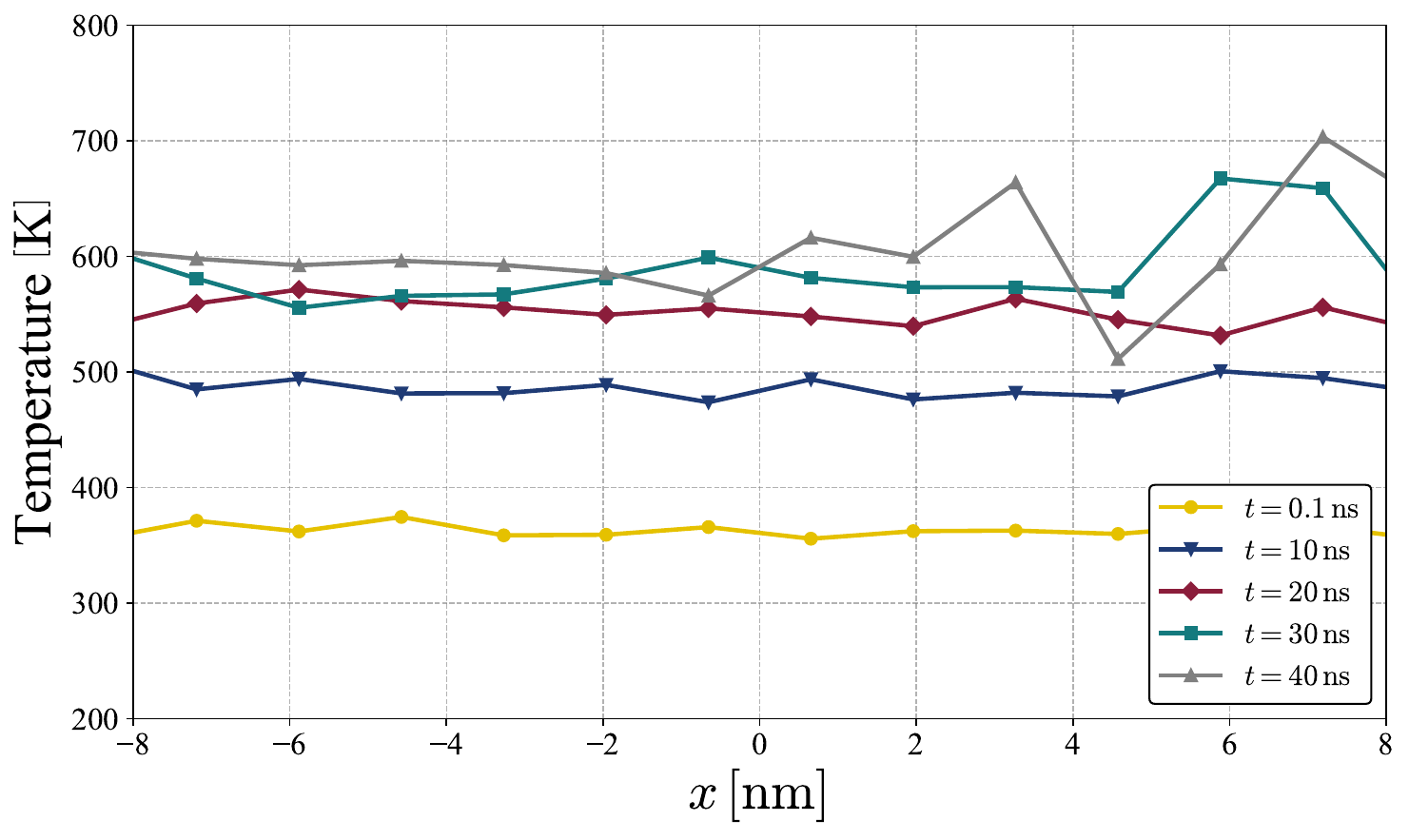}
        \subcaption{Case 5}
    \end{subfigure}

    \caption{
    Spatial temperature profiles for five simulation cases.
    Each subfigure shows the evolution of temperature across the vapor–liquid domain at successive times.
    Although the imposed conditions differ across Cases~1–5, 
    all cases exhibit a consistent pattern: heat transfer from the hot ambient region into the cooler liquid core,
    followed by the establishment of a stable thermal gradient that drives continuous evaporation at the interface.
    }
    \label{fig:temperature}
\end{figure}

We present the temporal evolution of surface tension for the C$_{12}$/N$_2$ system in 
Fig.~\ref{fig:surface_tension}. Across all four heating conditions, the surface tension 
initially decreases during the heat-up period as the liquid core warms, and subsequently 
reaches a quasi-steady value during the evaporation phase. Higher ambient temperatures 
(from $T_r = 0.7$ to $T_r = 0.95$) lead to lower surface tension levels and a steeper 
initial decline. Despite this reduction, a finite surface tension persists throughout 
Cases~1--5, indicating the continued presence of a vapor--liquid interface.

We examine the spatial and temporal evolution of temperature for all five cases in 
Fig.~\ref{fig:temperature}. Each subfigure shows the temperature distribution across the 
vapor–liquid domain at successive times. In every case, the temperature in the liquid 
region initially rises as heat is transferred from the surrounding hot ambient gas. 
As the system evolves, a stable thermal gradient develops between the heated vapor 
region and the cooler liquid core, and this gradient is maintained throughout the 
evaporation process. Although the imposed heating conditions differ among Cases~1–5, 
the overall thermal behavior remains consistent.

\begin{figure}[htp!]
    \centering
    \begin{subfigure}[b]{0.99\linewidth}
        \includegraphics[width=\linewidth]{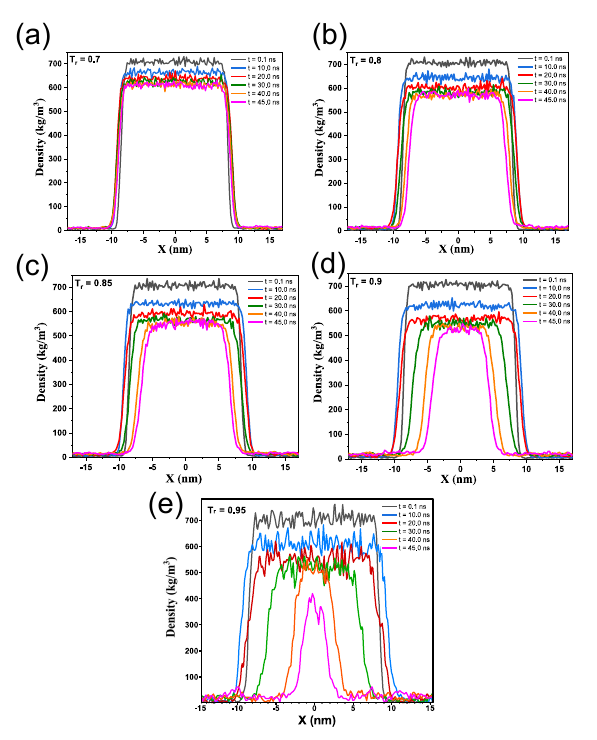}
        \label{fig:density_case1}
    \end{subfigure}
    \vspace{-16 pt}
\caption{Spatial density profiles of the binary system n-dodecane/N$_2$ at five ambient conditions: 
(a) $T_r = 0.70$, (b) $T_r = 0.80$, (c) $T_r = 0.85$, (d) $T_r = 0.90$, and (e) $T_r = 0.95$. 
As $T_r$ increases, the liquid-phase density decreases, the interface width broadens, and the vapor–liquid interface becomes increasingly diffuse, indicating a gradual weakening of phase separation near the critical region.}
    \label{fig:surface_density_all_cases}
\end{figure}

\begin{figure}[hpt!]
    \centering
    \begin{subfigure}[b]{0.99\linewidth}
        \includegraphics[width=\linewidth]{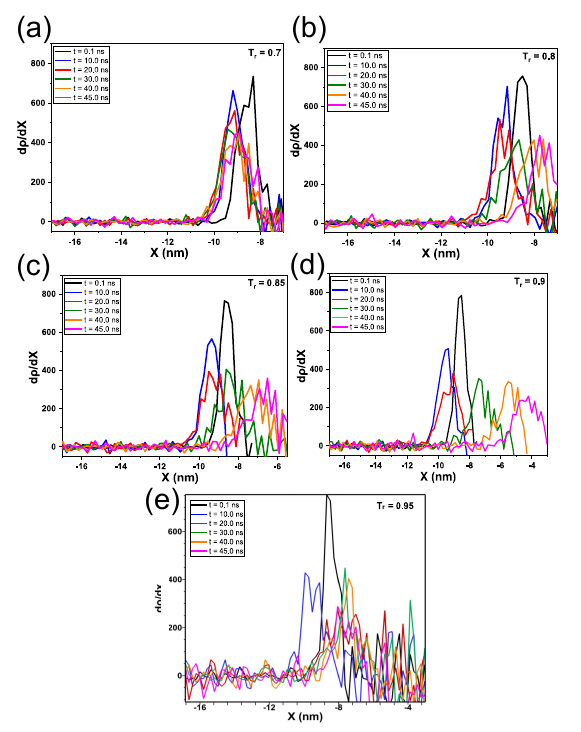}
        \label{fig:densitygrad_case1}
    \end{subfigure}
    \vspace{-16 pt}
    \caption{Spatial profiles of the density gradient ($\mathrm{d}\rho/\mathrm{d}x$) near the left vapor–liquid interface for the binary system C$_{12}$/N$_2$ at five ambient conditions: 
    (a) $T_r = 0.70$, (b) $T_r = 0.80$, (c) $T_r = 0.85$, (d) $T_r = 0.90$, and (e) $T_r = 0.95$. 
    As $T_r$ increases, the peak density gradient decreases and the interfacial region broadens, indicating a more gradual transition between the vapor and liquid phases.}
    \label{fig:density_gradient_all_cases}
\end{figure}

To check the spatial distribution of density, we plot the evolution of spatial density profiles in the $x$-direction over time for the binary mixture of C$_{12}$/N$_2$ in Fig.~\ref{fig:surface_density_all_cases}. For all five test cases, as the liquid core is progressively heated by the surrounding ambient gas, the density of the liquid region decreases and then remains nearly constant during the steady-state evaporation phase. The temporal change in the width of the density profile reveals an initial swelling of the liquid core in the early stages of evaporation, followed by a gradual regression. When the ambient temperature is $T_r = 0.7$, the vapor–liquid interface remains well-defined even up to 45.0~ns. However, as the ambient temperature increases to $T_r = 0.95$, the interface becomes noticeably thinner and rapidly diminishes.

To examine the interface position and its temporal evolution under varying thermal conditions, 
the spatial density gradient in the \(x\)-direction was evaluated. 
Figure~\ref{fig:density_gradient_all_cases} shows the density–gradient profiles 
\(\mathrm{d}\rho/\mathrm{d}x\) near the left interface for all test cases. 
Larger gradient magnitudes indicate sharper interfaces, whereas smaller gradients correspond to broader transition regions. 
The peak locations exhibit a leftward shift at \(t = 10~\mathrm{ns}\) and \(20~\mathrm{ns}\), 
followed by a gradual rightward movement after \(t = 30~\mathrm{ns}\), 
consistent with the transient expansion and subsequent recession of the liquid region observed in the density fields.
As the temperature increases, the vapor–liquid transition becomes increasingly diffuse, and for \(T_r = 0.95\) the contrast diminishes to the point where separating the liquid core from the surrounding medium becomes difficult, signaling a shift toward a more uniform fluid state.

\begin{figure}[hbt!]
    \centering
    \begin{subfigure}[b]{0.48\linewidth}
        \centering
        \includegraphics[width=\linewidth]{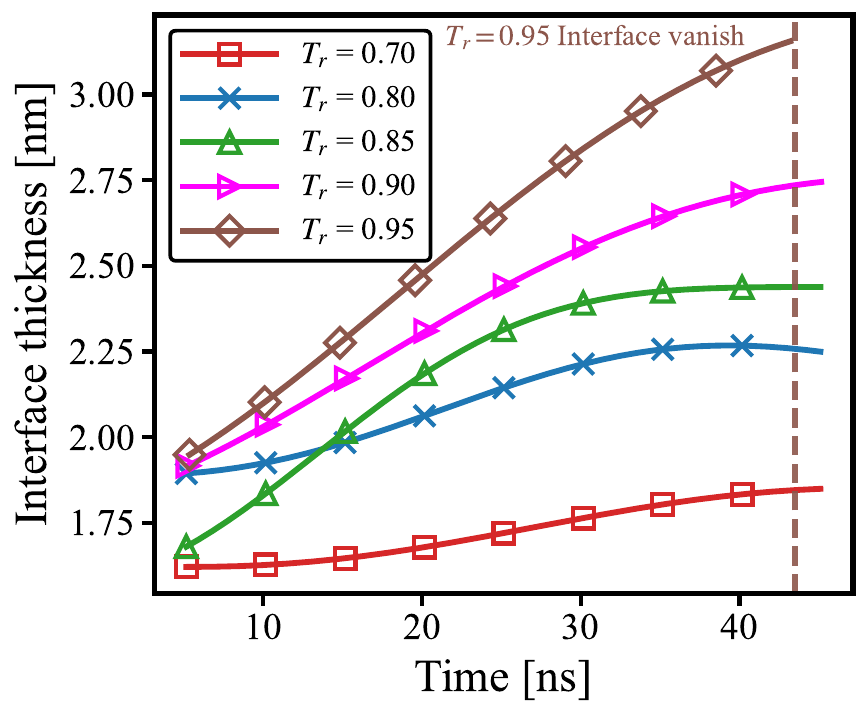}
        \caption{Interface thickness}
        \label{fig:interface_thickness}
    \end{subfigure}
    \hfill
    \begin{subfigure}[b]{0.48\linewidth}
        \centering
        \includegraphics[width=\linewidth]{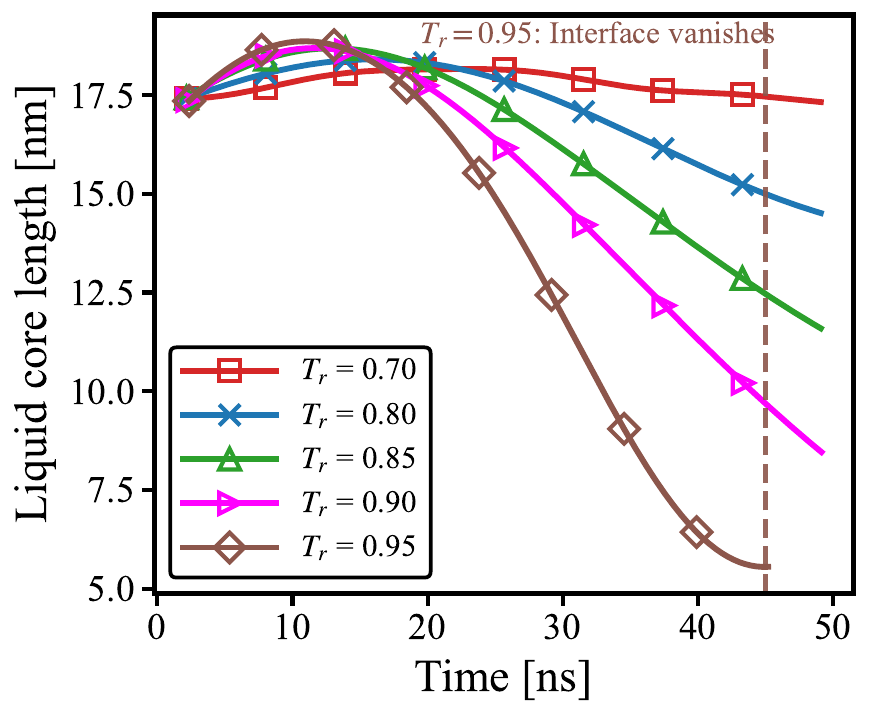}
        \caption{Liquid core length}
        \label{fig:liquid_core}
    \end{subfigure}
    \caption{Temporal evolution of (a) vapor--liquid interface thickness and (b) liquid core length for the binary C$_{12}$/N$_2$ system under different ambient temperatures. 
    The interface becomes progressively broader over time and eventually stabilizes, with higher ambient temperatures leading to thicker interfaces. 
    The liquid core length decreases due to evaporation, and the shrinkage rate increases with ambient temperature.}
    \label{fig:interface_core_summary}
\end{figure}

To quantify how the vapor–liquid transition region evolves during evaporation, 
the interface thickness was evaluated using the 90--10 rule~\cite{lekner1977surface} 
and averaged over the two interfaces, as shown in Fig.~\ref{fig:interface_thickness}. 
For all reduced temperatures, the interface thickness increases monotonically during the
simulation, indicating a progressive broadening of the transition region as evaporation proceeds.  
The rate of thickening clearly depends on the ambient temperature. 
At lower reduced temperatures (\(T_r = 0.70\) and \(0.80\)), the increase is relatively modest, 
whereas intermediate and high reduced temperatures (\(T_r = 0.85\) and \(0.90\)) exhibit a steeper growth and a larger final thickness. 
A slight reduction in the growth rate is observed after approximately \(20~\mathrm{ns}\), 
but the interface does not reach a strict plateau and continues to broaden throughout the simulation window.  
For \(T_r = 0.95\), the interface thickness grows almost linearly up to about \(40~\mathrm{ns}\), 
just before the interface disappears due to the depletion of liquid C$_{12}$. 
This behavior is consistent with the temperature and density-gradient trends, which both reflect enhanced molecular mixing at higher ambient temperatures.

Figure~\ref{fig:liquid_core} shows the temporal evolution of the C$_{12}$ liquid–core length for
Cases~1--5. The liquid–core length is defined as the distance between the midpoints of the left and right
interfacial thicknesses obtained from the 90--10 criterion. Across all cases, the core undergoes a mild
initial expansion during the heat–up stage, consistent with the early-time density variations observed in
Figs.~\ref{fig:surface_density_all_cases} and~\ref{fig:density_gradient_all_cases}. 
After this early expansion, the liquid core monotonically decreases in size as evaporation proceeds.
The rate of regression is strongly correlated with the ambient temperature. 
For the lower-temperature cases (\(T_r = 0.70\) and \(0.80\)), the shrinkage occurs gradually, and a substantial
liquid region remains even at the end of the simulation. 
In contrast, the intermediate- and high-temperature cases (\(T_r = 0.85\) and \(0.90\)) exhibit a much faster
reduction in liquid–core length, reflecting accelerated mass loss driven by enhanced thermal transport. 
At the highest temperature (\(T_r = 0.95\)), the liquid core decreases rapidly and vanishes entirely near
\(t \approx 46\,\mathrm{ns}\), marking the complete depletion of the condensed phase under strong ambient heating.

\subsection{Interfacial molecular mass transport}

This section examines the molecular mechanisms governing mass exchange across the
vapor–liquid interface during non–equilibrium evaporation. 
Two complementary post–processing strategies are employed to quantify the mass flux:
(i) the fixed box method, which evaluates the net mass loss from a control region surrounding the
liquid phase, and (ii) the two–boundary method, which identifies interfacial locations dynamically and 
tracks molecular exchange across the instantaneous vapor–liquid boundaries.
Using these approaches, we compute the evaporation flux, condensation flux, and reflection flux of
individual species. 
The resulting fluxes are then used to determine the temperature–dependent evaporation coefficient,
providing quantitative insight into interfacial transport behavior under varying thermal conditions.
We begin with the fixed box formulation.

\subsubsection{Fixed box method}

In this study, the fixed boundaries used to form the box were considered at $x = -9$~nm and $+9$~nm. For this computational setup, $S_\mathrm{flux}$ involves two surfaces each having an area of $64~\mathrm{nm}^2$. Thus, $S_\mathrm{flux}$ equals $128~\mathrm{nm}^2$. $\Delta t$ between two consecutive observations was considered to be 5~ns. At each observation instance, 10 samples were considered spread over a time of 0.1~ns.

\begin{figure}[hbt!]
    \centering
    \begin{subfigure}[b]{0.49\linewidth}
        \includegraphics[width=\linewidth]{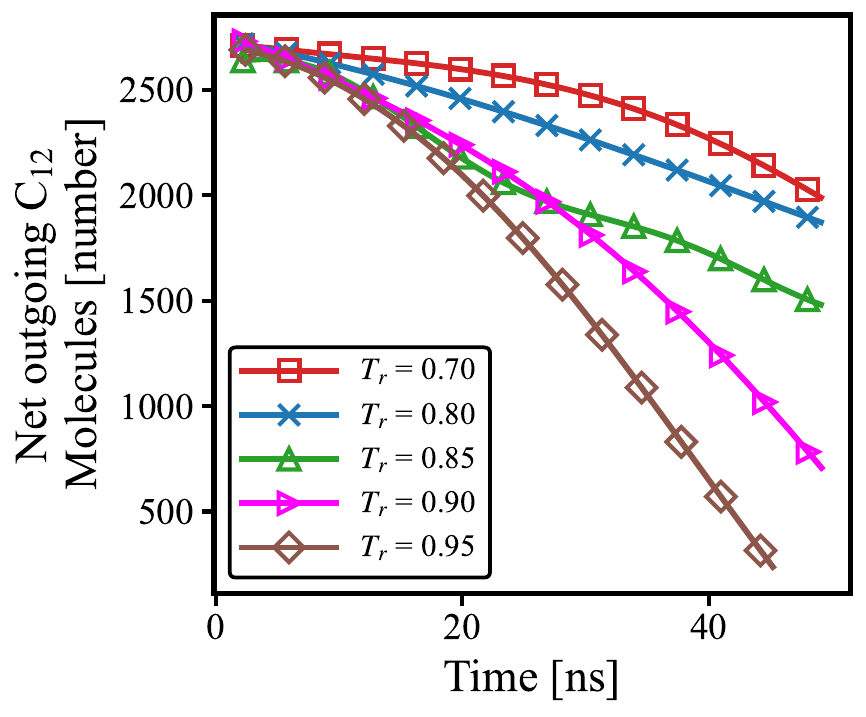}
        \subcaption{Molecule count}
        \label{fig:c12_count}
    \end{subfigure}
    \hfill
    \begin{subfigure}[b]{0.49\linewidth}
        \includegraphics[width=\linewidth]{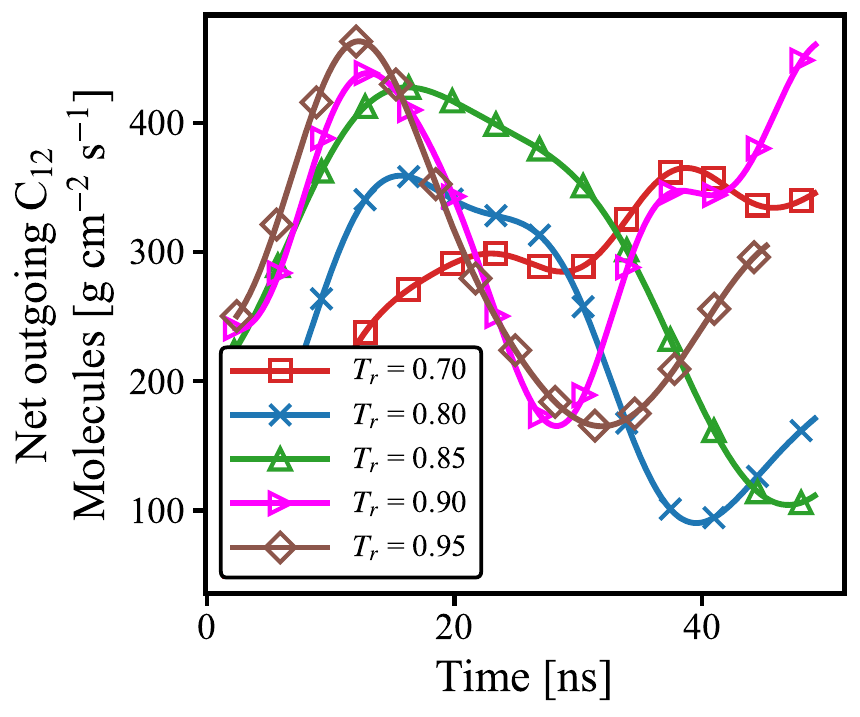}
        \subcaption{Mass flux}
        \label{fig:evap_flux}
    \end{subfigure}
\caption{Interfacial mass flux estimation of n-dodecane using the fixed box method. 
The left figure shows the temporal variation of the n-dodecane molecular count within a fixed control volume near the vapor--liquid interface, while the right figure presents the corresponding net mass flux obtained from its time derivative. 
However, since the interface position continuously shifts over time, the fixed box method fails to accurately estimate the true interfacial flux.}
    \label{fig:fixed_box_flux_summary}
\end{figure}

In Fig.~\ref{fig:fixed_box_flux_summary}(a), we report the temporal variation of the number of
C$_{12}$ molecules contained within the fixed control volume. 
For all five cases, the molecular population decreases monotonically as evaporation proceeds. 
At the lower reduced temperature ($T_r = 0.70$), the decrease is nearly linear, indicating a 
quasi-uniform rate of mass loss throughout the simulation. 
As the ambient temperature increases, the depletion of C$_{12}$ becomes 
progressively faster and exhibits a marked nonlinear decay. 
This behavior reflects the enhanced thermal driving at elevated temperatures, which accelerates 
interfacial mass transfer and leads to a more rapid contraction of the liquid phase.

Figure~\ref{fig:fixed_box_flux_summary}(b) presents the C$_{12}$ mass flux computed from
the time variation of the molecular population in the fixed control volume.
The flux shows an initial increase during the heat-up period and a gradual decrease as the
liquid region recedes.  
Lower ambient temperatures such as \(T_r=0.70\) produce a broader initial rise due to weaker thermal driving.
The flux curves also exhibit irregular variations rather than a clear trend.
This arises from the fixed-box formulation, because the vapor–liquid interface moves
while the control volume remains fixed.
Molecules near the interface are therefore not consistently assigned, which introduces
fluctuations in the computed flux.
The effect becomes more visible at higher ambient temperatures where the interface motion is larger.

\begin{figure}[hbt!]
    \centering
    \begin{subfigure}[b]{0.46\linewidth}
        \includegraphics[width=\linewidth]{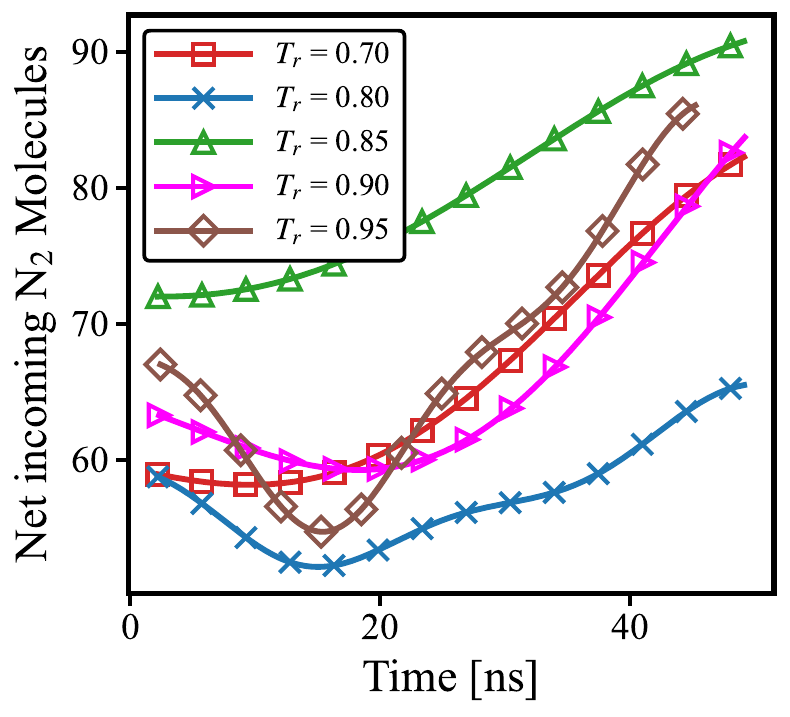}
        \subcaption{Molecule count}
        \label{fig:n2_count}
    \end{subfigure}
    \hfill
    \begin{subfigure}[b]{0.505\linewidth}
        \includegraphics[width=\linewidth]{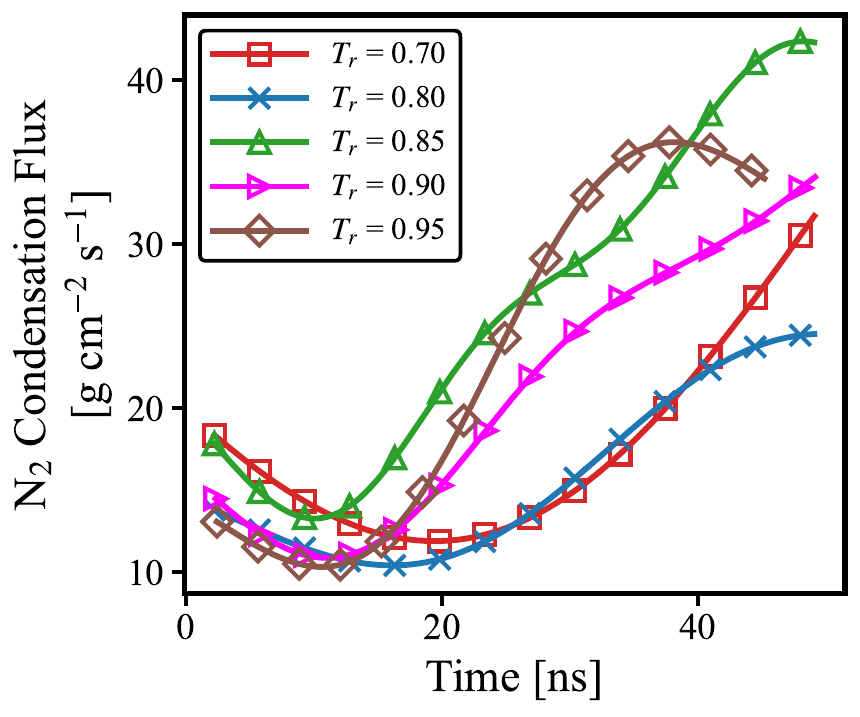}
        \subcaption{Condensation flux}
        \label{fig:n2_flux}
    \end{subfigure}
    \caption{Interfacial condensation flux estimation of nitrogen using the fixed box method. 
    The left figure shows the temporal evolution of the number of nitrogen molecules within a fixed control volume near the vapor--liquid interface, while the right figure presents the corresponding net mass flux entering the liquid phase obtained by differentiating the molecular count. 
    However, because the interface position changes dynamically over time, the fixed box method fails to provide an accurate estimation of the condensation flux.}
    \label{fig:fixed_box_flux_n2}
\end{figure}

Although evaporation is the dominant transport process, nitrogen also enters the liquid region due
to the concentration difference across the interface.
As shown in Fig.~\ref{fig:fixed_box_flux_n2}(a), the number of N$_2$ molecules inside the fixed
volume first decreases because of the sharp concentration jump at $t=0$, then gradually increases
as evaporation progresses.
The corresponding mass flux in Fig.~\ref{fig:fixed_box_flux_n2}(b) follows the same pattern: an
initial drop followed by a sustained rise.
After the heat-up period, higher ambient temperatures accelerate the reduction of C$_{12}$
within the fixed boundaries, and this faster retreat is accompanied by a larger inward transfer
of N$_2$.

\subsection{Two boundary method}

The two-boundary method utilizes dynamic vapor and liquid boundaries that define the 90–10 interface thickness on either side of the liquid core. As the liquid core expands and subsequently regresses during evaporation, the positions of these boundaries vary accordingly. Flux calculations were performed independently for the left and right interfaces, and the reported values represent their average. For each interface, the flux area $S_{\text{flux}}$ is 64~nm\textsuperscript{2}. Each data point is based on 10 samples collected over a 0.1~ns interval, with calculations performed every 5~ns; thus, $\Delta t = 0.1~\text{ns}$.

\begin{figure}[hbt!]
    \centering
    \begin{subfigure}[b]{0.41\linewidth}
        \includegraphics[width=\linewidth]{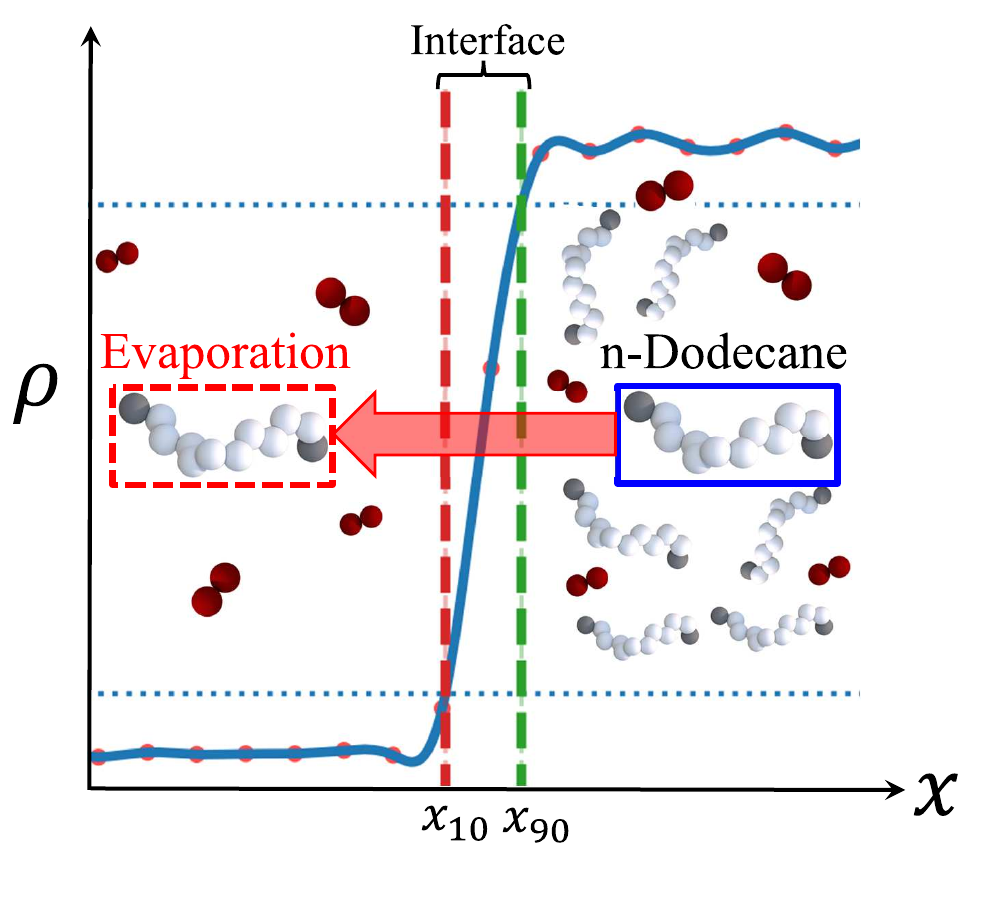}
        \caption{Schematic view of evaporation}
        
    \end{subfigure}
    \begin{subfigure}[b]{0.42\linewidth}
        \includegraphics[width=\linewidth]{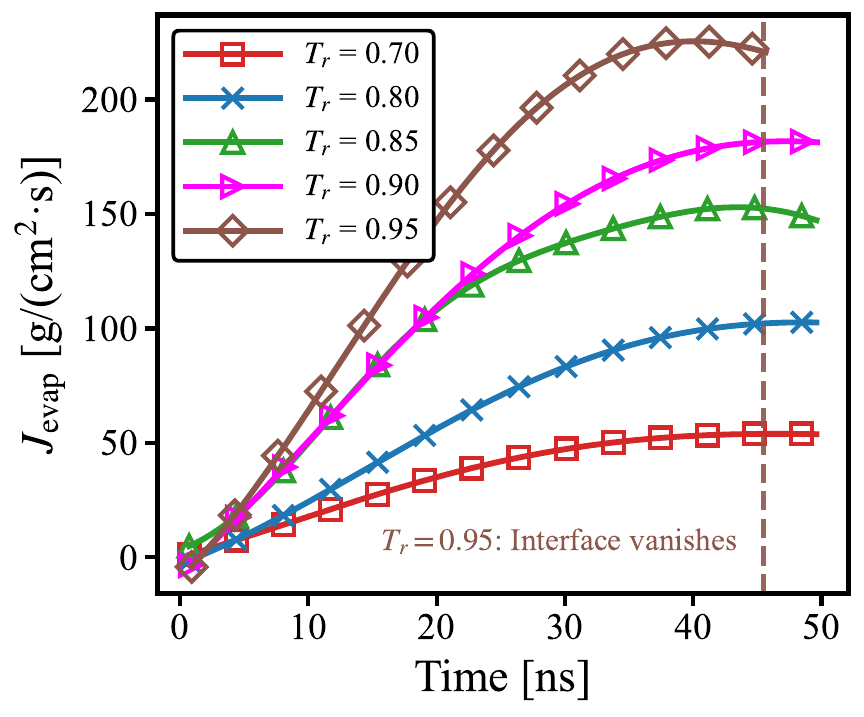}
        \caption{$J^{C_{12}}_{{\text{evap}}}$}
        \label{fig:J_Evap_C12}
    \end{subfigure} \\
    \begin{subfigure}[b]{0.41\linewidth}
        \includegraphics[width=\linewidth]{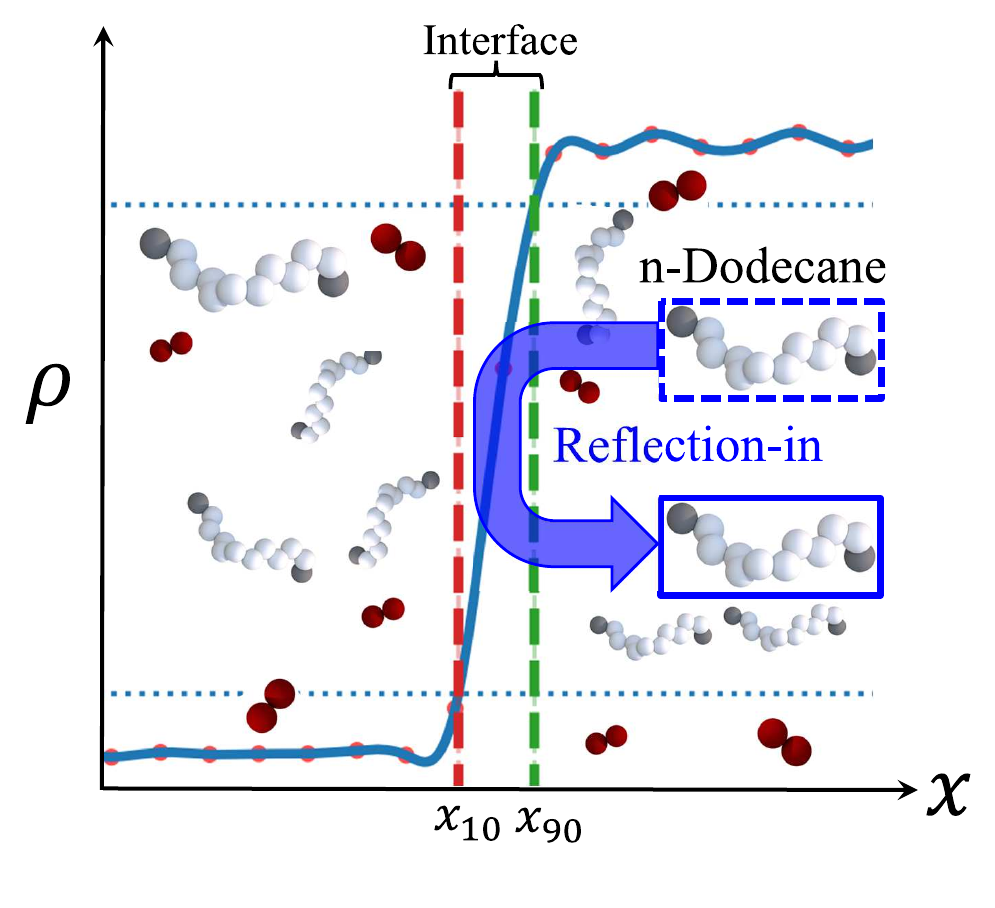}
        \caption{Schematic view of reflection-in}
        \label{fig:cond_flux_two_boundary}
    \end{subfigure}
    \begin{subfigure}[b]{0.42\linewidth}
        \includegraphics[width=\linewidth]{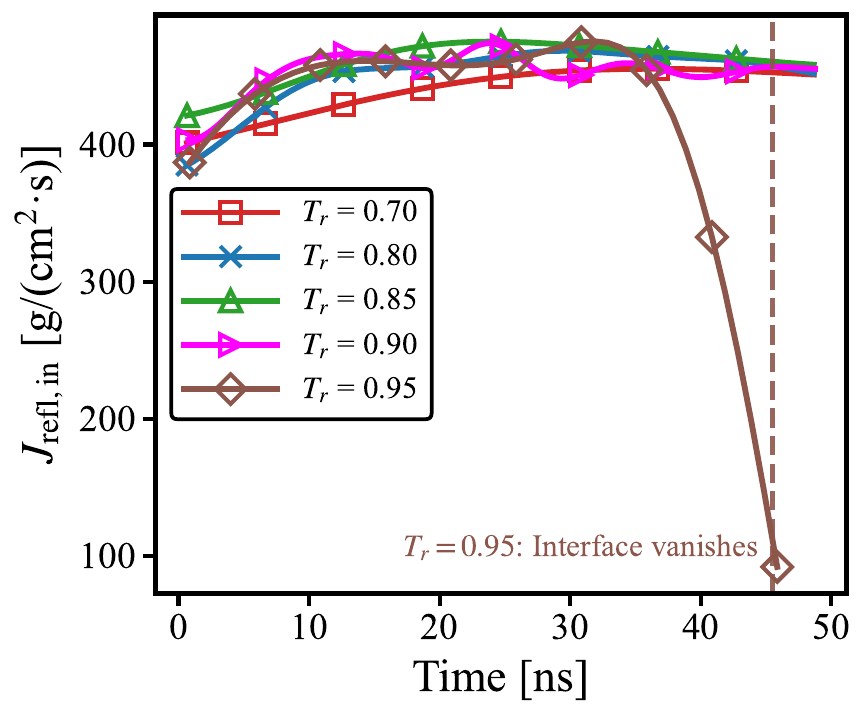}
        \caption{$J^{C_{12}}_{\text{ref, in}}$}
        \label{fig:J_Ref_In_C12}
    \end{subfigure} 
    \begin{subfigure}[b]{0.41\linewidth}
        \includegraphics[width=\linewidth]{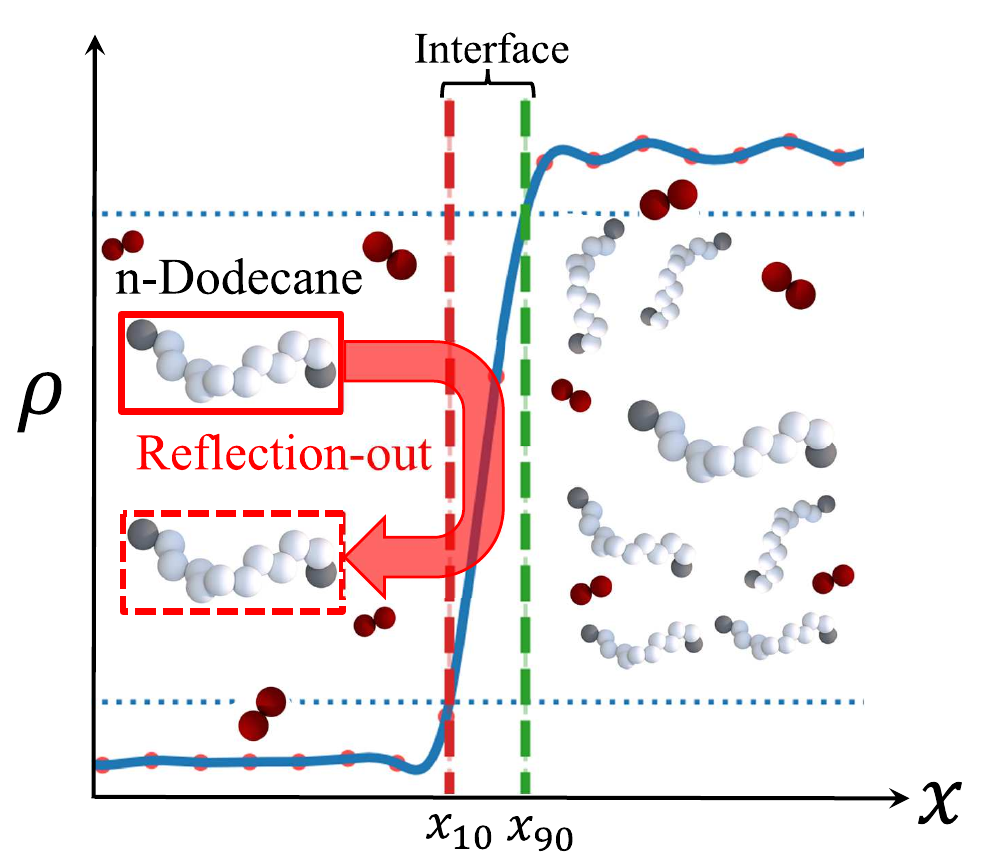}
        \caption{Schematic view of reflection-out}
        \label{fig:cond_flux_two_boundary}
    \end{subfigure}
    \begin{subfigure}[b]{0.42\linewidth}
        \includegraphics[width=\linewidth]{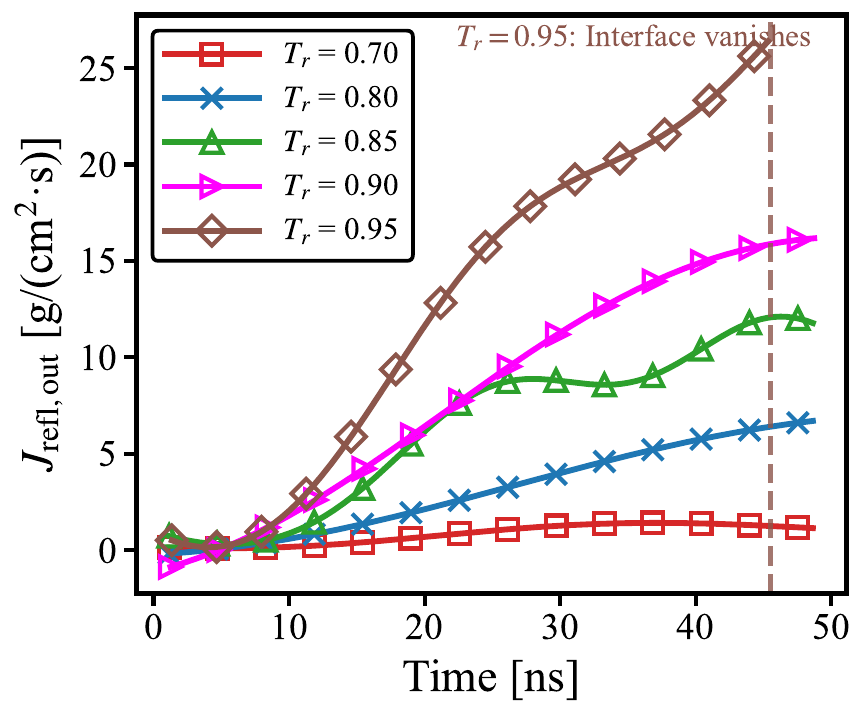}
        \caption{$J^{C_{12}}_{{\text{ref, out}}}$}
        \label{fig:J_Ref_Out_C12}
    \end{subfigure}
    \vspace{-7 pt}
\caption{Interfacial molecular mass transport of n-dodecane analyzed from MD simulations. 
The flux profiles were obtained directly from molecular dynamics data and smoothed to highlight the temporal trends. 
Each plot presents the evaporation, reflection-in, and reflection-out fluxes alongside their schematic representations.}
    \label{fig:C12_flux_summary_gpr}
\end{figure}

\begin{figure}[hbt!]
    \centering
    \begin{subfigure}[b]{0.41\linewidth}
        \includegraphics[width=\linewidth]{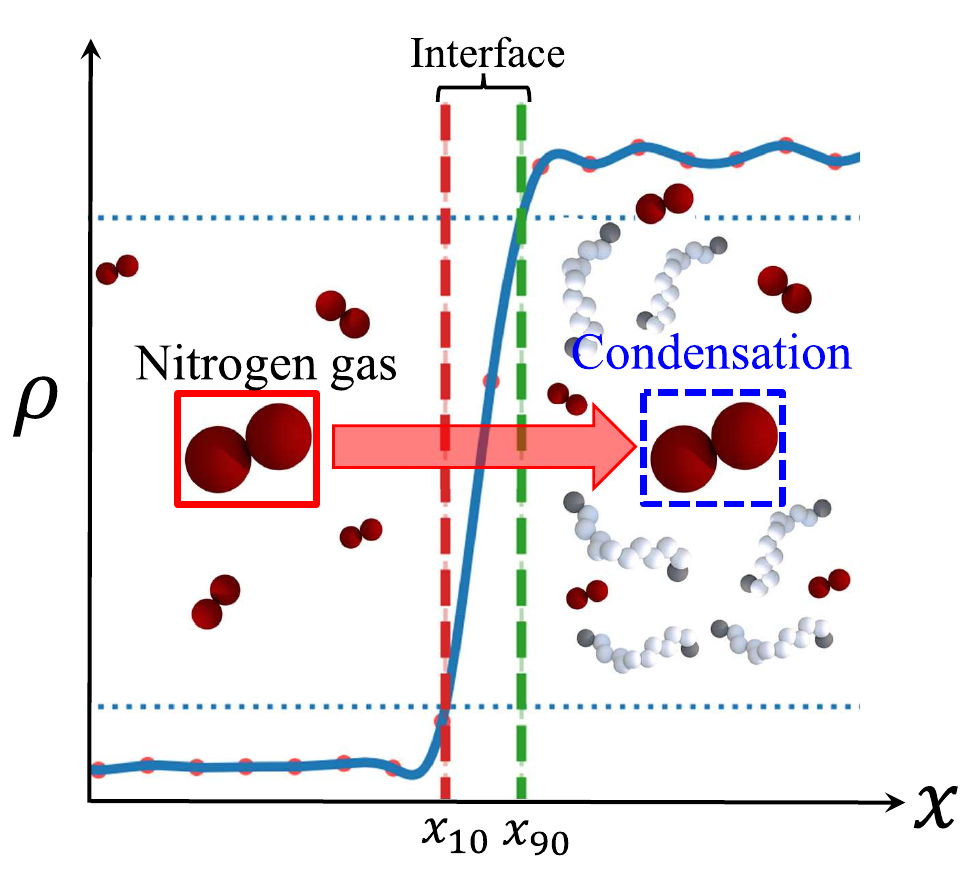}
        \caption{Schematic view of condensation}
        \label{fig:cond_flux_schematic}
    \end{subfigure}
    \begin{subfigure}[b]{0.42\linewidth}
        \includegraphics[width=\linewidth]{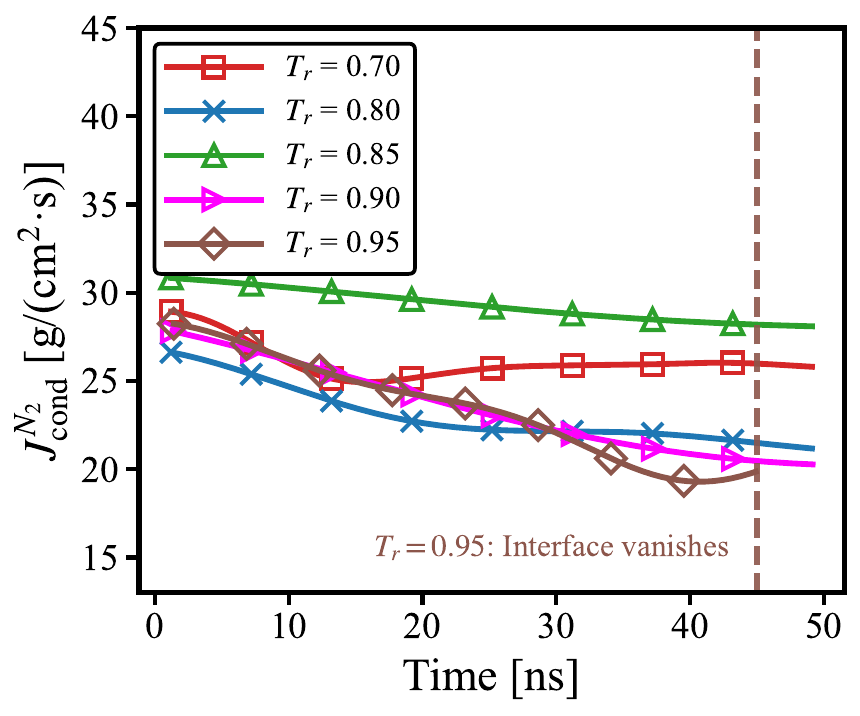}
        \caption{$J^{N_2}_{\mathrm{cond}}$}
        \label{fig:cond_flux_gpr}
    \end{subfigure} \\[6pt]
    \begin{subfigure}[b]{0.41\linewidth}
        \includegraphics[width=\linewidth]{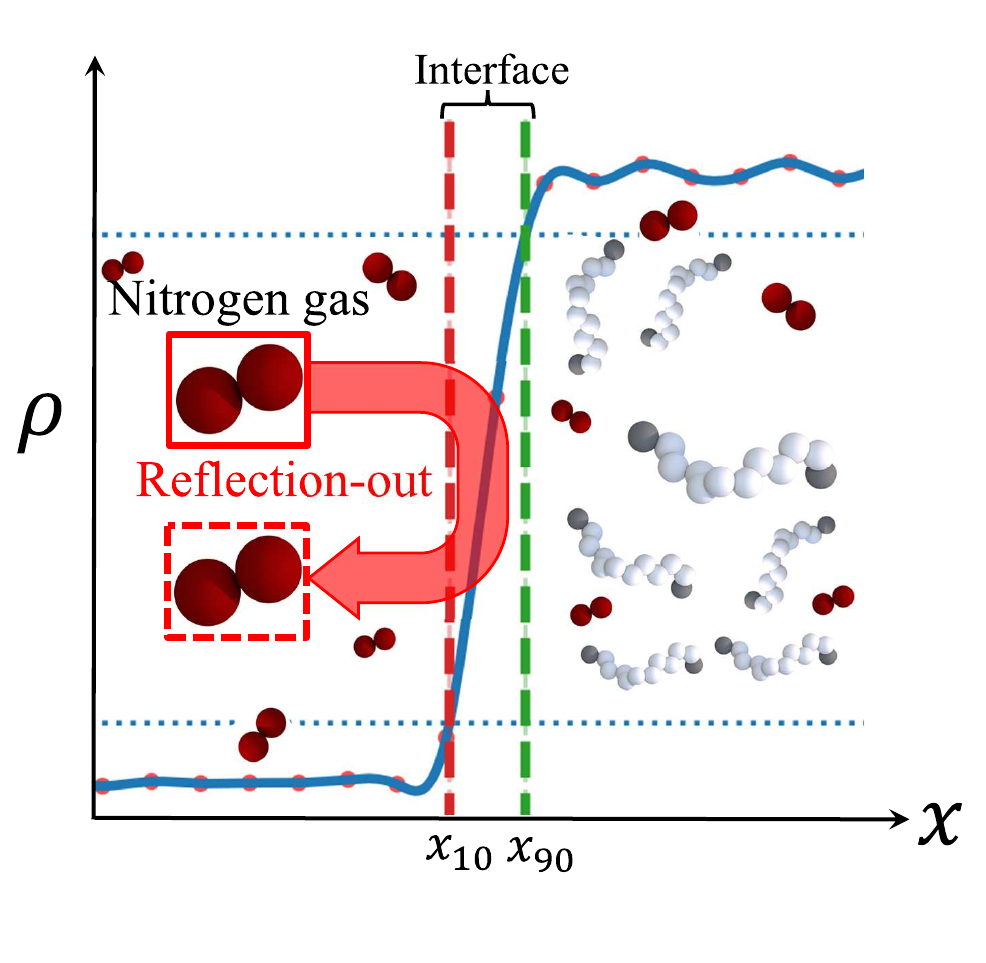}
        \caption{Schematic view of reflection-out}
        \label{fig:refl_flux_schematic}
    \end{subfigure}
    \begin{subfigure}[b]{0.42\linewidth}
        \includegraphics[width=\linewidth]{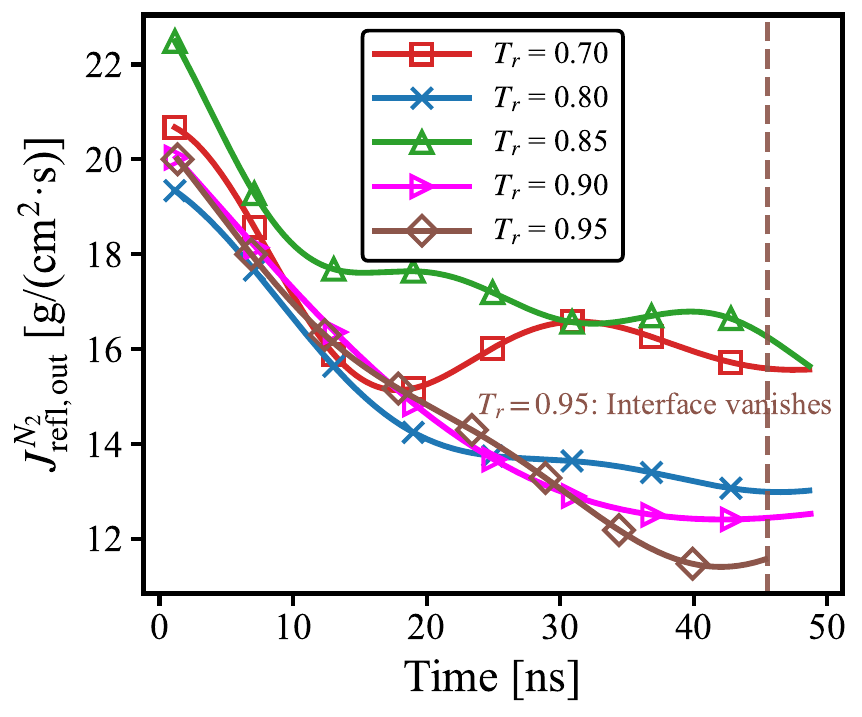}
        \caption{$J^{N_2}_{\mathrm{refl,out}}$}
        \label{fig:refl_flux_gpr}
    \end{subfigure}
    \vspace{-6pt}
\caption{Interfacial molecular mass transport of nitrogen (N$_2$) analyzed from MD simulations. 
Each pair of plots shows schematic views and the corresponding flux profiles for the condensation and reflection-out processes.}
    \label{fig:N2_flux_summary_gpr}
\end{figure}

\begin{figure}[hbt!]
    \centering
    \begin{subfigure}[H]{0.48\linewidth}
        \includegraphics[width=\linewidth]{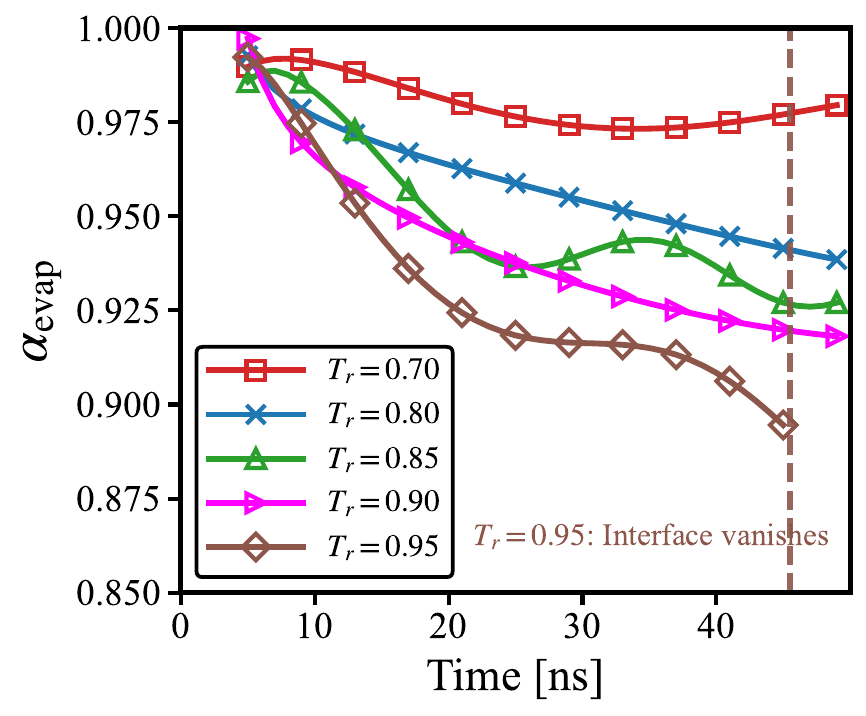}
        \subcaption{$\alpha_{\text{evap}}$}
    \end{subfigure}
    \begin{subfigure}[H]{0.48\linewidth}
        \includegraphics[width=\linewidth]{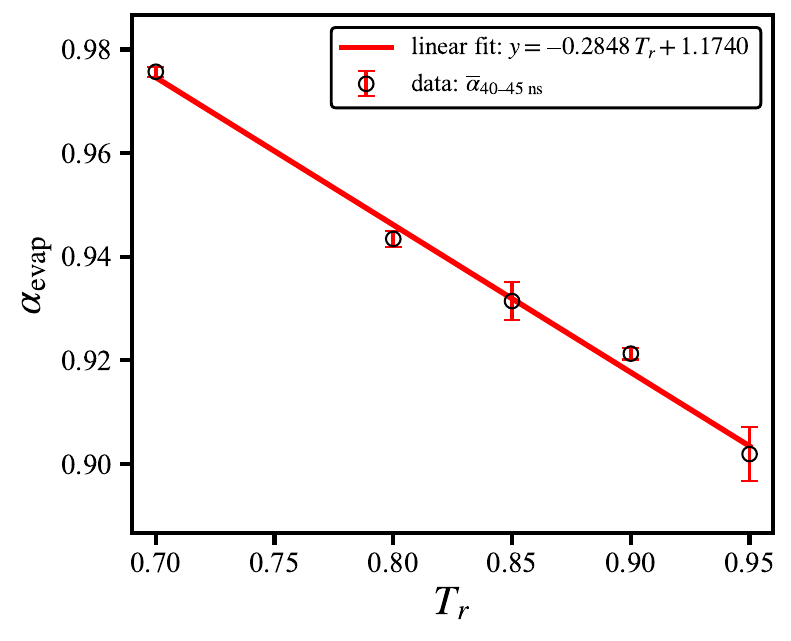}
        \subcaption{Linear regression}
    \end{subfigure}
    \caption{Evaporation coefficient analysis of the n-dodecane/nitrogen system at different reduced temperatures. 
    (a) shows the temporal variation of $\alpha_{\mathrm{evap}}$, 
    while (b) presents its averaged value in the $40$--$45~\mathrm{ns}$ range with a linear fit versus $T_r$.}
    \label{fig:evap_coeff_regression}
\end{figure}

Regarding the evaporation flux, Fig.~\ref{fig:C12_flux_summary_gpr}~(b) presents its temporal evolution for n-dodecane under varying ambient temperature conditions. In all cases, despite minor fluctuations due to MD noise, the evaporation flux generally increases over time and exhibits higher magnitudes at elevated ambient temperatures.

Figure~\ref{fig:C12_flux_summary_gpr}(d) and (f) show the reflection-related fluxes of C$_{12}$ at the
vapor-side boundary. As the liquid region warms and the interface widens, molecular motion becomes
more active within the transition layer, increasing the probability that outgoing molecules collide
and return toward the liquid. This trend becomes more pronounced at higher ambient temperatures,
where stronger evaporation enhances the frequency of intermolecular encounters near the interface.

The interfacial behavior of nitrogen is analyzed in 
Fig.~\ref{fig:N2_flux_summary_gpr}(b)--(d), which show the condensation
and reflection-out fluxes.  
For cases with $T_r \ge 0.85$, the condensation flux begins to increase
noticeably after $t \approx 15$~ns.  
This trend is consistent with the interface displacement observed in 
Fig.~\ref{fig:density_gradient_all_cases}, where the liquid region
contracts more rapidly at higher ambient temperatures.
As the interface moves toward the center of the domain, a larger fraction
of N$_2$ molecules from the vapor side reaches the dynamically shifting
boundary and enters the liquid region, leading to an enhanced condensation flux.

At the same time, the reflection-out flux in 
Fig.~\ref{fig:N2_flux_summary_gpr}(d) decreases gradually for all
temperature cases.
This reduction reflects the diminishing density gradient on the vapor side as
evaporation progresses: fewer N$_2$ molecules experience sufficient
collisions near the interfacial region to reverse direction and return to
the vapor phase.
The combined behavior of the condensation and reflection-out fluxes
captures how the interfacial transport of nitrogen becomes increasingly
governed by the motion and broadening of the interface under stronger
thermal driving.

The evaporation coefficient obtained from Eq.~\eqref{eq:evaporation_coeff} is
shown in Fig.~\ref{fig:evap_coeff_regression}(a)--(b).  
The coefficient $\alpha_{\mathrm{evap}}$ denotes the fraction of the outward
C$_{12}$ flux associated with true evaporation rather than interfacial
recrossing.  
For all test cases, $\alpha_{\mathrm{evap}}$ starts near unity at early times,
indicating that nearly all outward-moving C$_{12}$ molecules successfully enter
the vapor phase during the initial stages of heating.  
As the simulation progresses, $\alpha_{\mathrm{evap}}$ decreases monotonically,
with a larger reduction observed at higher reduced temperatures.  
By the end of the simulation window ($t \approx 45$~ns), the coefficient ranges
from approximately 0.97 for $T_r=0.70$ to about 0.87 for $T_r=0.95$.

Figure~\ref{fig:evap_coeff_regression}(b) summarizes the dependence of the
late-time evaporation coefficient on the reduced temperature.  
The averaged values exhibit a nearly linear decrease with increasing $T_r$, and
are well described by the regression model
\begin{equation}
    \alpha(T_r) = -0.2848\,T_r + 1.1740 .
\end{equation}
This relation provides a convenient parametrization of the evaporation
coefficient for use in continuum-scale boundary conditions, particularly when
modeling multicomponent evaporation under elevated-temperature conditions where
interfacial recrossing becomes more prevalent.
Compared with the fixed boundary formulation, the two-boundary method captures
interfacial mass transfer by following individual molecular crossings of the
time-dependent vapor and liquid interfaces.  
This classification into evaporation, reflection, and condensation yields a more
consistent description of interfacial transport, especially at elevated temperatures
where the transition region broadens and fixed control volumes no longer represent
the interface accurately.

\section{Conclusion and future work}
\label{sec6:conclusion}
MD simulations were used to study the vapor and liquid interfacial properties of a Type-III binary mixture of n-dodecane and nitrogen under net evaporation conditions. The non-equilibrium MD simulations carried out shed light on temperature dependence of macroscopic properties like surface tension, density gradient, interfacial thickness. In addition to looking at macroscopic properties, the main goal of the study was to lay out a framework to estimate the molecular mass flux through the interface. To that end, two methods were used. Firstly, a fixed boundary method which estimates the net mass flux in either direction (evaporating or condensing). Secondly, a two-boundary method which was used to count the evaporation and outgoing reflecting n-dodecane molecules across the interface thickness. The outgoing reflection flux of n-dodecane molecules increased with temperature as the interface broadened and chances of collision increased. The evaporation coefficient estimated using these interfacial fluxes was observed to increase as the ambient temperature shot up. The same methodology can also be extended to calculate the condensing molecular flux and reflecting ambient molecules at any chosen location of interface.
The framework presented in chapters 3 and 4, specifically calculating velocity distribution functions and mass flux transport coefficients, can be used in conjunction to propose kinetic boundary conditions. Simulations like these helps to develop a better understanding of vapor and liquid interface of multi-component systems under non-equilibrium conditions. To that end, we are currently using this developed framework to work on a joint project with Dr. Grazia Lamanna’s research group at University of Stuttgart. The work revolves around developing a hybrid MD and kinetic theory approach to better the existing evaporation and condensation models. The simulation configuration of these simulations are different, as they involve spherical droplets of n-heptane and n-dodecane in an ambient of nitrogen.

\section*{Acknowledgments}
This work was supported by the National Science Foundation (NSF).

\section*{Declaration of Generative AI and AI-assisted technologies in the writing process}
During the preparation of this work the authors used ChatGPT and DeepL Write in order to proofread the manuscript and enhance its readability. After using this tool/service, the authors reviewed and edited the content as needed and take full responsibility for the content of the publication.

\bibliographystyle{abbrv}
\bibliography{mybib}


\section*{Appendix: Gaussian process regression–based uncertainty quantification of interfacial dynamics and molecular mass fluxes}

\begin{figure}[hbt!]
    \centering
    \begin{subfigure}[hbtp!]{0.9\linewidth}
        \includegraphics[width=\linewidth]{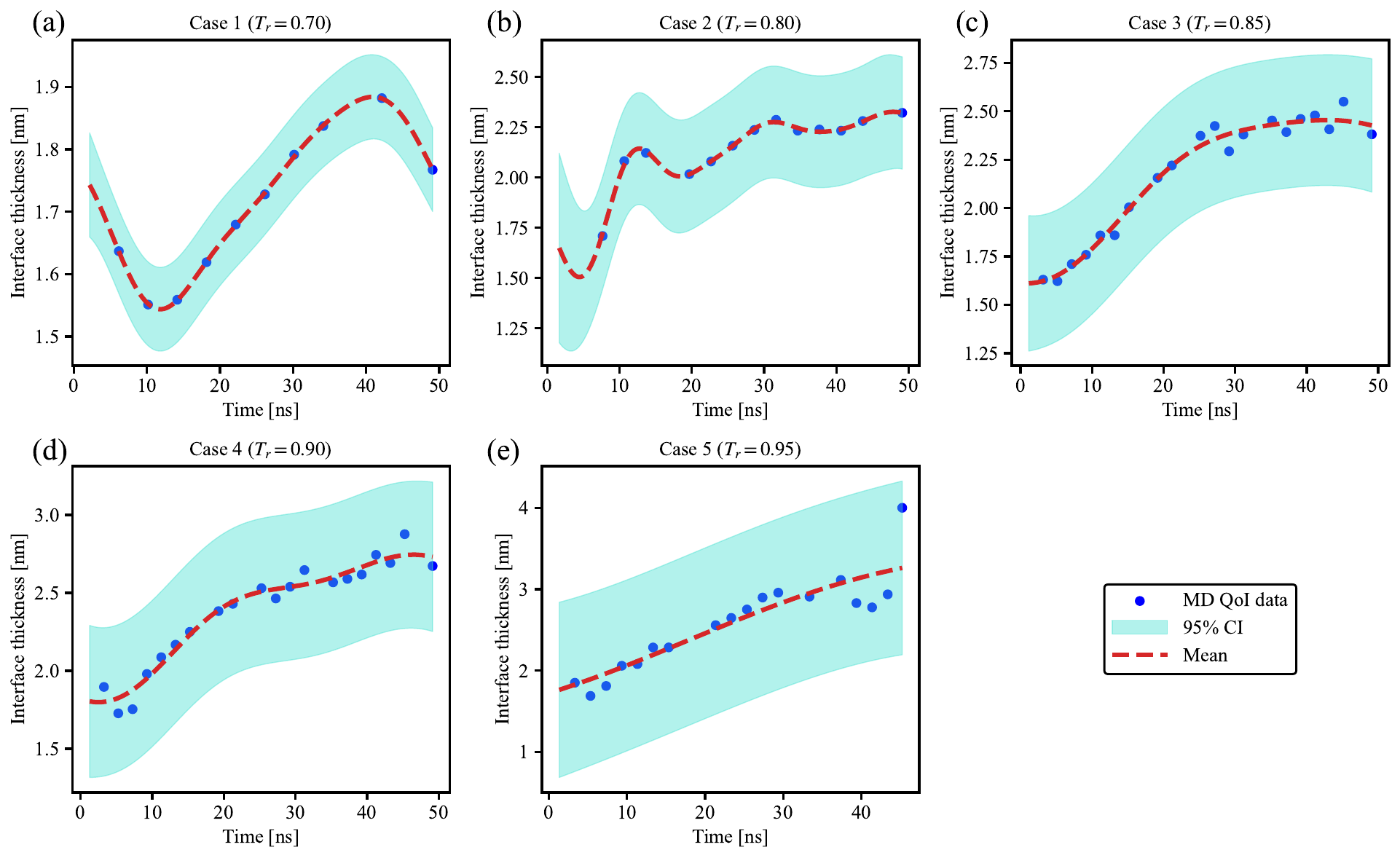}
    \end{subfigure}
    \caption{Interface thickness: The shaded regions indicate 95\% confidence intervals obtained from GPR, 
    and the red dashed lines represent the predicted mean trends.}
    \label{fig:interface_thickness_gpr}
\end{figure}

\begin{figure}[hbt!]
    \centering
    \begin{subfigure}[hbtp!]{0.9\linewidth}
        \includegraphics[width=\linewidth]{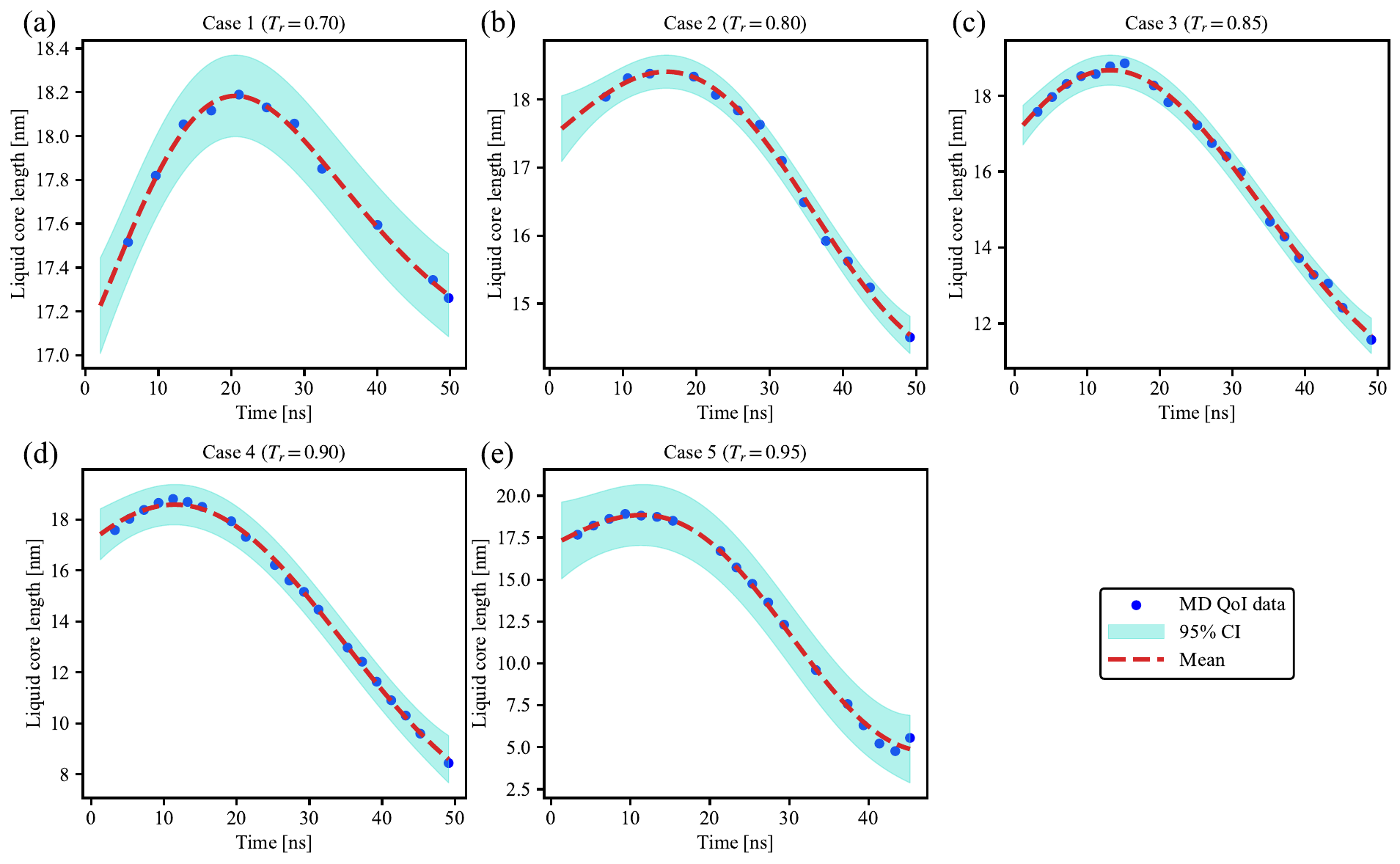}
    \end{subfigure}
    \caption{Liquid core length: The shaded regions denote 95\% confidence intervals obtained from GPR, 
    and the red dashed lines represent the predicted mean trends. 
    The core region was defined as the distance between the left and right interfacial centroids determined from the 10--90\% density transition.}
    \label{fig:liquid_core_length_gpr}
\end{figure}

\begin{figure}[hbt!]
    \centering
    \begin{subfigure}[hbtp!]{0.9\linewidth}
        \includegraphics[width=\linewidth]{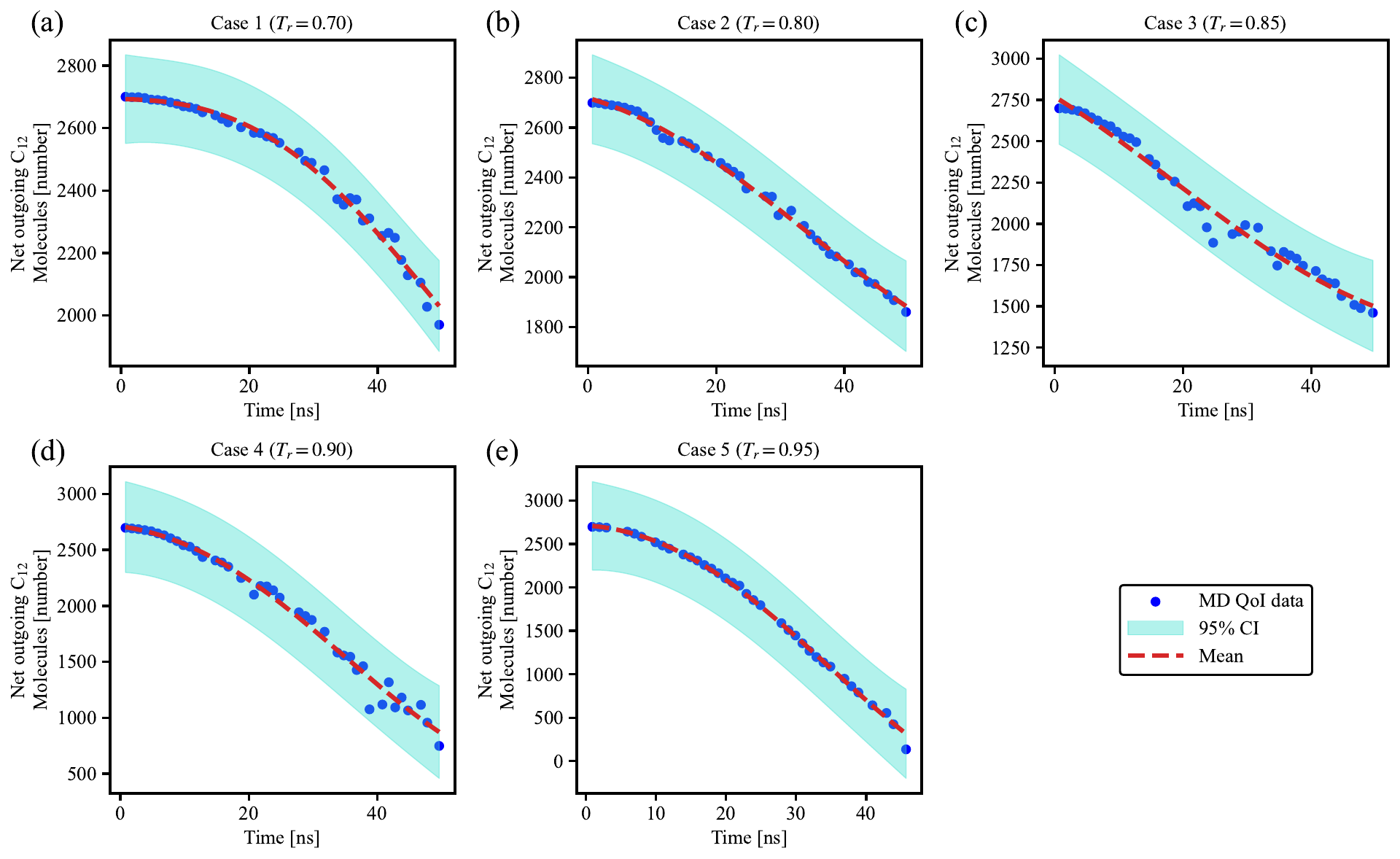}
    \end{subfigure}
    \caption{Time evolution of the number of C$_{12}$ molecules within the fixed control volume.
    The shaded regions denote the 95\% confidence intervals obtained from GPR, 
    and the red dashed lines represent the predicted mean trends.}
    \label{fig:C12_Number_GPR_fixed_2}
\end{figure}

\begin{figure}[hbt!]
    \centering
    \begin{subfigure}[hbtp!]{0.9\linewidth}
        \includegraphics[width=\linewidth]{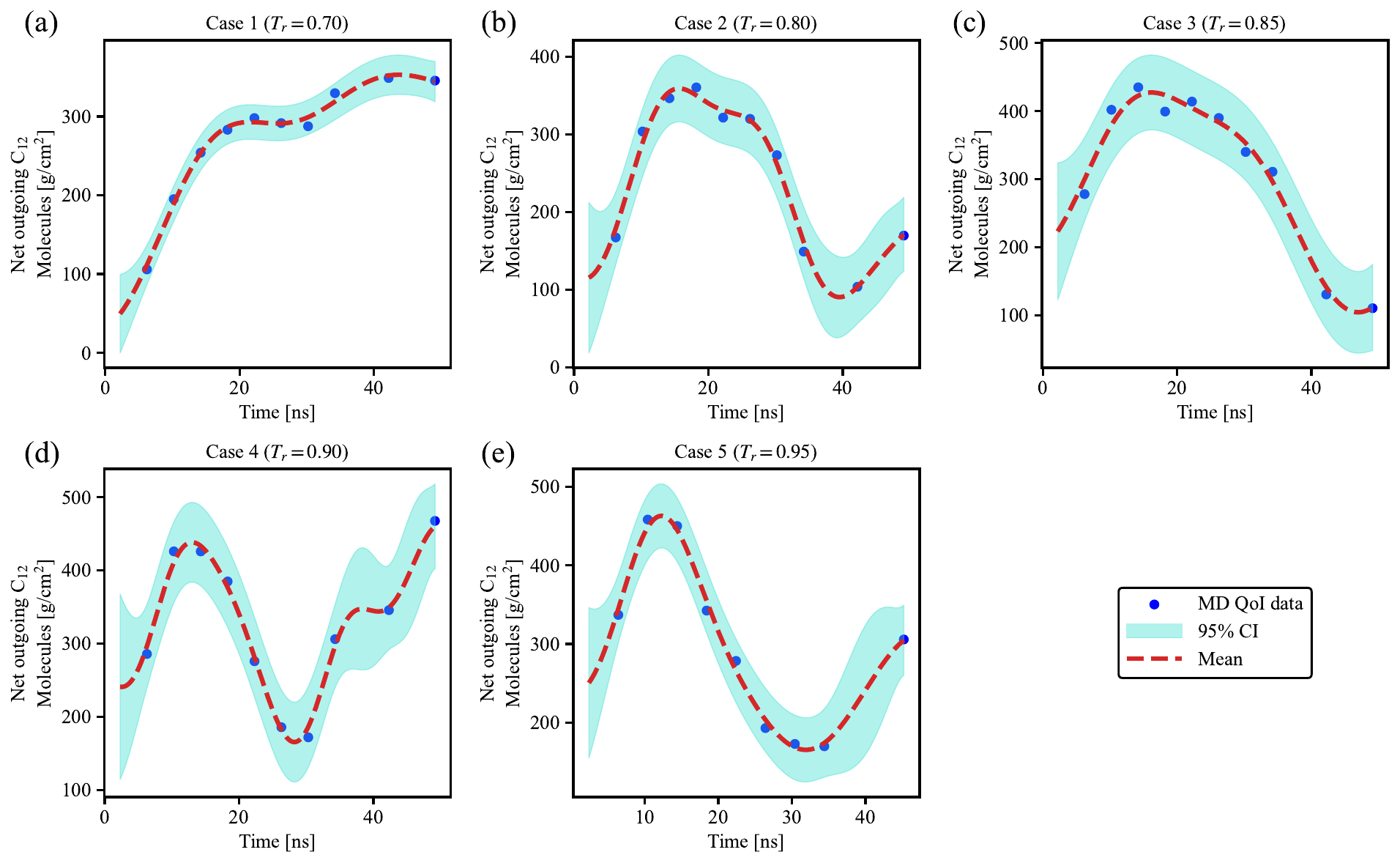}
    \end{subfigure}
    \caption{Evaporation flux by fixed box method: The shaded regions denote the 95\% confidence intervals obtained from GPR, 
    and the red dashed lines represent the predicted mean evaporation trends over time. 
    The flux values were derived from the molecular crossing events at the vapor–liquid interface in the n-dodecane/nitrogen system.}
    \label{fig:evaporation_flux_gpr}
\end{figure}

\begin{figure}[hbt!]
    \centering
    \begin{subfigure}[hbtp!]{0.9\linewidth}
        \includegraphics[width=\linewidth]{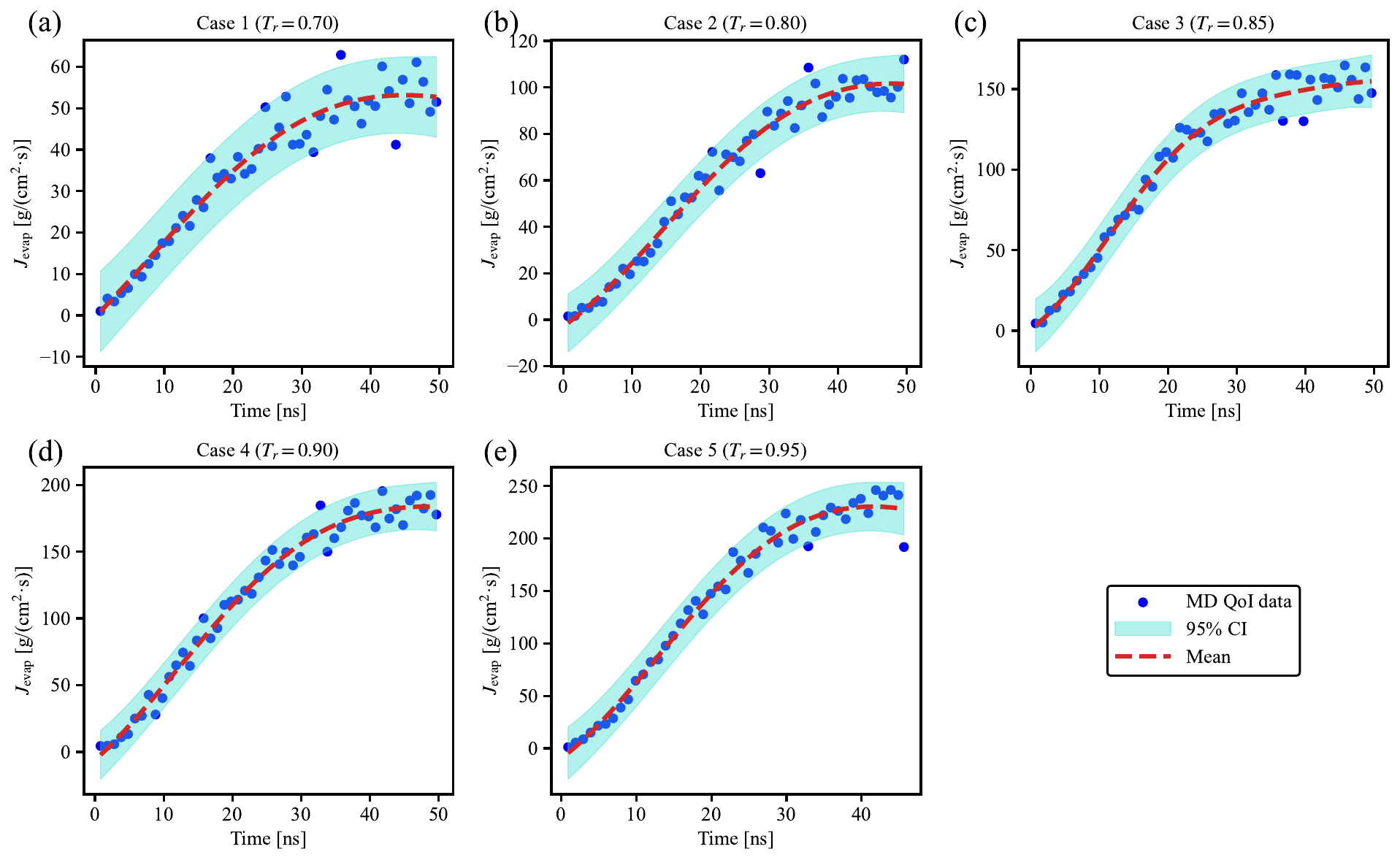}
    \end{subfigure}
    \caption{Evaporation flux estimated by the two-boundary method for n-dodecane.
    The red dashed lines indicate the GPR-predicted mean trends over time, and the shaded regions represent the 95\% confidence intervals.
    The flux values were obtained by counting molecular crossing events across the vapor--liquid interface in the n-dodecane/nitrogen system.}
    \label{fig:evap_flux_two_boundary}
\end{figure}

\begin{figure}[hbt!]
    \centering
    \begin{subfigure}[hbtp!]{0.9\linewidth}
        \includegraphics[width=\linewidth]{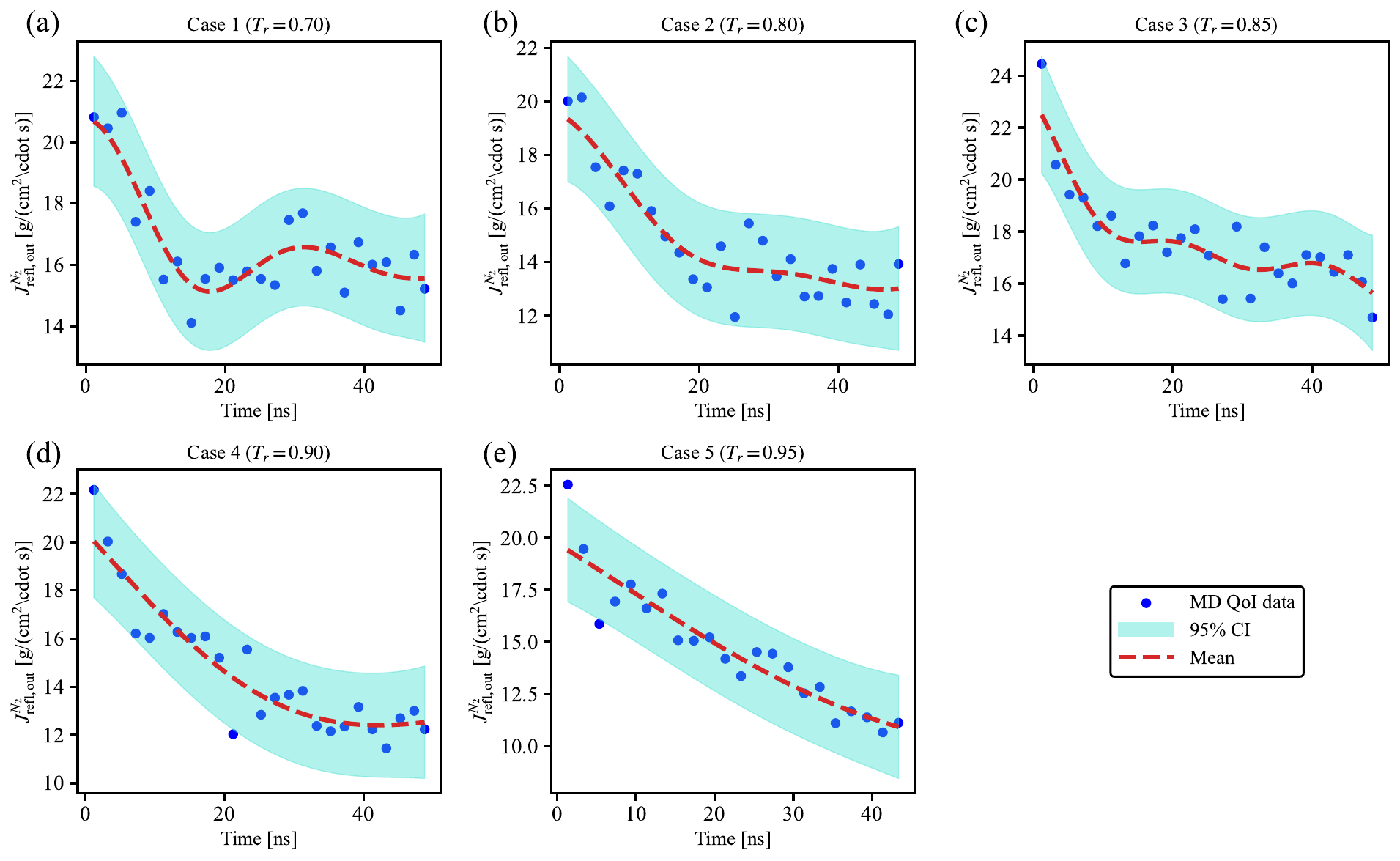}
    \end{subfigure}
    \caption{Reflection-out flux of nitrogen (N$_2$) estimated using the two-boundary method.
    The red dashed lines indicate the GPR-predicted mean trends over time, and the shaded regions denote the 95\% confidence intervals.
    The flux values were calculated by tracking nitrogen molecules reflected from the vapor--liquid interface back into the vapor phase in the n-dodecane/nitrogen system.}
    \label{fig:ref_out_flux_n2_two_boundary}
\end{figure}

This appendix presents the Gaussian Process--based uncertainty quantification (UQ)
framework developed for post-processing the NEMD data associated with interfacial
transport phenomena. The primary objective is to provide a statistically consistent
characterization of the temporal evolution of several key quantities of interest (QoIs),
including the interfacial thickness shown in Fig.~\ref{fig:interface_thickness_gpr},
the liquid core length in Fig.~\ref{fig:liquid_core_length_gpr}, the molecular
population within the fixed boundaries in Fig.~\ref{fig:C12_Number_GPR_fixed_2}, and
the molecular mass fluxes obtained from both the fixed-box method
(Fig.~\ref{fig:evaporation_flux_gpr}) and the two-boundary method
(Figs.~\ref{fig:evap_flux_two_boundary} and \ref{fig:ref_out_flux_n2_two_boundary}).

Instantaneous flux signals extracted from NEMD trajectories exhibit significant
stochastic fluctuations arising from finite sampling, thermal agitation, and the
dynamic motion of the diffuse interface.  
To mitigate this noise, GPR is employed as a
nonparametric Bayesian surrogate that infers smooth posterior mean trends
$\mu(t)$ while simultaneously estimating the predictive uncertainty
through the time-dependent variance $\sigma^2(t)$.  
For each QoI, the GPR model yields a posterior mean and a corresponding
95\% confidence interval, $\mu(t) \pm 1.96\,\sigma(t)$, visualized as shaded
bands in the figures throughout this appendix.
These intervals quantify the statistical variability of molecular-scale processes,
allowing a rigorous interpretation of interfacial behavior beyond instantaneous
fluctuating signals.

The regression quality for each QoI is summarized in
Tables~\ref{table:gpr_liquid_core}--\ref{table:gpr_ref_out_flux_n2_two_boundary}
using standard metrics such as the mean-squared error (MSE), mean absolute error (MAE),
and coefficient of determination ($R^2$).  
These results demonstrate that the GPR models consistently provide accurate and
stable reconstructions across all thermodynamic conditions,
capturing both the smooth temporal evolution and the uncertainty structure of the
interfacial dynamics.  
The probabilistic representation established here offers a pathway for
propagating molecular-level variability into continuum-scale simulations,
enabling more robust and data-informed boundary conditions for CFD models of
multicomponent evaporation and condensation.

\begin{figure}[hbt!]
    \centering
    \begin{subfigure}[hbtp!]{0.9\linewidth}
        \includegraphics[width=\linewidth]{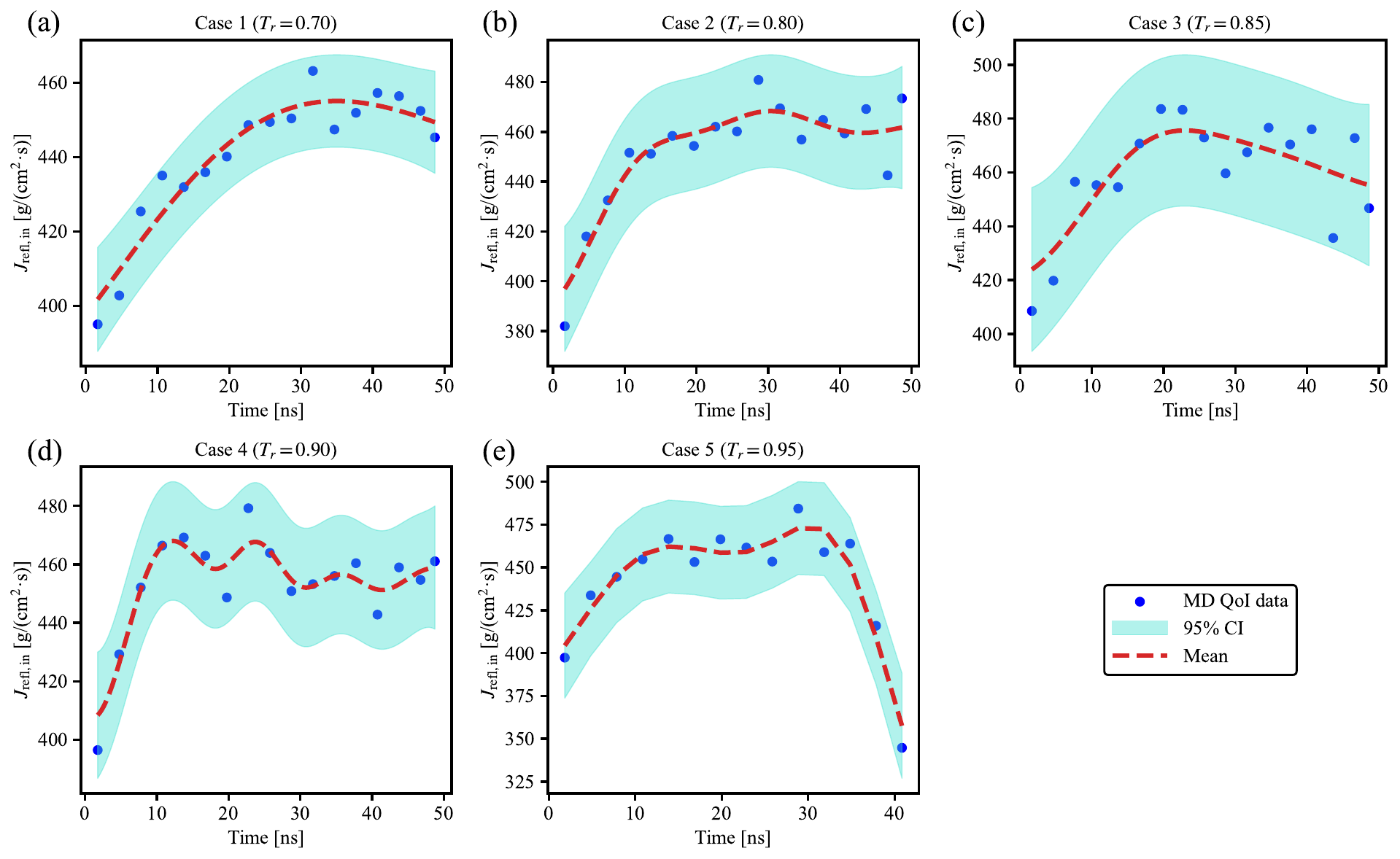}
    \end{subfigure}
    \caption{Reflection-in flux of n-dodecane estimated using the two-boundary method.
    The red dashed lines indicate the GPR-predicted mean trends over time, and the shaded regions denote the 95\% confidence intervals.
    The flux values were calculated by tracking n-dodecane molecules crossing from the vapor phase into the liquid region in the n-dodecane/nitrogen system.}
    \label{fig:ref_in_flux_two_boundary}
\end{figure}

\begin{figure}[hbt!]
    \centering
    \begin{subfigure}[hbtp!]{0.9\linewidth}
        \includegraphics[width=\linewidth]{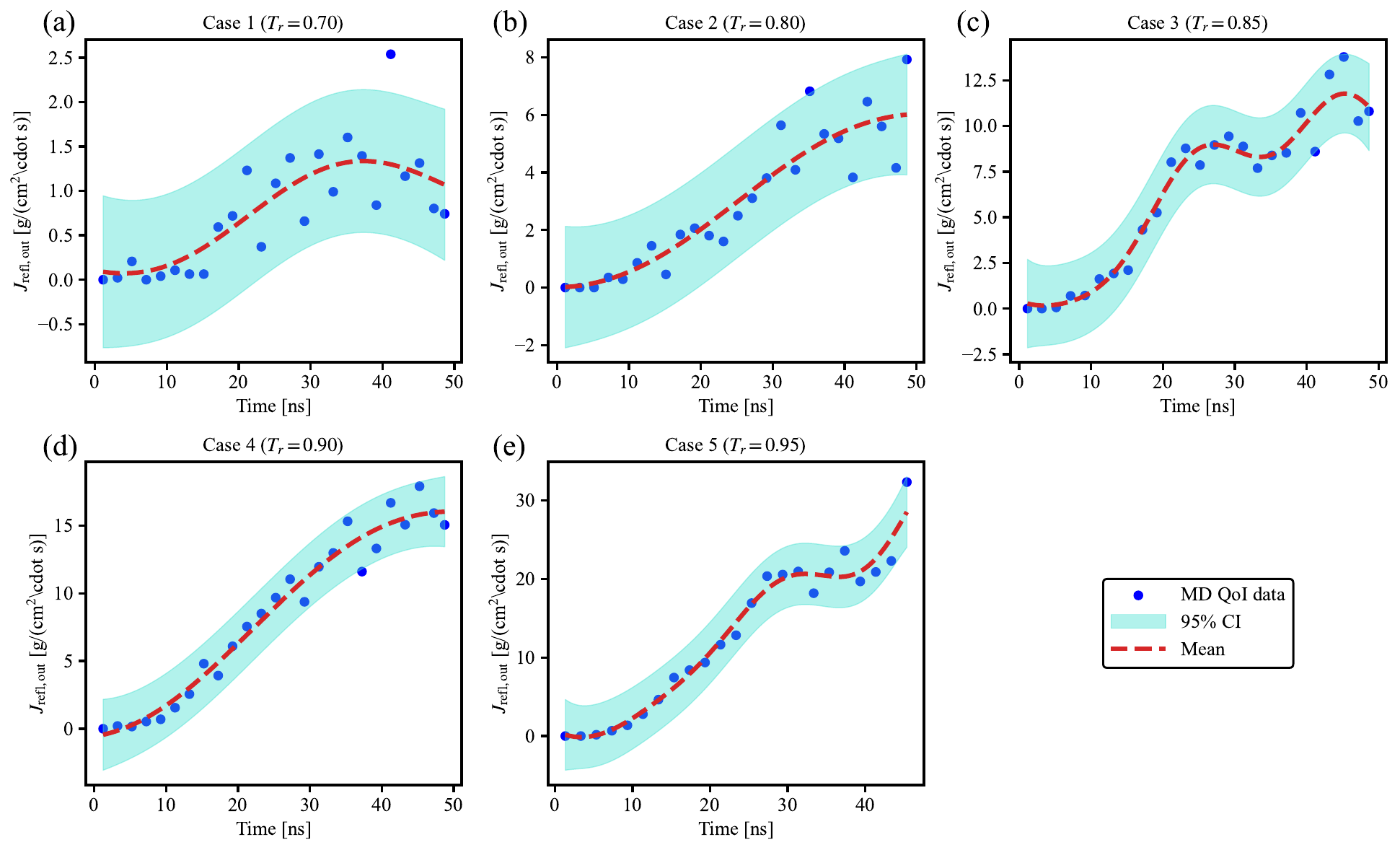}
    \end{subfigure}
    \caption{Reflection-out flux of n-dodecane estimated using the two-boundary method.
    The red dashed lines indicate the GPR-predicted mean trends over time, and the shaded regions denote the 95\% confidence intervals.
    The flux values were calculated by tracking n-dodecane molecules reflected back from the vapor--liquid interface into the vapor region in the n-dodecane/nitrogen system.}
    \label{fig:ref_out_flux_two_boundary}
\end{figure}

\begin{figure}[hbt!]
    \centering
    \begin{subfigure}[hbtp!]{0.9\linewidth}
        \includegraphics[width=\linewidth]{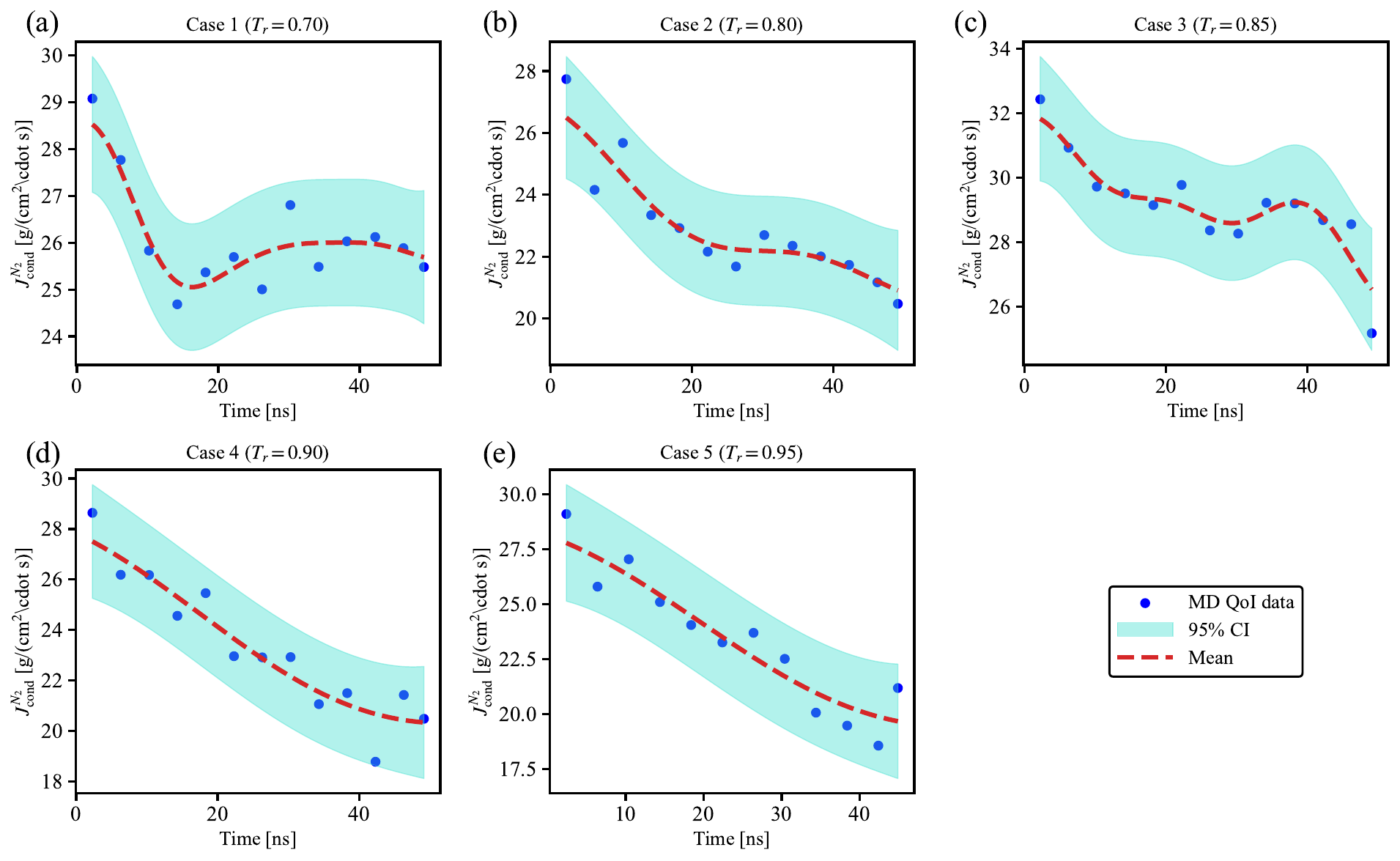}
    \end{subfigure}
    \caption{Condensation flux of nitrogen (N$_2$) estimated using the two-boundary method.
    The red dashed lines indicate the GPR-predicted mean trends over time, and the shaded regions denote the 95\% confidence intervals.
    The flux values were calculated by tracking nitrogen molecules crossing from the vapor phase into the liquid region in the n-dodecane/nitrogen system.}
    \label{fig:cond_flux_n2_two_boundary}
\end{figure}

\clearpage

\newpage

\begin{table}[hbtp!]
\renewcommand{\arraystretch}{1.05}
\centering
\caption{Regression performance for the liquid core length of the n-dodecane/nitrogen system obtained from GPR.}
\begin{tabular}{l ccc}
\hline
Case & MSE & MAE & $R^2$ \\
\hline
Case 1 & 0.001926 & 0.037102 & 0.976717 \\
Case 2 & 0.005299 & 0.058403 & 0.996779 \\
Case 3 & 0.010605 & 0.078638 & 0.998055 \\
Case 4 & 0.015792 & 0.104383 & 0.998627 \\
Case 5 & 0.061458 & 0.163599 & 0.997806 \\
\hline
\end{tabular}
\label{table:gpr_liquid_core}
\end{table}

\begin{table}[hbtp!]
\renewcommand{\arraystretch}{1.05}
\centering
\caption{Regression performance for the interface thickness of the n-dodecane/nitrogen system obtained from GPR.}
\begin{tabular}{l ccc}
\hline
Case & MSE & MAE & $R^2$ \\
\hline
Case 1 & 0.000002 & 0.001245 & 0.999817 \\
Case 2 & 0.000089 & 0.007474 & 0.996344 \\
Case 3 & 0.001912 & 0.033727 & 0.980031 \\
Case 4 & 0.004072 & 0.051873 & 0.962739 \\
Case 5 & 0.056332 & 0.157315 & 0.816416 \\
\hline
\end{tabular}
\label{table:gpr_interface_thickness}
\end{table}

\begin{table}[hbtp!]
\renewcommand{\arraystretch}{1.05}
\centering
\caption{Regression performance for the molecular count of the n-dodecane/nitrogen system obtained from GPR.}
\begin{tabular}{l ccc}
\hline
Case & MSE & MAE & $R^2$ \\
\hline
Case 1 & 138.198351 & 7.895030 & 0.996967 \\
Case 2 & 77.902067 & 7.504591 & 0.998715 \\
Case 3 & 973.972757 & 21.180434 & 0.992618 \\
Case 4 & 192.000070 & 11.153437 & 0.999448 \\
Case 5 & 63.716490 & 6.611614 & 0.999880 \\
\hline
\end{tabular}
\label{table:gpr_molecular_count}
\end{table}

\begin{table}[hbtp!]
\renewcommand{\arraystretch}{1.05}
\centering
\caption{Regression performance for the evaporation flux by fixed box method for n-dodecane obtained from GPR.}
\begin{tabular}{l ccc}
\hline
Case & MSE & MAE & $R^2$ \\
\hline
Case 1 & 0.117974 & 0.294379 & 0.999976 \\
Case 2 & 86.757479 & 8.032648 & 0.989121 \\
Case 3 & 234.729630 & 13.599301 & 0.980753 \\
Case 4 & 86.781509 & 8.491977 & 0.990555 \\
Case 5 & 69.723062 & 7.440809 & 0.993453 \\
\hline
\end{tabular}
\label{table:gpr_evaporation_flux}
\end{table}

\begin{table}[hbtp!]
\renewcommand{\arraystretch}{1.05}
\centering
\caption{Regression performance for the evaporation flux of the n-dodecane/nitrogen system obtained from GPR using the two-boundary method.}
\begin{tabular}{l ccc}
\hline
Case & MSE & MAE & $R^2$ \\
\hline
Case 1 & 19.051219 & 3.152099 & 0.937055 \\
Case 2 & 30.255707 & 4.091664 & 0.974893 \\
Case 3 & 48.709394 & 5.489531 & 0.980138 \\
Case 4 & 63.146182 & 6.364353 & 0.983236 \\
Case 5 & 117.000707 & 8.016234 & 0.981409 \\
\hline
\end{tabular}
\label{table:gpr_evap_flux_two_boundary}
\end{table}

\begin{table}[hbtp!]
\renewcommand{\arraystretch}{1.05}
\centering
\caption{Regression performance for the reflection-in flux of the n-dodecane/nitrogen system obtained from GPR using the two-boundary method.}
\begin{tabular}{l ccc}
\hline
Case & MSE & MAE & $R^2$ \\
\hline
Case 1 & 27.621918 & 4.379525 & 0.914116 \\
Case 2 & 68.697298 & 6.347815 & 0.868342 \\
Case 3 & 128.661533 & 9.663576 & 0.702006 \\
Case 4 & 36.172143 & 4.444343 & 0.886066 \\
Case 5 & 76.918433 & 7.827930 & 0.935729 \\
\hline
\end{tabular}
\label{table:gpr_ref_in_flux_two_boundary}
\end{table}

\begin{table}[hbtp!]
\renewcommand{\arraystretch}{1.05}
\centering
\caption{Regression performance for the reflection-out flux of the n-dodecane/nitrogen system obtained from GPR using the two-boundary method.}
\begin{tabular}{l ccc}
\hline
Case & MSE & MAE & $R^2$ \\
\hline
Case 1 & 0.130841 & 0.265529 & 0.674163 \\
Case 2 & 0.790421 & 0.619499 & 0.858285 \\
Case 3 & 0.703975 & 0.611341 & 0.960852 \\
Case 4 & 1.109477 & 0.824800 & 0.968328 \\
Case 5 & 2.313456 & 1.065821 & 0.973249 \\
\hline
\end{tabular}
\label{table:gpr_ref_out_flux_two_boundary}
\end{table}

\begin{table}[hbtp!]
\renewcommand{\arraystretch}{1.05}
\centering
\caption{Regression performance for the nitrogen (N$_2$) condensation flux of the n-dodecane/nitrogen system obtained from GPR using the two-boundary method.}
\begin{tabular}{l ccc}
\hline
Case & MSE & MAE & $R^2$ \\
\hline
Case 1 & 0.184892 & 0.350018 & 0.857803 \\
Case 2 & 0.450173 & 0.485476 & 0.873644 \\
Case 3 & 0.324326 & 0.410654 & 0.870258 \\
Case 4 & 0.719225 & 0.697571 & 0.898261 \\
Case 5 & 0.980816 & 0.903653 & 0.893153 \\
\hline
\end{tabular}
\label{table:gpr_cond_flux_n2_two_boundary}
\end{table}

\begin{table}[hbtp!]
\renewcommand{\arraystretch}{1.05}
\centering
\caption{Regression performance for the nitrogen (N$_2$) reflection-out flux of the n-dodecane/nitrogen system obtained from GPR using the two-boundary method.}
\begin{tabular}{l ccc}
\hline
Case & MSE & MAE & $R^2$ \\
\hline
Case 1 & 0.623620 & 0.671554 & 0.807463 \\
Case 2 & 0.837604 & 0.803987 & 0.832503 \\
Case 3 & 0.660111 & 0.648703 & 0.820624 \\
Case 4 & 0.950086 & 0.726188 & 0.867768 \\
Case 5 & 1.108016 & 0.745573 & 0.867901 \\
\hline
\end{tabular}
\label{table:gpr_ref_out_flux_n2_two_boundary}
\end{table}

\end{document}